\documentclass[11pt]{article}
\usepackage{amsmath,amsfonts,amssymb,graphicx,theorem,multirow}
\usepackage[usenames]{color}
\usepackage{pstricks,cite} 
\usepackage{caption}

\usepackage[backref=false]{hyperref}
\numberwithin{equation}{section}
\usepackage[capitalise,noabbrev]{cleveref} 
\hypersetup{
colorlinks=true,
citecolor=red,
linkcolor=darkblue,
urlcolor=darkblue
}

\long\def\ignore#1{}
\definecolor{darkblue}{rgb}{0,0,.8}
\definecolor{red}{rgb}{1,0,0}
\definecolor{purple}{rgb}{1,0,1}
\definecolor{coloroflink}{rgb}{0.7,0,1}
\definecolor{coloroflink}{rgb}{0.180392, 0.545098, 0.341176}
\definecolor{darkpurple}{rgb}{1,.2,1}
\definecolor{pink}{rgb}{1,.7,.7}
\definecolor{lightblue}{rgb}{.61,.61,1}
\definecolor{midblue}{rgb}{.7,.7,1}
\definecolor{lightlightblue}{rgb}{.9,.9,1}
\definecolor{lightestblue}{rgb}{.96,.96,1}
\definecolor{lightpurple}{rgb}{1,.65,1}
\definecolor{darkgreen}{rgb}{0.180392, 0.545098, 0.341176}
\definecolor{mygray}{rgb}{.75,.75,.75}
\definecolor{lightlightgray}{rgb}{.85,.85,.85}

\theorembodyfont{\itshape} 
\theoremheaderfont{\scshape}
\theoremstyle{plain}  

\newtheorem{Theoreme}{Theorem}
\newtheorem{Proposition}{Proposition}

\newcommand{\nc}{\newcommand}
\nc{\bib}{\bibitem}
\nc{\be}{\begin{equation}}
\nc{\ee}{\end{equation}}
\nc{\nn}{\nonumber\\ }

\nc{\proof}{{\scshape Proof \;}} 				
\nc{\eproof}{{\hfill \rule{0.5em}{0.5em}\medskip}}

\def\arxiv#1#2{\href{http://arxiv.org/abs/#1}{\textsf{arXiv:#1 #2}}}
\nc{\chit}{\raisebox{0.25ex}{$\chi$}}
\nc{\chih}{\raisebox{0.25ex}{$\hat\chi$}}
\nc{\ir}{\mathrm{i}}
\nc{\tl}{\mathsf{TL}}
\nc{\stan}{\mathsf{V}}

\nc{\ctwotimes}[1]{(\mathbb C^2)^{\otimes{\hspace{0.025cm}#1}}}
\nc{\Db}{\mbox{\boldmath $D$}}
\nc{\Jb}{\mbox{\boldmath $J$}}
\nc{\funkyDb}{\boldsymbol{\mathcal D}}
\nc{\Tb}{\mbox{\boldmath $T$}}
\nc{\Ib}{\mbox{\boldmath $I$}}
\def\ch{{\rm ch}}
\nc{\Cj}{C^{\phantom{\dagger}}}
\nc{\Cjd}{C^{\dagger}}
\nc{\Eta}{\eta^{\phantom{\dagger}}}
\nc{\Etad}{\eta^{\dagger}}
\nc{\Zeta}{\zeta^{\phantom{\dagger}}}
\nc{\Zetad}{\zeta^{\dagger}}
\nc{\PPhi}{\Phi^{\phantom{\dagger}}}
\nc{\PPhid}{\Phi^{\dagger}}
\nc{\PPsi}{\Psi^{\phantom{\dagger}}}
\nc{\PPsid}{\Psi^{\dagger}}
\nc{\pphi}{\phi^{\phantom{\dagger}}}
\nc{\pphid}{\phi^{\dagger}}
\nc{\ppsi}{\psi^{\phantom{\dagger}}}
\nc{\ppsid}{\psi^{\dagger}}
\nc{\Lf}{L^{\textrm{\tiny 1f}}}
\nc{\Lff}{L^{\textrm{\tiny 2f}}}
\nc{\LFF}{L^{\textrm{\tiny FF}}}
\nc{\LSF}{L^{\textrm{\tiny SF}}}
\nc{\Ldim}{L^{\textrm{\tiny dim}}}
\nc{\Deltaf}{\Delta^{\! \textrm{\tiny 1f}}}
\nc{\Deltaff}{\Delta^{\! \textrm{\tiny 2f}}}
\nc{\DeltaFF}{\Delta}

\renewcommand{\ge}{\geqslant}
\renewcommand{\le}{\leqslant}

\DeclareMathOperator{\arcsinh}{arcsinh}

\def\facegrid#1#2{
\psframe[fillstyle=solid,fillcolor=lightlightblue,linewidth=0pt]#1#2
\psgrid[gridlabels=0pt,subgriddiv=1]#1#2}

\def\loopa{
\psframe[linewidth=.25pt](0,0)(1,1)
\psarc[linewidth=1.5pt,linecolor=blue](1,0){.5}{90}{180}
\psarc[linewidth=1.5pt,linecolor=blue](0,1){.5}{-90}{0}
}
\def\loopb{
\psframe[linewidth=.25pt](0,0)(1,1)
\psarc[linewidth=1.5pt,linecolor=blue](0,0){.5}{0}{90}
\psarc[linewidth=1.5pt,linecolor=blue](1,1){.5}{180}{270}
}

\def\loopc{
\psarc[linewidth=1.5pt,linecolor=blue](0,0){.707}{45}{135}
\psarc[linewidth=1.5pt,linecolor=blue](0,2){.707}{-135}{-45}
}
\def\loopd{
\psarc[linewidth=1.5pt,linecolor=blue](1,1){.707}{135}{-135}
\psarc[linewidth=1.5pt,linecolor=blue](-1,1){.707}{-45}{45}
}

\def\dimvcoord{
\pspolygon[fillstyle=solid,linewidth=0pt,linecolor=white]
(-0.2,0)(-0.2,1.)(-0.199605,1.01256)(-0.198423,1.02507)(-0.196457,1.03748)(-0.193717,1.04974)(-0.190211,1.0618)(-0.185955,1.07362)(-0.180965,1.08516)(-0.175261,1.09635)(-0.168866,1.10717)(-0.161803,1.11756)(-0.154103,1.12748)(-0.145794,1.13691)(-0.136909,1.14579)(-0.127485,1.1541)(-0.117557,1.1618)(-0.107165,1.16887)(-0.0963507,1.17526)(-0.0851559,1.18097)(-0.0736249,1.18596)(-0.0618034,1.19021)(-0.049738,1.19372)(-0.0374763,1.19646)(-0.0250666,1.19842)(-0.0125581,1.19961)(0.,1.2)(0.0125581,1.19961)(0.0250666,1.19842)(0.0374763,1.19646)(0.049738,1.19372)(0.0618034,1.19021)(0.0736249,1.18596)(0.0851559,1.18097)(0.0963507,1.17526)(0.107165,1.16887)(0.117557,1.1618)(0.127485,1.1541)(0.136909,1.14579)(0.145794,1.13691)(0.154103,1.12748)(0.161803,1.11756)(0.168866,1.10717)(0.175261,1.09635)(0.180965,1.08516)(0.185955,1.07362)(0.190211,1.0618)(0.193717,1.04974)(0.196457,1.03748)(0.198423,1.02507)(0.199605,1.01256)(0.2,1.)(0.2,0.)(0.199605,-0.0125581)(0.198423,-0.0250666)(0.196457,-0.0374763)(0.193717,-0.049738)(0.190211,-0.0618034)(0.185955,-0.0736249)(0.180965,-0.0851559)(0.175261,-0.0963507)(0.168866,-0.107165)(0.161803,-0.117557)(0.154103,-0.127485)(0.145794,-0.136909)(0.136909,-0.145794)(0.127485,-0.154103)(0.117557,-0.161803)(0.107165,-0.168866)(0.0963507,-0.175261)(0.0851559,-0.180965)(0.0736249,-0.185955)(0.0618034,-0.190211)(0.049738,-0.193717)(0.0374763,-0.196457)(0.0250666,-0.198423)(0.0125581,-0.199605)(0.,-0.2)(-0.0125581,-0.199605)(-0.0250666,-0.198423)(-0.0374763,-0.196457)(-0.049738,-0.193717)(-0.0618034,-0.190211)(-0.0736249,-0.185955)(-0.0851559,-0.180965)(-0.0963507,-0.175261)(-0.107165,-0.168866)(-0.117557,-0.161803)(-0.127485,-0.154103)(-0.136909,-0.145794)(-0.145794,-0.136909)(-0.154103,-0.127485)(-0.161803,-0.117557)(-0.168866,-0.107165)(-0.175261,-0.0963507)(-0.180965,-0.0851559)(-0.185955,-0.0736249)(-0.190211,-0.0618034)(-0.193717,-0.049738)(-0.196457,-0.0374763)(-0.198423,-0.0250666)(-0.199605,-0.0125581)(-0.2,0.)
}

\def\dimhcoord{
\pspolygon[fillstyle=solid,linewidth=0pt,linecolor=white]
(0,-0.2)(1.,-0.2)(1.01256,-0.199605)(1.02507,-0.198423)(1.03748,-0.196457)(1.04974,-0.193717)(1.0618,-0.190211)(1.07362,-0.185955)(1.08516,-0.180965)(1.09635,-0.175261)(1.10717,-0.168866)(1.11756,-0.161803)(1.12748,-0.154103)(1.13691,-0.145794)(1.14579,-0.136909)(1.1541,-0.127485)(1.1618,-0.117557)(1.16887,-0.107165)(1.17526,-0.0963507)(1.18097,-0.0851559)(1.18596,-0.0736249)(1.19021,-0.0618034)(1.19372,-0.049738)(1.19646,-0.0374763)(1.19842,-0.0250666)(1.19961,-0.0125581)(1.2,0.)(1.19961,0.0125581)(1.19842,0.0250666)(1.19646,0.0374763)(1.19372,0.049738)(1.19021,0.0618034)(1.18596,0.0736249)(1.18097,0.0851559)(1.17526,0.0963507)(1.16887,0.107165)(1.1618,0.117557)(1.1541,0.127485)(1.14579,0.136909)(1.13691,0.145794)(1.12748,0.154103)(1.11756,0.161803)(1.10717,0.168866)(1.09635,0.175261)(1.08516,0.180965)(1.07362,0.185955)(1.0618,0.190211)(1.04974,0.193717)(1.03748,0.196457)(1.02507,0.198423)(1.01256,0.199605)(1.,0.2)(0.,0.2)(-0.0125581,0.199605)(-0.0250666,0.198423)(-0.0374763,0.196457)(-0.049738,0.193717)(-0.0618034,0.190211)(-0.0736249,0.185955)(-0.0851559,0.180965)(-0.0963507,0.175261)(-0.107165,0.168866)(-0.117557,0.161803)(-0.127485,0.154103)(-0.136909,0.145794)(-0.145794,0.136909)(-0.154103,0.127485)(-0.161803,0.117557)(-0.168866,0.107165)(-0.175261,0.0963507)(-0.180965,0.0851559)(-0.185955,0.0736249)(-0.190211,0.0618034)(-0.193717,0.049738)(-0.196457,0.0374763)(-0.198423,0.0250666)(-0.199605,0.0125581)(-0.2,0.)(-0.199605,-0.0125581)(-0.198423,-0.0250666)(-0.196457,-0.0374763)(-0.193717,-0.049738)(-0.190211,-0.0618034)(-0.185955,-0.0736249)(-0.180965,-0.0851559)(-0.175261,-0.0963507)(-0.168866,-0.107165)(-0.161803,-0.117557)(-0.154103,-0.127485)(-0.145794,-0.136909)(-0.136909,-0.145794)(-0.127485,-0.154103)(-0.117557,-0.161803)(-0.107165,-0.168866)(-0.0963507,-0.175261)(-0.0851559,-0.180965)(-0.0736249,-0.185955)(-0.0618034,-0.190211)(-0.049738,-0.193717)(-0.0374763,-0.196457)(-0.0250666,-0.198423)(-0.0125581,-0.199605)(0.,-0.2)
}

\def\halfdimcoordup{
\pspolygon[fillstyle=solid,linewidth=0pt,linecolor=white](-0.2,0)(-0.2,0.5)(0.2,0.5)(0.2,0.)(0.199605,-0.0125581)(0.198423,-0.0250666)(0.196457,-0.0374763)(0.193717,-0.049738)(0.190211,-0.0618034)(0.185955,-0.0736249)(0.180965,-0.0851559)(0.175261,-0.0963507)(0.168866,-0.107165)(0.161803,-0.117557)(0.154103,-0.127485)(0.145794,-0.136909)(0.136909,-0.145794)(0.127485,-0.154103)(0.117557,-0.161803)(0.107165,-0.168866)(0.0963507,-0.175261)(0.0851559,-0.180965)(0.0736249,-0.185955)(0.0618034,-0.190211)(0.049738,-0.193717)(0.0374763,-0.196457)(0.0250666,-0.198423)(0.0125581,-0.199605)(0.,-0.2)(-0.0125581,-0.199605)(-0.0250666,-0.198423)(-0.0374763,-0.196457)(-0.049738,-0.193717)(-0.0618034,-0.190211)(-0.0736249,-0.185955)(-0.0851559,-0.180965)(-0.0963507,-0.175261)(-0.107165,-0.168866)(-0.117557,-0.161803)(-0.127485,-0.154103)(-0.136909,-0.145794)(-0.145794,-0.136909)(-0.154103,-0.127485)(-0.161803,-0.117557)(-0.168866,-0.107165)(-0.175261,-0.0963507)(-0.180965,-0.0851559)(-0.185955,-0.0736249)(-0.190211,-0.0618034)(-0.193717,-0.049738)(-0.196457,-0.0374763)(-0.198423,-0.0250666)(-0.199605,-0.0125581)(-0.2,0.)
}

\def\halfdimcoorddown{
\rput(0,-1){\pspolygon[fillstyle=solid,linewidth=0pt,linecolor=white](-0.2,0.5)
(-0.2,1.)(-0.199605,1.01256)(-0.198423,1.02507)(-0.196457,1.03748)(-0.193717,1.04974)(-0.190211,1.0618)(-0.185955,1.07362)(-0.180965,1.08516)(-0.175261,1.09635)(-0.168866,1.10717)(-0.161803,1.11756)(-0.154103,1.12748)(-0.145794,1.13691)(-0.136909,1.14579)(-0.127485,1.1541)(-0.117557,1.1618)(-0.107165,1.16887)(-0.0963507,1.17526)(-0.0851559,1.18097)(-0.0736249,1.18596)(-0.0618034,1.19021)(-0.049738,1.19372)(-0.0374763,1.19646)(-0.0250666,1.19842)(-0.0125581,1.19961)(0.,1.2)(0.0125581,1.19961)(0.0250666,1.19842)(0.0374763,1.19646)(0.049738,1.19372)(0.0618034,1.19021)(0.0736249,1.18596)(0.0851559,1.18097)(0.0963507,1.17526)(0.107165,1.16887)(0.117557,1.1618)(0.127485,1.1541)(0.136909,1.14579)(0.145794,1.13691)(0.154103,1.12748)(0.161803,1.11756)(0.168866,1.10717)(0.175261,1.09635)(0.180965,1.08516)(0.185955,1.07362)(0.190211,1.0618)(0.193717,1.04974)(0.196457,1.03748)(0.198423,1.02507)(0.199605,1.01256)(0.2,1.)(0.2,0.5)}
}

\def\dimerv{
\psset{fillcolor=lightblue}\dimvcoord
}

\def\dimerh{
\psset{fillcolor=lightpurple}\dimhcoord
}

\def\halfdimervup{
\psset{fillcolor=lightblue}\halfdimcoordup
}

\def\halfdimervdown{
\psset{fillcolor=lightblue}\halfdimcoorddown
}

\def\dimervblue{\psset{fillcolor=darkgreen}\dimvcoord}
\def\dimerhblue{\psset{fillcolor=darkgreen}\dimhcoord}
\def\dimervred{\psset{fillcolor=red}\dimvcoord}
\def\dimerhred{\psset{fillcolor=red}\dimhcoord}

\def\halfdimerupred{\psset{fillcolor=red}\halfdimcoordup}
\def\halfdimerupblue{\psset{fillcolor=darkgreen}\halfdimcoordup}
\def\halfdimerdownred{\psset{fillcolor=red}\halfdimcoorddown}
\def\halfdimerdownblue{\psset{fillcolor=darkgreen}\halfdimcoorddown}

\def\uspin{
\psline[linecolor=blue,linewidth=2.0pt,arrowscale=0.8,arrowinset=0.1]{-}(0,0.3)(0,0.7)
\psline[linecolor=blue,linewidth=2.0pt,arrowscale=0.8,arrowinset=0.1]{-}(-0.2,0.5)(0.2,0.5)
}

\def\dspin{
\psline[linecolor=gray,linewidth=2.0pt,arrowscale=0.8,arrowinset=0.1]{-}(-0.2,0.5)(0.2,0.5)
}

\def\rarrowblue{\psline[linecolor=darkgreen,linewidth=1.2pt,arrowscale=1.3,arrowinset=0.1]{->}(0,0)(1.35,0)
\psline[linecolor=darkgreen,linewidth=1.2pt,arrowscale=0.8,arrowinset=0.1]{-}(0.8,0)(2,0)}
\def\larrowblue{\psline[linecolor=darkgreen,linewidth=1.2pt,arrowscale=1.3,arrowinset=0.1]{->}(0,0)(-1.35,0)
\psline[linecolor=darkgreen,linewidth=1.2pt,arrowscale=0.8,arrowinset=0.1]{-}(-0.8,0)(-2,0)}
\def\uarrowblue{\psline[linecolor=darkgreen,linewidth=1.2pt,arrowscale=1.3,arrowinset=0.1]{->}(0,0)(0,1.35)
\psline[linecolor=darkgreen,linewidth=1.2pt,arrowscale=1.3,arrowinset=0.1]{-}(0,0.8)(0,2)}
\def\darrowblue{\psline[linecolor=darkgreen,linewidth=1.2pt,arrowscale=1.3,arrowinset=0.1]{->}(0,0)(0,-1.35)
\psline[linecolor=darkgreen,linewidth=1.2pt,arrowscale=1.3,arrowinset=0.1]{-}(0,-0.8)(0,-2)}

\def\rarrowred{\psline[linecolor=red,linewidth=1.2pt,arrowscale=1.3,arrowinset=0.1]{->}(0,0)(1.35,0)
\psline[linecolor=red,linewidth=1.2pt,arrowscale=0.8,arrowinset=0.1]{-}(0.8,0)(2,0)}
\def\larrowred{\psline[linecolor=red,linewidth=1.2pt,arrowscale=1.3,arrowinset=0.1]{->}(0,0)(-1.35,0)
\psline[linecolor=red,linewidth=1.2pt,arrowscale=0.8,arrowinset=0.1]{-}(-0.8,0)(-2,0)}
\def\uarrowred{\psline[linecolor=red,linewidth=1.2pt,arrowscale=1.3,arrowinset=0.1]{->}(0,0)(0,1.35)
\psline[linecolor=red,linewidth=1.2pt,arrowscale=1.3,arrowinset=0.1]{-}(0,0.8)(0,2)}
\def\darrowred{\psline[linecolor=red,linewidth=1.2pt,arrowscale=1.3,arrowinset=0.1]{->}(0,0)(0,-1.35)
\psline[linecolor=red,linewidth=1.2pt,arrowscale=1.0,arrowinset=0.1]{-}(0,-0.8)(0,-2)}

\def\rnoarrowblue{\psline[linecolor=darkgreen,linewidth=1.2pt,arrowscale=1.0,arrowinset=0.1]{-}(0,0)(1.25,0)
\psline[linecolor=darkgreen,linewidth=1.2pt,arrowscale=0.8,arrowinset=0.1]{-}(0.8,0)(2,0)}
\def\lnoarrowblue{\psline[linecolor=darkgreen,linewidth=1.2pt,arrowscale=1.0,arrowinset=0.1]{-}(0,0)(-1.25,0)
\psline[linecolor=darkgreen,linewidth=1.2pt,arrowscale=0.8,arrowinset=0.1]{-}(-0.8,0)(-2,0)}
\def\unoarrowblue{\psline[linecolor=darkgreen,linewidth=1.2pt,arrowscale=1.0,arrowinset=0.1]{-}(0,0)(0,1.25)
\psline[linecolor=darkgreen,linewidth=1.2pt,arrowscale=1.0,arrowinset=0.1]{-}(0,0.8)(0,2)}
\def\dnoarrowblue{\psline[linecolor=darkgreen,linewidth=1.2pt,arrowscale=1.0,arrowinset=0.1]{-}(0,0)(0,-1.25)
\psline[linecolor=darkgreen,linewidth=1.2pt,arrowscale=1.0,arrowinset=0.1]{-}(0,-0.8)(0,-2)}

\def\rnoarrowred{\psline[linecolor=red,linewidth=1.2pt,arrowscale=1.0,arrowinset=0.1]{-}(0,0)(1.25,0)
\psline[linecolor=red,linewidth=1.2pt,arrowscale=0.8,arrowinset=0.1]{-}(0.8,0)(2,0)}
\def\lnoarrowred{\psline[linecolor=red,linewidth=1.2pt,arrowscale=1.0,arrowinset=0.1]{-}(0,0)(-1.25,0)
\psline[linecolor=red,linewidth=1.2pt,arrowscale=0.8,arrowinset=0.1]{-}(-0.8,0)(-2,0)}
\def\unoarrowred{\psline[linecolor=red,linewidth=1.2pt,arrowscale=1.0,arrowinset=0.1]{-}(0,0)(0,1.25)
\psline[linecolor=red,linewidth=1.2pt,arrowscale=1.0,arrowinset=0.1]{-}(0,0.8)(0,2)}
\def\dnoarrowred{\psline[linecolor=red,linewidth=1.2pt,arrowscale=1.0,arrowinset=0.1]{-}(0,0)(0,-1.25)
\psline[linecolor=red,linewidth=1.2pt,arrowscale=1.0,arrowinset=0.1]{-}(0,-0.8)(0,-2)}

\begin{document}

\topmargin -15mm
\oddsidemargin 05mm

%%%%%%%%%%%%%%%%%%%%%
%
% Title page
%
%%%%%%%%%%%%%%%%%%%%%

\title{\mbox{}\vspace{-.2in}
\bf 
Integrability and conformal data of the dimer model
}

\date{}
\maketitle

\begin{center}
{\vspace{-5mm}\Large Alexi Morin-Duchesne$^\ast$,\, J{\o}rgen Rasmussen$^\S$,\, Philippe Ruelle$^\ast$}
\\[.5cm]
{\em {}$^\ast$Institut de Recherche en Math\'ematique et Physique\\ Universit\'e catholique de Louvain, Louvain-la-Neuve, B-1348, Belgium}
\\[.2cm]
{\em {}$^\S$School of Mathematics and Physics, University of Queensland}\\
{\em St Lucia, Brisbane, Queensland 4072, Australia}
\\[.4cm] 
{\tt alexi.morin-duchesne\,@\,uclouvain.be}
\qquad
{\tt j.rasmussen\,@\,uq.edu.au}
\quad
{\tt philippe.ruelle\,@\,uclouvain.be}
\end{center}

%%%%%%%%%%%%%%%%%%%%%
%
% Abstract
%
%%%%%%%%%%%%%%%%%%%%%

\vspace{0.5cm}

\begin{abstract}
The central charge of the dimer model on the square lattice is still being debated in the literature. In this paper, we provide evidence supporting the consistency of a $c=-2$ description. Using Lieb's transfer matrix and its description in terms of the Temperley-Lieb algebra $\mathsf{TL}_n$ at $\beta = 0$, we provide a new solution of the dimer model in terms of the model of critical dense polymers on a tilted lattice and offer an understanding of the lattice integrability of the dimer model. The dimer transfer matrix is analysed in the scaling limit and the result for $L_0-\frac c{24}$ is expressed in terms of fermions. Higher Virasoro modes are likewise constructed as limits of elements of $\mathsf{TL}_n$ and are found to yield a $c=-2$ realisation of the Virasoro algebra, familiar from fermionic $bc$ ghost systems. In this realisation, the dimer Fock spaces are shown to decompose, as Virasoro modules, into direct sums of Feigin-Fuchs modules, themselves exhibiting reducible yet indecomposable
structures. In the scaling limit, the eigenvalues of the lattice integrals of motion are found to agree exactly with those of the $c=-2$ conformal integrals of motion. Consistent with the expression for $L_0-\frac c{24}$ obtained from the transfer matrix, we also construct higher Virasoro modes with $c=1$ and find that the dimer Fock space is completely reducible under their action. However, the transfer matrix is found not to be a generating function for the $c=1$ integrals of motion. Although this indicates that Lieb's transfer matrix description is incompatible with the $c=1$ interpretation, 
it does not rule out the existence of an alternative, $c=1$ compatible, transfer matrix description of the dimer model.
\vspace{0.25cm}

\noindent Keywords: dimer model, critical dense polymers, Temperley-Lieb algebra, integrability, conformal field theory, conformal integrals of motion, fermionic $bc$ ghosts, Feigin-Fuchs modules. \end{abstract}

%%%%%%%%%%%%%%%%%%%%%
%
% TOC
%
%%%%%%%%%%%%%%%%%%%%%

\newpage

\tableofcontents
\clearpage

%%%%%%%%%%%%%%%%%%%%
\section{Introduction}
\label{sec:Introduction}
%%%%%%%%%%%%%%%%%%%%

The classical dimer model, also known as the perfect matching or domino tiling problem, has a rather long history. It was originally introduced in 1937 by Fowler and Rushbrook as a model for the adsorption of diatomic molecules on a substrate \cite{FR37}. Its mathematical formulation is extremely simple and asks to enumerate the number of ways the edges of a finite graph can be covered by dimers (dominos) so that each vertex is covered by exactly one dimer. This is the so-called close-packed dimer model. A common generalisation allows for monomers (or vacancies), corresponding to vertices 
not covered by dimers, and leads to the general monomer-dimer problem. Except for a brief mention later in this introduction, this paper is exclusively concerned with the close-packed dimer model, formulated on a square lattice grid.

The first important results regarding the statistical properties of the model, such as partition functions and correlation functions, were obtained in the sixties, mostly in 
the two-dimensional incarnation of the model \cite{Kas61,TF61,Fi61,FiSt63,Ha66,Fe67}. Dating from the same period, the paper \cite{Lieb67} by Lieb reformulates and solves the dimer model in terms of a transfer matrix; it will play a crucial role in this paper. 

After these pioneering works, the dimer model has been revisited and generalised in a number of ways by both the physics and mathematics communities, improving our understanding of the model and of its numerous relations with other problems. Recent developments include arctic circle type phenomena \cite{JPS98,Jo05,CP07}, dimers and amoebae \cite{KOS06}, correlations of monomers \cite{PR08,Du11,AF14}, the development of quantum dimer models \cite{RK88,MSP02,MSC01}, trimer and more generally $p$-mer models \cite{GDJ07}, and double dimer models \cite{KW11,Ke14}. A recent review of mathematical aspects of the dimer model can be found in \cite{Ke09}.

The two-dimensional dimer model belongs to the class of exactly solvable models, since, in most instances, it can be solved exactly and explicitly, by various techniques. Often, such models are in fact integrable, in a technical sense, where the integrability is related to the presence of algebraic structures such as the Yang-Baxter equation, the presence of a spectral parameter in the transfer matrix or, in the scaling limit, the existence of an infinite number of conserved charges. It appears that the dimer model has not been explicitly demonstrated to be integrable in any of these senses. 
Our first goal is thus to introduce a spectral parameter in the transfer matrix, in the case where the model is defined on a strip. The construction relies on two essential ingredients: 
(i) the fact that the dimer model carries a representation of the Temperley-Lieb algebra with fugacity $\beta=0$, as shown recently in \cite{MDRR14}, and 
(ii) the relation, based on Temperley's correspondence~\cite{T74}, of its transfer matrix $T(\alpha)$ with a family of generalised polymer models described by a two-parameter transfer tangle $\Db(u,v)$ pertaining to the same Temperley-Lieb algebra.

As indicated, the dimer transfer matrix $T(\alpha)$ depends on a parameter $\alpha$, but this parameter is not spectral: The corresponding family of transfer matrices is not self-commuting. This is contrasted by the family of transfer tangles $\Db(u,v)$ where $u$ is a spectral parameter (while $v$ is not). The crucial point is that we can rewrite the square of $T(\alpha)$ as the transfer tangle $\Db(u(\alpha),v(\alpha))$ in a spin representation for particular values of $u=u(\alpha)$ and $v=v(\alpha)$. 
From this, it directly follows that the {\em one}-parameter family of matrices $\Db(u,v(\alpha))$, labelled by $u$, commutes with $T(\alpha)$. The coefficients in the series expansion in powers of 
$\sin{(2u)}$ then yield {\em lattice integrals of motion}, a self-commuting family of operators all commuting with $T(\alpha)$, establishing the integrability of the dimer model. Their number is finite for finite 
system size $n$, grows linearly with $n$ and becomes infinite in the scaling limit. Furthermore, as we will show, the connection with the polymer transfer tangles provides a new solution of the dimer model.

This integrability of the dimer model also gives a way to understand the large $n$ expansion of the eigenvalues of (the logarithm of) $T(\alpha)$, which here takes the general form \cite{IH02,Sa01,IH01}
\be
E = {\rm Eig} \big(\!-\tfrac12\log T^2(\alpha)\big) = n f_{\textrm{bulk}} + f_{\textrm{bdy}} + \sum_{p=1}^\infty \frac{a_p}{n^{2p-1}},
\label{eq:Elargen}
\ee
where $f_{\textrm{bulk}}$ and $f_{\textrm{bdy}}$ are respectively the bulk and boundary free energies. 
Although the eigenvalues of $T(\alpha)$ are real, they can be positive or negative. It is therefore easier to work with 
$\tfrac12\log T^2(\alpha)$ instead of $\log T(\alpha)$, as in \eqref{eq:Elargen}.

From an exact computation, we will find that the coefficients $a_p$ in \eqref{eq:Elargen} take the form 
\be
a_p = \frac{2^p\pi^{2p-1}}{(2p)!} \lambda^{}_{2p-1}(\alpha) \, I_{2p-1},
\label{eq:ap}
\ee
where the (polynomial) functions $\lambda_{2p-1}(\alpha)$, $p \ge 1$, are \emph{the same for all energy values}. The peculiar relationship between $T^2(\alpha)$ and the matrix realisation of the transfer tangle $\Db(u(\alpha),v(\alpha))$ shows that the $I_{2p-1}$ are related to the eigenvalues of lattice integrals of motion. On general grounds \cite{BLZ96,BLZ97}, this is exactly what one would expect. Restricting to the conformal part of the spectrum, it is also expected that the numbers $I_{2p-1}$ become universal in the continuum limit, in the sense that they correspond to the eigenvalues of universal (conformal) integrals of motion ${\mathbf I}_{2p-1}$. The first integral of motion is simply ${\mathbf I}_1 = L_0 - \frac {c}{24}$, while the higher-order ones are computable polynomials of degree $p$ in the Virasoro modes \cite{SY88}. The $1/n$ expansion was also obtained in~\cite{Nigro12} and conjecturally related to the conformal integrals of motion.

In the particular case of the ground-state energy $E_0$, the first coefficient \eqref{eq:ap} is given by 
\be
a_1^{(0)} = \pi \lambda_1(\alpha) \, \big(\Delta_0 -  \frac {c}{24}\big) = -\frac{\pi}{24} \lambda_1(\alpha) c_{\rm eff},
\ee
where $\Delta_0$ is the conformal weight of the state to which the ground state converges in the scaling limit. The factor $\lambda_1(\alpha)$ is usually referred to as the {\em speed of sound}, while the combination $c_{\rm eff} = c - 24\Delta_0$ is the {\em effective central charge}.

As the previous discussion already assumes implictly, it is widely believed that the dimer model on a square lattice\footnote{In contrast, the dimer model is not conformally invariant on non-bipartite 
graphs such as the isotropic triangular lattice, where correlations decay exponentially \cite{FMS02}.} is conformally invariant. However, there has been some discussions about how that symmetry is realised, and our second main theme concerns this issue. Many works have been devoted to various aspects of this question, often with different conclusions. In fact, two principal proposals have been put forward: a conformal description with $c=-2$ and one with $c=1$. Let us briefly review some of the arguments in favour of one or the other. 

A number of contributions have concentrated on the analysis of finite-size corrections, for different boundary conditions and different geometries. This issue is quite subtle because of the fact, known from old results by Ferdinand \cite{Fe67}, that the dimer model is very sensitive to the parities of the box dimensions in which it is defined (as recalled above, it is even more sensitive to the lattice type). This peculiarity requires one to be extremely cautious in the analysis of finite-size data, in particular when inferring the central charge of the model. The problem is even more acute when the model is defined on a rectangle, in which case the free energy contains, at order $p=1$, a second universal term $\frac1n{\log{n}}$. Each corner of the rectangle contributes a term of that form that depends not only on the central charge \cite{CaPe88}, but also on the conformal weight of a boundary condition changing field in case the boundary conditions on either side of the corner are different \cite{KlVa92}. Somewhat surprisingly, these two finite-size corrections in the free energy can be consistently interpreted in a conformal scheme based on either candidate for the central charge. Namely, those relying on the equivalence with spanning trees support $c=-2$, on a cylinder \cite{IPRH05} and on a rectangle \cite{Ru07,IKGW14}, whereas the use of the height function to describe dimer configurations yields $c=1$ \cite{Al14}. As indicated in the case of the rectangular geometry, the main difference between the two approaches stems from the presence or absence of appropriate changes of boundary conditions, contributing (or not) to the free energy and affecting the way the lattice result is to be interpreted.

The description in terms of spanning trees and its close connection with the Abelian sandpile model \cite{MD92}, widely believed to be described by a $c=-2$ logarithmic conformal field theory \cite{Ru13}, substantiates the $c=-2$ interpretation. 
Combinatorial properties of spanning trees and spanning webs have also been considered within the $c=-2$ framework in \cite{BGPT08,BPPR14}.
Another hint is the connection between the dimer model and the critical dense polymer model \cite{PR07} which will be reinforced in this paper: As mentioned above, the dimer transfer matrix squared can be written in terms of a transfer tangle pertaining to the Temperley-Lieb algebra with fugacity $\beta=0$ \cite{MDRR14}, which is the value canonically associated with $c=-2$. However, a striking result supporting the alternative view is the proof \cite{Ke01} that the fluctuations of the height function are described, in the scaling limit, by a massless free scalar field, strongly pointing to $c=1$. The use of the Kasteleyn matrix and Pfaffian techniques, 
essential tools in the classical solution of the dimer model \cite{Kas61} and still much in use today \cite{AF14}, is closely related to free fermionic theories \cite{FMS02,DOR09} and adds further support to $c=1$. The identification of certain operators \cite{PLF07}, including the insertion of monomers and dimers, as vertex operators in a free boson (Gaussian) theory points in the same direction. 

Results from the computation of dimer correlation functions similarly allow for dual representations.
Dimer correlations can be naturally accounted for in a $c=-2$ conformal theory \cite{Ru13} (the operator inserting a dimer can be explicitly written in terms of symplectic fermions), but can also be easily interpreted as $c=1$ correlators \cite{Al14}. This is in sharp contrast to the corresponding results for monomer correlations. The correlation functions for an arbitrary (even) number of boundary monomers have been computed and shown to be equal to the correlators of a complex (Dirac) free fermion \cite{PR08}, indicating
$c=1$. Likewise, the bulk monomer correlators, computed in \cite{Du11}, are consistent with the monomer field identification proposed in \cite{PLF07} and do not appear to be naturally accounted for within a $c=-2$ framework.

Together, these observations suggest that there exist two conformal descriptions of the dimer model: one with $c=-2$ and one with $c=1$. The descriptions would presumably be related to different degrees of freedom and come with their own limitations. In this sense, the name {\em dimer model} can be ambiguous and cover different situations. Depending on whether one keeps track of spanning tree connections, includes height degrees of freedom, monomers or still other features, one of the two descriptions may impose itself.\footnote{The analysis made in \cite{GDJ07} suggests that including trimers requires a conformal description with central charge $c=2$.
However, by extending the paradigm of dual interpretations, there might be more than one conformal description.}  

Our objective here is to present a self-consistent close-packed dimer model fully compatible with a conformal description based on $c=-2$. Lieb's transfer matrix is central to our approach.

Relying on spin degrees of freedom, the transfer matrix can be written as a spin chain, in terms of Pauli matrices, and diagonalised by standard techniques. The conformal spectrum can then be easily computed, but the spectrum, or even the characters of the representations involved in the model, is in general not enough to identify the central charge, even less so to determine the structure of the representations. However, the finite-size transfer matrix is closely tied to the Temperley-Lieb algebra and allows us to push our examination much further.

It was shown in \cite{MDRR14} that the transfer matrix (actually its square) can be written in terms of spin operators (Pauli matrices) satisfying the Temperley-Lieb algebra $\tl_n(\beta)$ at fugacity $\beta=0$. An immediate consequence is that the (spin) configuration space becomes a module over $\tl_n(0)$. Partitioning the full space into sectors according to the value $v$ of the total magnetisation gives rise to submodules, whose structures have been completely determined \cite{MDRR14}.

The presence of the algebra $\tl_n(0)$ is key to assessing the conformal structure. Indeed, one can construct finite-size operators, as universal linear combinations of $\tl_n(0)$ elements, which are believed to converge, in the scaling limit, to the Virasoro modes~\cite{KS94}. In this way, the $\tl_n(0)$ modules, present at finite size, induce a conformal module structure in the infinite-dimensional space obtained in the scaling limit. We explicitly carry out this program in the present context and find, after a Jordan-Wigner transformation, operator modes $L_k$ written in terms of fermion modes. We then verify that their commutation relations are those of the Virasoro algebra with central charge $c=-2$. 

The scaling limit yields an infinite number of infinite-dimensional sectors $E^v$, labelled by a quantum number $v$, each one being a Virasoro module. We determine the structures of these modules by establishing an isomorphism between the $E^v$ and specific Feigin-Fuchs modules \cite{FF82}. The result of this analysis is that each module $E^v$, whose character is that of a single Verma module, contains an infinite number of (irreducible) composition factors, with a structure that depends on whether the scaling limit is taken over a sequence of odd or even lattice widths. In one case, the modules $E^v$ are direct sums of irreducible modules, while in the other, the composition factors form a single reducible yet indecomposable module (see \cref{thm:structurethm} in Section~\ref{sec:structure}). None of the modules exhibits non-trivial Jordan blocks for $L_0$.

Finally, we examine the consistency of this conformal picture by linking it to the coefficients $a_p$ in the large $n$ expansion of the energy levels. From \eqref{eq:ap}, these coefficients are proportional to the eigenvalues of integrals of motion ${\mathbf I}_{2p-1}$, which are, as discussed above, computable polynomials in the Virasoro modes. The action of the integrals of motion is completely determined by the conformal structure of the modules $E^v$. Their eigenvalues $I_{2p-1}$ can then be explicitly evaluated and used to compute the coefficients $a_p$. In other words, the eigenstates of the transfer matrix, which converge to conformal states populating a certain level in a module $E^v$, should be such that their associated coefficients $a_p$, for each $p$, are related by \eqref{eq:ap} to the eigenvalues of ${\mathbf I}_{2p-1}$ acting on that level. Comparing the so-obtained coefficients with their actual values computed from the lattice eigenvalues yields, in principle, an infinite number of verifications of the value of the central charge for each module $E^v$. In practice, it is rather hard to perform them all, because the integrals of motion are not 
all explicitly known, or because the calculations are too cumbersome, as is the case for high levels. For the three lowest-lying levels in each module $E^v$, and for the first two non-trivial integrals of motion, ${\mathbf I}_3$ and ${\mathbf I}_5$, we find that there is complete agreement.

These verifications constitute substantial evidence confirming that Lieb's transfer matrix approach to the dimer model is consistent with a $c=-2$ conformal description. It is well known that different realisations of the Virasoro modes can be constructed in terms of free fermion modes, with different values of $c$. We take advantage of this freedom to construct, in terms of the same fermion modes as in the $c=-2$ description, an alternative 
realisation of the Virasoro algebra, this one with $c=1$. Under this new conformal action, the full state space is partitioned into different subspaces $\hat E^x$, according to the values of a new quantum number $x$ preserved by the Virasoro algebra. We determine the full structure of the subspaces $\hat E^x$ as conformal modules and find a very different structure. Indeed, in all but one case, the $\hat E^x$ are irreducible Verma modules. However, this alternative point of view turns out to be inconsistent with the transfer matrix description, in the sense that the eigenvalues of the integrals of motion do not reproduce the coefficients $a_p$ computed from the lattice.

It is natural to speculate that a consistent conformal framework with $c=1$ could nevertheless
emerge from the use of a different transfer matrix, relying on different degrees of freedom. We hope to return to this question in the future.

The paper is organised as follows. In \cref{sec:dimersintro}, we recall the basics of the dimer model on a strip and its description in terms of Lieb's transfer matrix. We give the details of its diagonalisation, which are useful in later sections. We also introduce the partitioning of the configuration space into sectors based on the total magnetisation and compute the corresponding partition functions. \cref{sec:Polymers} reviews aspects of the critical dense polymer model and a class of transfer tangles pertaining to the Temperley-Lieb algebra at $\beta=0$. In \cref{sec:dimersandloops}, we first perform a detailed comparison of the spectra of the dimer and critical dense polymer models. We explain their similarities using a series of maps relating perfect matchings to configurations of the critical dense polymer model on a lattice tilted by $45^\circ$. The integrability of the dimer model is also established in \cref{sec:dimersandloops}, where we exhibit a one-parameter family of matrices closely related to polymer transfer tangles and commuting with Lieb's transfer matrix. \cref{sec:CFT} examines the way conformal invariance is realised by investigating in detail the two options $c=-2$ and $c=1$. These turn out to involve Feigin-Fuchs modules of the Virasoro algebra, whose definitions are reviewed in \cref{sec:FFs}. The proofs of the structures of the conformal modules are deferred until \cref{sec:iso,sec:iso1}. \cref{sec:CFT} also provides a detailed comparison of the integrals of motion and the lattice data, demonstrating the self-consistency of the $c=-2$ description and the inconsistency of $c=1$ in the present context. Concluding remarks are presented in \cref{sec:Discussion}.

%%%%%%%%%%%%%%%%%%%%
\section{Dimer model}
\label{sec:dimersintro}
%%%%%%%%%%%%%%%%%%%%

Here, we review definitions and results for the dimer model on the square lattice. The presentation follows \cite{RR12} and \cite{MDRR14}.

%%%%%%%%%%%%%%%%%
\subsection{Statistical model}
\label{sec:dimersstatmodel}
%%%%%%%%%%%%%%%%%

A dimer is a bridge that covers two neighbouring sites of a lattice. We define the model on the square lattice, where the sites are organised in an $M \times N$ rectangular grid, with $M$ and $N$ counting the rows and columns, respectively. Each site is labeled by a pair of integers $(i,j)$ that gives its position in the grid, with $1\le i \le N$ and $1 \le j \le M$.  A {\it perfect matching} is a configuration of $\frac{MN}2$ dimers, where each site is covered exactly once. If $MN$ is odd, the set of perfect matchings is empty.

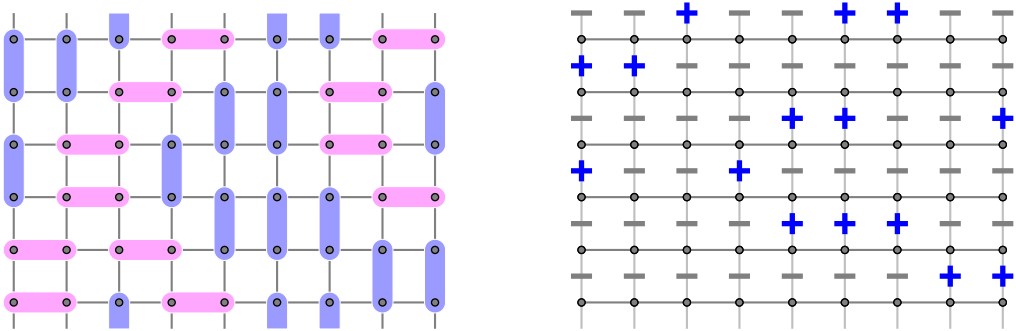
\begin{figure}[h!]
\psset{unit=.7cm}
\begin{center}
\begin{pspicture}(-0.2,-0.4)(8.2,5.7)
\multiput(0,0)(0,1){6}{\psline[linecolor=gray]{-}(0,0)(8,0)}\multiput(0,0)(1,0){9}{\psline[linecolor=gray]{-}(0,-0.5)(0,5.5)}
\rput(0,0){\dimerh}
\rput(0,1){\dimerh}
\rput(1,2){\dimerh}
\rput(1,3){\dimerh}
\rput(2,4){\dimerh}
\rput(3,5){\dimerh}
\rput(2,1){\dimerh}
\rput(3,0){\dimerh}
\rput(7,5){\dimerh}
\rput(6,3){\dimerh}
\rput(6,4){\dimerh}
\rput(7,2){\dimerh}
\rput(0,2){\dimerv}
\rput(0,4){\dimerv}
\rput(1,4){\dimerv}
\rput(3,2){\dimerv}
\rput(4,1){\dimerv}
\rput(4,3){\dimerv}
\rput(5,3){\dimerv}
\rput(5,1){\dimerv}
\rput(8,3){\dimerv}
\rput(6,1){\dimerv}
\rput(7,0){\dimerv}
\rput(8,0){\dimerv}
\rput(2,0){\halfdimervdown}
\rput(2,5){\halfdimervup}
\rput(5,0){\halfdimervdown}
\rput(5,5){\halfdimervup}
\rput(6,0){\halfdimervdown}
\rput(6,5){\halfdimervup}
\multiput(0,0)(0,1){6}{\multiput(0,0)(1,0){9}{\pscircle[linewidth=0.025,fillstyle=solid,fillcolor=gray](0,0){0.08}}}
\end{pspicture} 
\qquad \qquad
\begin{pspicture}(-0.2,-0.4)(8.2,5.7)
\multiput(0,0)(0,1){6}{\psline[linecolor=gray]{-}(0,0)(8,0)}\multiput(0,0)(1,0){9}{\psline[linecolor=lightgray]{-}(0,-0.5)(0,5.5)}
\multiput(0,0)(0,1){6}{\multiput(0,0)(1,0){9}{\pscircle[linewidth=0.025,fillstyle=solid,fillcolor=lightgray](0,0){0.08}}}
\rput(0,5){\dspin}\rput(1,5){\dspin}\rput(2,5){\uspin}\rput(3,5){\dspin}\rput(4,5){\dspin}\rput(5,5){\uspin}\rput(6,5){\uspin}\rput(7,5){\dspin}\rput(8,5){\dspin}
\rput(0,4){\uspin}\rput(1,4){\uspin}\rput(2,4){\dspin}\rput(3,4){\dspin}\rput(4,4){\dspin}\rput(5,4){\dspin}\rput(6,4){\dspin}\rput(7,4){\dspin}\rput(8,4){\dspin}
\rput(0,3){\dspin}\rput(1,3){\dspin}\rput(2,3){\dspin}\rput(3,3){\dspin}\rput(4,3){\uspin}\rput(5,3){\uspin}\rput(6,3){\dspin}\rput(7,3){\dspin}\rput(8,3){\uspin}
\rput(0,2){\uspin}\rput(1,2){\dspin}\rput(2,2){\dspin}\rput(3,2){\uspin}\rput(4,2){\dspin}\rput(5,2){\dspin}\rput(6,2){\dspin}\rput(7,2){\dspin}\rput(8,2){\dspin}
\rput(0,1){\dspin}\rput(1,1){\dspin}\rput(2,1){\dspin}\rput(3,1){\dspin}\rput(4,1){\uspin}\rput(5,1){\uspin}\rput(6,1){\uspin}\rput(7,1){\dspin}\rput(8,1){\dspin}
\rput(0,0){\dspin}\rput(1,0){\dspin}\rput(2,0){\dspin}\rput(3,0){\dspin}\rput(4,0){\dspin}\rput(5,0){\dspin}\rput(6,0){\dspin}\rput(7,0){\uspin}\rput(8,0){\uspin}
\multiput(0,0)(0,1){6}{\multiput(0,0)(1,0){9}{\pscircle[linewidth=0.025,fillstyle=solid,fillcolor=gray](0,0){0.08}}}
\end{pspicture}
\caption{A perfect matching on the $6\times9$ cylinder and its corresponding spin configuration.}
\label{fig:dimerconf}
\end{center}
\end{figure}

We fold the square lattice on a horizontal cylinder, meaning that the sites $(i,1)$ and $(i,M)$, for $1 \le i \le N$, are neighbours and can be covered by the same dimer. In contrast, the sites $(1,j)$ and $(N,j)$ are not neighbours and cannot be covered by the same dimer (unless $N=2$). An example of a perfect matching on the $6\times 9$ cylinder is given in \cref{fig:dimerconf}.

The model depends on a free parameter $\alpha \in \mathbb R$, which measures the preference for the horizontal dimers over the vertical ones. Perfect matchings are thus weighted by $\alpha^h$, where $h$ is the number of horizontal dimers. The partition function is 
\be
Z^{\textrm{dim}} = \sum_{\substack{\textrm{perfect} \\[0.05cm] \textrm{matchings} }}\!\! \alpha^h.
\ee

%%%%%%%%%%%%%%%%%
\subsection{Transfer matrix}
%%%%%%%%%%%%%%%%%

Lieb's solution~\cite{Lieb67} of the dimer model uses a transfer matrix approach. The construction builds on a map from perfect matchings to two-dimensional spin configurations and ultimately expresses the cylinder partition function as the trace of the $M$-th power of a spin-chain transfer matrix. 

A vertical edge is said to be {\it occupied} if a vertical dimer covers the sites at its ends. It is then assigned an up spin. Otherwise, it is {\it unoccupied} and assigned a down spin. This assignment of up and down spins to vertical edges maps a perfect matching to a two-dimensional spin configuration. An example is given in \cref{fig:dimerconf}. For a row of vertical edges (of length $N$), we refer to its edge occupations as a {\it row configuration}. Locally, the map sends a row configuration to an element of the canonical basis of $(\mathbb C^2)^{\otimes N}$, the vector space spanned by $N$ $\tfrac12$-spins.
For example,
\be
\psset{unit=.7cm}
\begin{pspicture}[shift=-0.7](-0.2,-0.3)(3.2,1.5)
\psline[linecolor=gray]{-}(0,0)(3,0)\psline[linecolor=gray]{-}(0,1)(3,1)
\psline[linecolor=gray]{-}(0,0)(0,1)\psline[linecolor=gray]{-}(1,0)(1,1)\psline[linecolor=gray]{-}(2,0)(2,1)\psline[linecolor=gray]{-}(3,0)(3,1)
\rput(0,0){\dimerv}
\rput(1,1){\halfdimervup}
\rput(1,0){\dimerh}
\rput(2,1){\dimerh}
\rput(3,0){\halfdimervdown}
\multiput(0,0)(1,0){4}{\pscircle[linewidth=0.025,fillstyle=solid,fillcolor=gray](0,0){0.08}}
\multiput(0,1)(1,0){4}{\pscircle[linewidth=0.025,fillstyle=solid,fillcolor=gray](0,0){0.08}}
\end{pspicture}
\quad \rightarrow \quad 
\begin{pspicture}[shift=-0.6](-0.2,-0.2)(3.2,1.2)
\psline[linecolor=gray]{-}(0,0)(3,0)\psline[linecolor=gray]{-}(0,1)(3,1)
\psline[linecolor=gray]{-}(0,0)(0,1)\psline[linecolor=gray]{-}(1,0)(1,1)\psline[linecolor=gray]{-}(2,0)(2,1)\psline[linecolor=gray]{-}(3,0)(3,1)
\rput(0,0){\uspin}
\rput(1,0){\dspin}
\rput(2,0){\dspin}
\rput(3,0){\dspin}
\multiput(0,0)(1,0){4}{\pscircle[linewidth=0.025,fillstyle=solid,fillcolor=gray](0,0){0.08}}
\multiput(0,1)(1,0){4}{\pscircle[linewidth=0.025,fillstyle=solid,fillcolor=gray](0,0){0.08}}
\end{pspicture} \ \in \ctwotimes 4  .  
\ee
This map thus encodes the degrees of freedom of the vertical dimers. The occupation of a vertical edge is nevertheless fixed by the dimer attached to the site immediately below it. A site covered either by a horizontal dimer or by the top half of a vertical dimer always lies below an unoccupied edge.

The canonical basis of $(\mathbb C^2)^{\otimes N}$ is made of the elements $| \mathsf{s} \rangle = | s_1 s_2 \dots s_N \rangle$ with $s_i \in \{+, -\} \simeq \{\left(\begin{smallmatrix}1\\0\end{smallmatrix}\right),\left(\begin{smallmatrix}0\\1\end{smallmatrix}\right)\}$. We denote by $|{\sf d} \rangle$ the state with all spins pointing down: 
\be
 |{\sf d} \rangle=|- - \,\cdots\,-\rangle.
\ee
The Pauli matrices acting on the $j$-th component in $(\mathbb C^2)^{\otimes N}$ are
\be
\sigma^a_j = \underbrace{\mathbb I_2 \otimes \dots \otimes \mathbb I_2}_{j-1} \otimes \; \sigma^a \otimes \underbrace{\mathbb I_2 \otimes \dots \otimes \mathbb I_2}_{N-j}, \qquad (a \in \{x,y,z,+,-\}), 
\ee 
with
\begin{subequations}
\begin{alignat}{3}
\sigma^x &= \begin{pmatrix} 0 & 1 \\ 1 & 0 \end{pmatrix}, \quad \
\sigma^y &&= \begin{pmatrix} 0 & - \ir \\ \ir & 0 \end{pmatrix}, \quad \
\sigma^z &&= \begin{pmatrix} 1 & 0 \\ 0 & -1 \end{pmatrix},\\[0.15cm]
\sigma^+ &= \begin{pmatrix} 0 & 1 \\ 0 & 0 \end{pmatrix}, \quad \ 
\sigma^- &&= \begin{pmatrix} 0 & 0 \\ 1 & 0 \end{pmatrix}, \quad \ \ \ \,
\mathbb I_2 &&= \begin{pmatrix} 1 & 0 \\ 0 & 1 \end{pmatrix}.
\end{alignat}
\end{subequations}

We now introduce Lieb's transfer matrix $\bar T(\alpha)$. For any $| \mathsf{s} \rangle$, $\bar T(\alpha)| \mathsf{s} \rangle$ is a linear combination of the states corresponding to the possible row configurations immediately above the row described by $| \mathsf{s} \rangle$. $\bar T(\alpha)$ is the product of two operators. The first one reverses all the spins,
\be 
V_1 = \prod_{j=1}^N \sigma^x_j\label{eq:V1}.
\ee
In terms of row configurations, acting with $V_1$ on $| \mathsf s \rangle$ produces a state $| \mathsf{s'}\rangle$ corresponding to a row configuration
where vertical edges are occupied whenever possible, that is, unless the top half of a vertical dimer appears just below. For instance,
\be
\psset{unit=.7cm}
\begin{pspicture}[shift=-0.9](-0.2,-0.5)(3.2,1.5)
\psline[linecolor=gray]{-}(0,0)(3,0)\psline[linecolor=gray]{-}(0,1)(3,1)
\psline[linecolor=gray]{-}(0,0)(0,1)\psline[linecolor=gray]{-}(1,0)(1,1)\psline[linecolor=gray]{-}(2,0)(2,1)\psline[linecolor=gray]{-}(3,0)(3,1)
\rput(0,0){\dimerv}
\rput(1,0){\dimerh}
\rput(3,0){\halfdimervdown}
\multiput(0,0)(1,0){4}{\pscircle[linewidth=0.025,fillstyle=solid,fillcolor=gray](0,0){0.08}}
\multiput(0,1)(1,0){4}{\pscircle[linewidth=0.025,fillstyle=solid,fillcolor=gray](0,0){0.08}}
\rput(0,0){\rput(0,0){\uspin}\rput(1,0){\dspin}\rput(2,0){\dspin}\rput(3,0){\dspin}}
\end{pspicture}
\qquad \overset{V_1}{\longrightarrow} \qquad 
\begin{pspicture}[shift=-0.9](-0.2,-0.5)(3.2,1.5)
\psline[linecolor=gray]{-}(0,0)(3,0)\psline[linecolor=gray]{-}(0,1)(3,1)
\psline[linecolor=gray]{-}(0,0)(0,1)\psline[linecolor=gray]{-}(1,0)(1,1)\psline[linecolor=gray]{-}(2,0)(2,1)\psline[linecolor=gray]{-}(3,0)(3,1)
\rput(0,0){\halfdimervdown}
\rput(1,0){\dimerv}
\rput(2,0){\dimerv}
\rput(3,0){\dimerv}
\multiput(0,0)(1,0){4}{\pscircle[linewidth=0.025,fillstyle=solid,fillcolor=gray](0,0){0.08}}
\multiput(0,1)(1,0){4}{\pscircle[linewidth=0.025,fillstyle=solid,fillcolor=gray](0,0){0.08}}
\rput(0,0){\rput(0,0){\dspin}\rput(1,0){\uspin}\rput(2,0){\uspin}\rput(3,0){\uspin}}
\end{pspicture}\ .
\label{eq:transferexample}
\ee

Of course, some perfect matchings also include horizontal dimers. If two adjacent spins in $|\mathsf s'\rangle$ are up, the state $|\mathsf s''\rangle$, where these two spins are flipped, corresponds to a row configuration of another perfect matching, where the two vertical dimers are replaced by a horizontal dimer that covers the two sites below. The second operation should thus replace $|\mathsf s'\rangle$ by a linear combination consistent with all possible insertions of horizontal dimers. This is implemented by acting on $|\mathsf s'\rangle$ with
\be 
V_3 = \prod_{j=1}^{N-1} (\mathbb I+\alpha\; \sigma^-_j\sigma^-_{j+1}) = \exp\Big(\sum_{j=1}^{N-1}\alpha\; \sigma^-_j\sigma^-_{j+1}\Big).
\ee
For instance,
\be
\psset{unit=.7cm}
\begin{pspicture}[shift=-0.9](-0.2,-0.5)(3.2,1.5)
\psline[linecolor=gray]{-}(0,0)(3,0)\psline[linecolor=gray]{-}(0,1)(3,1)
\psline[linecolor=gray]{-}(0,0)(0,1)\psline[linecolor=gray]{-}(1,0)(1,1)\psline[linecolor=gray]{-}(2,0)(2,1)\psline[linecolor=gray]{-}(3,0)(3,1)
\rput(0,0){\halfdimervdown}
\rput(1,0){\dimerv}
\rput(2,0){\dimerv}
\rput(3,0){\dimerv}
\multiput(0,0)(1,0){4}{\pscircle[linewidth=0.025,fillstyle=solid,fillcolor=gray](0,0){0.08}}
\multiput(0,1)(1,0){4}{\pscircle[linewidth=0.025,fillstyle=solid,fillcolor=gray](0,0){0.08}}
\rput(0,0){\rput(0,0){\dspin}\rput(1,0){\uspin}\rput(2,0){\uspin}\rput(3,0){\uspin}}
\end{pspicture}
\quad \overset{V_3}{\longrightarrow} \quad 
\begin{pspicture}[shift=-0.9](-0.2,-0.5)(3.2,1.5)
\psline[linecolor=gray]{-}(0,0)(3,0)\psline[linecolor=gray]{-}(0,1)(3,1)
\psline[linecolor=gray]{-}(0,0)(0,1)\psline[linecolor=gray]{-}(1,0)(1,1)\psline[linecolor=gray]{-}(2,0)(2,1)\psline[linecolor=gray]{-}(3,0)(3,1)
\rput(0,0){\halfdimervdown}
\rput(1,0){\dimerv}
\rput(2,0){\dimerv}
\rput(3,0){\dimerv}
\multiput(0,0)(1,0){4}{\pscircle[linewidth=0.025,fillstyle=solid,fillcolor=gray](0,0){0.08}}
\multiput(0,1)(1,0){4}{\pscircle[linewidth=0.025,fillstyle=solid,fillcolor=gray](0,0){0.08}}
\rput(0,0){\rput(0,0){\dspin}\rput(1,0){\uspin}\rput(2,0){\uspin}\rput(3,0){\uspin}}
\end{pspicture} \ \, +  \alpha \ \,
\begin{pspicture}[shift=-0.9](-0.2,-0.5)(3.2,1.5)
\psline[linecolor=gray]{-}(0,0)(3,0)\psline[linecolor=gray]{-}(0,1)(3,1)
\psline[linecolor=gray]{-}(0,0)(0,1)\psline[linecolor=gray]{-}(1,0)(1,1)\psline[linecolor=gray]{-}(2,0)(2,1)\psline[linecolor=gray]{-}(3,0)(3,1)
\rput(0,0){\halfdimervdown}
\rput(1,0){\dimerh}
\rput(3,0){\dimerv}
\multiput(0,0)(1,0){4}{\pscircle[linewidth=0.025,fillstyle=solid,fillcolor=gray](0,0){0.08}}
\multiput(0,1)(1,0){4}{\pscircle[linewidth=0.025,fillstyle=solid,fillcolor=gray](0,0){0.08}}
\rput(0,0){\rput(0,0){\dspin}\rput(1,0){\dspin}\rput(2,0){\dspin}\rput(3,0){\uspin}}
\end{pspicture}
 \ \, +  \alpha \ \,
\begin{pspicture}[shift=-0.9](-0.2,-0.5)(3.2,1.5)
\psline[linecolor=gray]{-}(0,0)(3,0)\psline[linecolor=gray]{-}(0,1)(3,1)
\psline[linecolor=gray]{-}(0,0)(0,1)\psline[linecolor=gray]{-}(1,0)(1,1)\psline[linecolor=gray]{-}(2,0)(2,1)\psline[linecolor=gray]{-}(3,0)(3,1)
\rput(0,0){\halfdimervdown}
\rput(1,0){\dimerv}
\rput(2,0){\dimerh}
\multiput(0,0)(1,0){4}{\pscircle[linewidth=0.025,fillstyle=solid,fillcolor=gray](0,0){0.08}}
\multiput(0,1)(1,0){4}{\pscircle[linewidth=0.025,fillstyle=solid,fillcolor=gray](0,0){0.08}}
\rput(0,0){\rput(0,0){\dspin}\rput(1,0){\uspin}\rput(2,0){\dspin}\rput(3,0){\dspin}}
\end{pspicture}\ .
\ee
The result indeed consists of all the row configurations that can appear above
$
\psset{unit=0.35}
\begin{pspicture}[shift=-0.5](-0.2,-0.5)(3.2,1.5)
\psline[linecolor=gray]{-}(0,0)(3,0)\psline[linecolor=gray]{-}(0,1)(3,1)
\psline[linecolor=gray]{-}(0,0)(0,1)\psline[linecolor=gray]{-}(1,0)(1,1)\psline[linecolor=gray]{-}(2,0)(2,1)\psline[linecolor=gray]{-}(3,0)(3,1)
\rput(0,0){\dimerv}
\rput(1,0){\dimerh}
\rput(3,0){\halfdimervdown}
\multiput(0,0)(1,0){4}{\pscircle[linewidth=0.025,fillstyle=solid,fillcolor=gray](0,0){0.08}}
\multiput(0,1)(1,0){4}{\pscircle[linewidth=0.025,fillstyle=solid,fillcolor=gray](0,0){0.08}}
\end{pspicture}\, 
$. Although not illustrated in the previous example, $V_3$ can flip more than one pair of adjacent up spins if $|s'\rangle$ permits it, thus allowing for more than one horizontal dimer in a given row of sites. It is worth noting that $V_3$ also keeps track of the weights $\alpha$ of the horizontal dimers. We arrive at the following expression for the transfer matrix:
\be
\bar T(\alpha) = V_3V_1 = \exp\Big(\sum_{j=1}^{N-1}\alpha\; \sigma^-_j\sigma^-_{j+1}\Big) \prod_{j=1}^N \sigma^x_j.
\ee

For $\alpha \in \mathbb R$, $\bar T(\alpha)$ is real and symmetric, thus making it diagonalisable with real eigenvalues. $\bar T(\alpha)$ is the appropriate transfer matrix for a strip domain
and can be used to compute the partition function on the horizontal
$M \times N$ cylinder (with vertical periodicity),
\be
Z^{\textrm{dim}} = \textrm{Tr}\ \bar T^M(\alpha).
\label{eq:Ztr}
\ee
The problem of computing $Z^{\textrm{dim}}$ is then reduced to the calculation of the eigenvalues of $\bar T(\alpha)$, a matrix of dimension $2^N$. 

We note that in general,
\be
[\bar T(\alpha), \bar T(\alpha')] \neq 0,
\ee
implying that a family of operators that commute with $\bar T(\alpha)$, and thus what we refer to as {\it lattice integrability}, is not naturally built-in. We will return to this question in \cref{sec:integ}.

%%%%%%%%%%%%%%%%%%%%%%%%%%%%%%%%%
\subsection{Diagonalisation of the transfer matrix}\label{sec:diago}
%%%%%%%%%%%%%%%%%%%%%%%%%%%%%%%%%

In preparation for \cref{sec:dimtrees,sec:dimloops}, we restrict our attention to cylinders where the number of rows $M$ is even. In this case, instead of $\bar T(\alpha)$, we can study its square $\bar T^2(\alpha)$, which is conveniently written as
\be \label{eq:T2+-}
\bar T^2(\alpha) = V_3^{} V_3^T =  \exp\Big(\sum_{j=1}^{N-1}\alpha\; \sigma^-_j\sigma^-_{j+1}\Big)  \exp\Big(\sum_{j=1}^{N-1}\alpha\; \sigma^+_j\sigma^+_{j+1}\Big).
\ee
The calculation of the spectra and eigenstates of the transfer matrix was carried out in \cite{Lieb67,RR12} using a Jordan-Wigner transformation. Here, we follow the same approach, but choose to first perform a unitary transformation that flips every odd spin,
\be\label{eq:T2+-v2}
T(\alpha) = U\,\bar T(\alpha)\, U^{-1},\qquad U = \prod_{\substack{1 \le j \le N \\[0.05cm] j = 1 \bmod 2}} \hspace{-0.2cm}\sigma^x_{j}.
\ee
This yields
\be\label{eq:newT}
T^2(\alpha) = \exp \Big( \sum_{j = 1}^{\lfloor \frac{N}2\rfloor} \alpha\,(\sigma_{2j-1}^+ \sigma_{2j}^- + \sigma_{2j}^- \sigma_{2j+1}^+) \Big) \exp \Big( \sum_{j = 1}^{\lfloor \frac{N}2\rfloor} \alpha\,(\sigma_{2j-1}^- \sigma_{2j}^+ + \sigma_{2j}^+ \sigma_{2j+1}^- )\Big),
\ee 
where the convention $\sigma_{N+1}^\pm = 0$ is used. 

The first step in diagonalising $T^2(\alpha)$ is to express it in terms of the operators
\be 
\Cj_j = (-1)^{j-1} \Big(\prod_{k=1}^{j-1}\sigma^z_k \Big)\, \sigma_j^-, \qquad  
\Cjd_j = (-1)^{j-1} \Big(\prod_{k=1}^{j-1}\sigma^z_k \Big)\, \sigma_j^+, \qquad 
(1 \le j \le N),
\ee
which satisfy the usual (fermionic) anticommutation relations,
\be
\{\Cj_j, \Cjd_k\} = \delta_{j,k}, \qquad \{\Cj_j, \Cj_k\} = \{\Cjd_j, \Cjd_k\} = 0.
\ee
With this transformation,
\be
T^2(\alpha) = \exp\Big( \sum_{j = 1}^{\lfloor \frac N 2 \rfloor} \alpha\,(\Cjd_{2j-1} \Cj_{2j}+\Cjd_{2j+1} \Cj_{2j}) \Big)\exp\Big(\sum_{j = 1}^{\lfloor \frac{N}2\rfloor} \alpha \,(\Cjd_{2j} \Cj_{2j-1}+\Cjd_{2j} \Cj_{2j+1}) \Big).
\ee

The next step is to perform a Fourier expansion of $\Cj_j$ and $\Cjd_j$,
\be
\Cj_j = \sqrt{\frac2{N+1}}  \sum_{q=1}^N \,\sin \Big( \frac{\pi q j}{N+1} \Big)\, \Eta_q, \qquad 
\Cjd_j = \sqrt{\frac2{N+1}} \sum_{q=1}^N \,\sin \Big( \frac{\pi q j}{N+1} \Big)\, \Etad_q.
\label{eq:Ceta}\ee
This transformation is unitary, so the anticommutation rules are preserved: 
\be
\{\Eta_p, \Etad_q\} = \delta_{p,q}, \qquad \{\Eta_p, \Eta_q\} = \{\Etad_p, \Etad_q\} = 0.
\ee
In terms of these Fourier coefficients, $T^2(\alpha)$ is expressed as a product of $\lfloor \tfrac{N+1}2\rfloor$ mutually commuting blocks,
\be
T^2(\alpha) = \!\prod_{q=1}^{\lfloor \tfrac{N+1}2\rfloor} \!T_q, \qquad T_q = \exp A^+_q \exp A^-_q, \qquad [T_p,T_q] = 0,
\ee
where
\be
A^\pm_q = \alpha \cos \Big( \frac{\pi q}{N+1}\Big) \Big[ \Etad_q \Eta_q - \Etad_{N+1-q}\Eta_{N+1-q} \pm \Big(\Etad_{N+1-q}\Eta_q - \Etad_q \Eta_{N+1-q}\Big)\Big].
\ee

The blocks $T_q$ can be diagonalised simultaneously. For $N$ odd, $T_{\frac{N+1}2} = \mathbb I$ is already diagonal. The diagonalisation of the other blocks is obtained via the (unitary) transformation
\begin{equation}
\begin{array}{l}
\Eta_q = \displaystyle\frac{1}{\sqrt{2(\mu_q^2+1)}} \Big((\mu_q + 1)\,\Zeta_q + (\mu_q - 1)\,\Zeta_{N+1-q} \Big),
\\[0.8cm]
\Etad_q = \displaystyle\frac{1}{\sqrt{2(\mu_q^2+1)}} \Big((\mu_q + 1)\,\Zetad_q + (\mu_q - 1)\,\Zetad_{N+1-q} \Big),
\end{array} \qquad (1 \le q \le N),
\label{eq:zetadef}
\end{equation}
where 
\be
\mu_q =  \alpha \cos \Big(\frac{\pi q}{N+1}\Big) + \sqrt{ 1+ \alpha^2 \cos^2 \Big(\frac{\pi q}{N+1}\Big)}.
\ee
Again, the commutation rules are preserved:
\be
\{\Zeta_p, \Zetad_q\} = \delta_{p,q}, \qquad \{\Zeta_p, \Zeta_q\} = \{\Zetad_p, \Zetad_q\} = 0.
\ee
Applying the transformation \eqref{eq:zetadef} to $T_q$ yields
\be
T_q = (\mu_q^2)^{\Zetad_q\Zeta_q - \Zetad_{N+1-q}\Zeta_{N+1-q}}.
\label{eq:expAexpA}
\ee
This $T_q$ is diagonal on the four-dimensional space spanned by $|{\sf d} \rangle,\ \Zetad_q|{\sf d} \rangle,\ \Zetad_{N+1-q}|{\sf d} \rangle$ and $\Zetad_{N+1-q} \Zetad_q |{\sf d} \rangle$, with respective eigenvalues $1$, $\mu_q^2$, $\mu_q^{-2}$ and $1$. We note that for $N$ odd, 
\be
\zeta^{}_{\frac{N+1}2} =\Eta_{\frac{N+1}2},\qquad \zeta^\dagger_{\frac{N+1}2} = \Etad_{\frac{N+1}2}.
\ee
The full transfer matrix squared is then obtained from the product of the blocks,
\be
T^2(\alpha) =\!\! \prod_{q=1}^{\lfloor \tfrac{N+1}2\rfloor} (\mu_q^2)^{\Zetad_q\Zeta_q - \Zetad_{N+1-q}\Zeta_{N+1-q}}.
\label{eq:Li}
\ee

The eigenvectors of $T^2(\alpha)$ take the form $|{\sf v}\rangle = \Zetad_{q_1} \Zetad_{q_2} \dots \Zetad_{q_r}|{\sf d}\rangle$, and the corresponding eigenvalues, denoted by $\Lambda$, are obtained from \eqref{eq:Li} by replacing $\Zetad_{q_i}\Zeta_{q_i}$ by $1$ for $i = 1, \dots, r$ and the other $\Zetad_{q}\Zeta_{q}$ by $0$. After a change of variable, $k = N+1-2q$, they can be written in the following convenient form,
\be
\Lambda = \hspace{-0.3cm}\prod_{\substack{k=1\\[0.1cm] k = N-1 \bmod 2}}^{N-1} \hspace{-0.3cm}\Big(\alpha \sin \theta_k + \sqrt{1+\alpha^2 \sin^2 \theta_k} \Big)^{2(1-\delta_k)}\hspace{-0.1cm},\qquad \delta_k = \nu_k + \tau_k, \qquad \theta_k = \frac {\pi k}{2(N+1)},
\label{eq:convL}
\ee 
where the freedom in the presence of the excitations $\Zetad_q$ is now encoded in the quantum numbers $\nu_k, \tau_k \in \{0,1\}$ defined by 
\be 
\begin{array}{l}\nu_k |{\sf v}\rangle = \Zetad_{\frac{N+1+k} 2}\Zeta_{\frac{N+1+k} 2}|{\sf v}\rangle, \\[0.3cm] \tau_k |{\sf v}\rangle= (1 - \Zetad_{\frac{N+1-k} 2}\Zeta_{\frac{N+1-k} 2})|{\sf v}\rangle,\end{array} \qquad (1 \le  k \le N-1,\, k = N-1 \bmod 2).\ee
For $N$ even, there are $2^N$ choices of vectors $(\nu, \tau)$. For $N$ odd, each of the $2^{N-1}$ vectors $(\nu, \tau)$ appears twice due to the block $T_{\frac{N+1}2} = \mathbb I$, and the total number of eigenvalues is again $2^N$, as required.

For $\alpha>0$, the largest eigenvalue\footnote{Because $\Lambda(-\alpha) = \Lambda(\alpha)^{-1}$, the maximal eigenvalue of $T^2(\alpha)$ for $\alpha <0$ instead corresponds to $\nu_k = \tau_k  = 1$ for all $k$.} of $T^2(\alpha)$, denoted by $\Lambda_0$, corresponds to $\nu_k = \tau_k  = 0$ for all $k$. It is non-degenerate for $N$ even, but appears twice for $N$ odd.  From \eqref{eq:convL}, the energies of the ground state and of the excited states are found to be
\begin{subequations}
\begin{align}
E^{\textrm{dim}}_0 &= -\tfrac12 \log \Lambda_0 = \hspace{0.2cm}-\hspace{-0.5cm} \sum_{\substack{k=1\\[0.05cm] k = N-1\bmod 2}}^{N-1} \hspace{-0.4cm}\arcsinh(\alpha \sin \theta_k)\label{eq:E0dim},\\
E^{\textrm{dim}} - E^{\textrm{dim}}_0 &= -\tfrac12 \log \Big( \frac{\Lambda}{\Lambda_0}\Big) = \hspace{-0.3cm} \sum_{\substack{k=1\\[0.05cm] k = N-1\bmod 2}}^{N-1} \hspace{-0.3cm}(\nu_k + \tau_k)\arcsinh(\alpha \sin \theta_k).\label{eq:Esdim}
\end{align}
\end{subequations}

%%%%%%%%%%%%%%%%%%
\subsection{Total magnetisation and variation index}\label{sec:Sz}
%%%%%%%%%%%%%%%%%%

The transfer matrix and its square, respectively, anticommutes and commutes with the total magnetisation $S^z$,
\be
 \{T(\alpha), S^z\} = 0, \qquad [T^2(\alpha), S^z] = 0, \qquad S^z = \tfrac12\sum_{j=1}^N \sigma^z_j.
\ee
This was first observed at the level of $\bar T(\alpha)$ and $\bar T^2(\alpha)$ in \cite{RR12}, where the operator
\be
\mathcal V = U^{-1}\, S^z\, U
\ee
was called the {\it variation index}. 
Each of the eigenspaces
\be
E_N^v = \ctwotimes{N}\big|_{S^z = v \cdot \mathbb I}\, ,  \qquad v \in \{-\tfrac N 2, -\tfrac {N-2} 2, \dots, \tfrac{N-2}2, \tfrac N 2\},
\ee 
with dimension
\be
\dim E_N^v = \binom{N}{\frac N2-v},
\ee
is thus invariant under the action of $T^2(\alpha)$. In particular, the existence of the two one-dimensional subspaces for $v = \pm \frac N 2$ is seen, at the level of the dimers, as coming from the (separate) invariance under $T^2(\alpha)$ of the two row configurations
\be
\psset{unit=.7cm}
\begin{pspicture}[shift=-0.9](-0.2,-0.5)(5.2,1.5)
\psline[linecolor=gray]{-}(0,0)(4.75,0)\psline[linecolor=gray]{-}(0,1)(4.75,1)
\psline[linecolor=gray]{-}(0,0)(0,1)\psline[linecolor=gray]{-}(1,0)(1,1)\psline[linecolor=gray]{-}(2,0)(2,1)\psline[linecolor=gray]{-}(3,0)(3,1)\psline[linecolor=gray]{-}(4,0)(4,1)
\rput(0,0){\dimerv}
\rput(1,0){\halfdimervdown}
\rput(1,1){\halfdimervup}
\rput(2,0){\dimerv}
\rput(3,0){\halfdimervdown}
\rput(3,1){\halfdimervup}
\rput(4,0){\dimerv}
\multiput(0,0)(1,0){5}{\pscircle[linewidth=0.025,fillstyle=solid,fillcolor=gray](0,0){0.08}}
\multiput(0,1)(1,0){5}{\pscircle[linewidth=0.025,fillstyle=solid,fillcolor=gray](0,0){0.08}}
\rput(5,0.5){$\dots$}
\end{pspicture}
\qquad {\rm and}\qquad
\begin{pspicture}[shift=-0.9](-0.2,-0.5)(5.2,1.5)
\psline[linecolor=gray]{-}(0,0)(4.75,0)\psline[linecolor=gray]{-}(0,1)(4.75,1)
\psline[linecolor=gray]{-}(0,0)(0,1)\psline[linecolor=gray]{-}(1,0)(1,1)\psline[linecolor=gray]{-}(2,0)(2,1)\psline[linecolor=gray]{-}(3,0)(3,1)\psline[linecolor=gray]{-}(4,0)(4,1)
\rput(0,0){\halfdimervdown}
\rput(0,1){\halfdimervup}
\rput(1,0){\dimerv}
\rput(2,0){\halfdimervdown}
\rput(2,1){\halfdimervup}
\rput(3,0){\dimerv}
\rput(4,0){\halfdimervdown}
\rput(4,1){\halfdimervup}
\multiput(0,0)(1,0){5}{\pscircle[linewidth=0.025,fillstyle=solid,fillcolor=gray](0,0){0.08}}
\multiput(0,1)(1,0){5}{\pscircle[linewidth=0.025,fillstyle=solid,fillcolor=gray](0,0){0.08}}
\rput(5,0.5){$\dots$}
\end{pspicture} \ \ .
\ee
The total magnetisation can equivalently be expressed in terms of the fermions as
\be
S^z = \sum_{j = 1}^N \,(\Cjd_j \Cj_j - \tfrac12\mathbb I) = \sum_{q = 1}^N \,(\Etad_q \Eta_q - \tfrac12\mathbb I) = \sum_{q = 1}^N \,(\Zetad_q \Zeta_q - \tfrac12\mathbb I).
\label{sz}
\ee
The one-dimensional spaces mentioned above are spanned by $| - - \, \cdots \, - \rangle$ for $v=-\frac N 2$ and $| + + \, \cdots \, + \rangle$ for $v=\frac N 2$. 

Only a subset of the eigenvalues of $T^2(\alpha)$ appear in each sector $E_N^v$. An eigenvalue is thus $v$-admissible if its eigenvector has eigenvalue $v$ under $S^z$: $\Zetad_{q_1} \Zetad_{q_2} \dots \Zetad_{q_r}|{\sf d}\rangle$ is in the sector $E_N^v$ with $v = r-\frac N2$. Equivalently, in terms of the quantum numbers $\nu_k$ and $\tau_k$, an eigenvalue $\Lambda$ as given in \eqref{eq:convL} is $v$-admissible if
\be
\sum_{\substack{k=1\\[0.1cm] k = N-1 \bmod 2}}^{N-1} \hspace{-0.4cm} (\nu_k - \tau_k) = 
\left\{\begin{array}{cl}
v & \ N {\rm \ even},\\[0.1cm]
v + \frac 1 2 {\rm \ or\ } v - \frac 1 2 & \ N {\rm \ odd}.
\end{array}\right.
\label{eq:vadm}
\ee
In particular, the two degenerate eigenvectors of $T_q$ discussed below \eqref{eq:expAexpA}, $|{\sf d} \rangle$ and $\Zetad_q \Zetad_{N+1-q}|{\sf d}\rangle$, belong to different magnetisation sectors.

%%%%%%%%%%%%%%%%%%%%%%%%%%%%
\subsection{Partition functions}
%%%%%%%%%%%%%%%%%%%%%%%%%%%%

In the scaling limit, $M$ and $N$ are taken very large with the ratio $M/N$ converging to a finite non-zero number $\delta$. The finite-size corrections \eqref{eq:Elargen} are extracted from \eqref{eq:convL} using the Euler-Maclaurin formula. The bulk and boundary free energies are given by\footnote{We note that for the dimer model, finite-size corrections of the form \eqref{eq:Elargen}, with only odd powers of $\frac1n$ in the sum, require that we set $n=N+1$. This explains why the boundary free energies given here and in \cite{RR12} are different.}
\be
f_{\textrm{bulk}} = -\frac 1 \pi \int_{0}^{\tfrac\pi2}\arcsinh(\alpha \sin t)\,{\rm d}t, \qquad 
f_{\textrm{bdy}} = \frac12 \arcsinh\alpha,
\ee
whereas the speed of sound and the effective central charge are 
\be 
\vartheta = \alpha, \qquad c_{\textrm{eff}} = 
\left\{
\begin{array}{cl}
-2 & \ N \, \textrm{odd,} \\[0.1cm]
1 & \ N \, \textrm{even.}
\end{array}
\right.
\ee

The full partition function can be split into contributions from the different $E_N^v$ sectors,
\be
Z_v^{\textrm{dim}} = \sum_{v\textrm{-admissible}\, \Lambda} \hspace{-0.4cm}\Lambda^{M/2}, \qquad Z^{\textrm{dim}} = \sum_v Z_v^{\textrm{dim}}.
\ee
The conformal partition functions $\tilde Z_v^{\textrm{dim}}(q)$ and $\tilde Z^{\textrm{dim}}(q)$ are defined as
\be
e^{M(N+1)f_{\textrm{bulk}}+Mf_{\textrm{bdy}}} Z_v^{\textrm{dim}} \; \xrightarrow{M,N \rightarrow \infty} \; \tilde Z_v^{\textrm{dim}}(q), \qquad e^{M(N+1)f_{\textrm{bulk}}+Mf_{\textrm{bdy}}} Z^{\textrm{dim}} \;\xrightarrow{M,N \rightarrow \infty} \; \tilde Z^{\textrm{dim}}(q).
\label{eq:ZZt}
\ee
It was shown in \cite{RR12} that the $\tilde Z_v^{\textrm{dim}}(q)$ are equal to Virasoro Verma module characters,
\be
\tilde Z_v^{\textrm{dim}}(q) = \frac{q^{v^2/2}}{\eta(q)}, \qquad q = e^{-\alpha \pi \delta},
\label{eq:Vermachar}
\ee
for both parities of $N$. The full conformal partition function $\tilde Z^{\textrm{dim}}(q)$ is obtained by summing $\tilde Z_v^{\textrm{dim}}(q)$ over $v \in \mathbb Z$ for $N$ even and $v \in \mathbb Z+\frac12$ for $N$ odd. This yields
\be
\tilde Z^{\textrm{dim}}(q) = \sum_v \tilde Z_v^{\textrm{dim}}(q) = \left\{\begin{array}{ll}
\displaystyle \frac {\theta_3(q)}{\eta(q)}& N\, \textrm{even,} \\[0.5cm]
\displaystyle \frac {\theta_2(q)}{\eta(q)}& N\, \textrm{odd,}
\end{array}\right.
\label{eq:Zdimfinal}
\ee
where $\theta_j(q)$ is the $j$-th standard Jacobi theta function and $\eta(q)$ is the Dedekind eta function:
\be
\theta_2(q) = \sum_{j\in \mathbb Z} q^{\frac12(j+\frac12)^2},\qquad 
\theta_3(q) = \sum_{j\in \mathbb Z} q^{{j^2}/2},\qquad 
\eta(q) = q^{\tfrac1{24}}\prod_{j=1}^\infty (1-q^j).
\ee

%%%%%%%%%%%%%%%%%%%%
\section{Critical dense polymers}
\label{sec:Polymers}
%%%%%%%%%%%%%%%%%%%%

This section reviews definitions and results for critical dense polymers \cite{PR07}.

%%%%%%%%%%%%%%%%%%%%%%
\subsection{Statistical model}
%%%%%%%%%%%%%%%%%%%%%%

The model of critical dense polymers is defined in terms of nonlocal observables called {\it loop segments}. We consider a lattice constructed as an $m \times n$ array of square faces, with $m$ even. Periodicity is imposed 
in the vertical direction, so the lattice is folded on a horizontal cylinder. 
Each of the $mn$ faces is decorated by either 
$\, 
\psset{unit=.35cm}
\begin{pspicture}[shift=-0.2](0,0)(1,1)
\facegrid{(0,0)}{(1,1)}
\loopa
\end{pspicture}
\,$
or
$\,
\psset{unit=.35cm}
\begin{pspicture}[shift=-0.2](0,0)(1,1)
\facegrid{(0,0)}{(1,1)}
\loopb
\end{pspicture}
\,$. 
The midpoint of a face edge is called a {\it node} and this is where the loop segments are attached and connected to those of neighbouring faces, thereby forming longer loop segments in the form of curves on the surface of the cylinder. At both ends of the cylinder, we draw additional loop segments in the form of semi-circles that connect each odd row with the even one just above it. The resulting diagram is a {\it loop configuration} $\sigma$.  \cref{fig:cdpconf} presents an explicit example. 

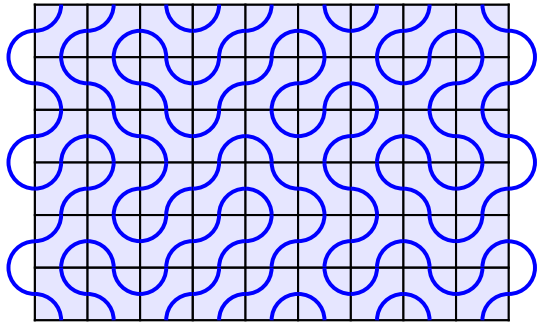
\begin{figure}[h!]
\psset{unit=.7cm}
\begin{center}
\begin{pspicture}(-0.5,0)(9.6,6)
\facegrid{(0,0)}{(9,6)}
\rput(0,5){\loopa}\rput(1,5){\loopa}\rput(2,5){\loopb}\rput(3,5){\loopa}\rput(4,5){\loopa}\rput(5,5){\loopb}\rput(6,5){\loopa}\rput(7,5){\loopa}\rput(8,5){\loopb}
\rput(0,4){\loopb}\rput(1,4){\loopb}\rput(2,4){\loopb}\rput(3,4){\loopa}\rput(4,4){\loopa}\rput(5,4){\loopa}\rput(6,4){\loopb}\rput(7,4){\loopa}\rput(8,4){\loopa}
\rput(0,3){\loopa}\rput(1,3){\loopb}\rput(2,3){\loopb}\rput(3,3){\loopa}\rput(4,3){\loopb}\rput(5,3){\loopa}\rput(6,3){\loopa}\rput(7,3){\loopb}\rput(8,3){\loopb}
\rput(0,2){\loopa}\rput(1,2){\loopa}\rput(2,2){\loopa}\rput(3,2){\loopa}\rput(4,2){\loopb}\rput(5,2){\loopb}\rput(6,2){\loopb}\rput(7,2){\loopb}\rput(8,2){\loopa}
\rput(0,1){\loopa}\rput(1,1){\loopb}\rput(2,1){\loopa}\rput(3,1){\loopa}\rput(4,1){\loopa}\rput(5,1){\loopa}\rput(6,1){\loopa}\rput(7,1){\loopb}\rput(8,1){\loopa}
\rput(0,0){\loopb}\rput(1,0){\loopb}\rput(2,0){\loopa}\rput(3,0){\loopa}\rput(4,0){\loopa}\rput(5,0){\loopb}\rput(6,0){\loopa}\rput(7,0){\loopb}\rput(8,0){\loopa}
\rput(0,0){\psarc[linewidth=1.5pt,linecolor=blue](0,1){.5}{90}{-90}}
\rput(0,2){\psarc[linewidth=1.5pt,linecolor=blue](0,1){.5}{90}{-90}}
\rput(0,4){\psarc[linewidth=1.5pt,linecolor=blue](0,1){.5}{90}{-90}}
\rput(9,0){\psarc[linewidth=1.5pt,linecolor=blue](0,1){.5}{-90}{90}}
\rput(9,2){\psarc[linewidth=1.5pt,linecolor=blue](0,1){.5}{-90}{90}}
\rput(9,4){\psarc[linewidth=1.5pt,linecolor=blue](0,1){.5}{-90}{90}}
\end{pspicture} 
\caption{A configuration of critical dense polymers on the $6\times 9$ cylinder.}
\label{fig:cdpconf}
\end{center}
\end{figure}

Every loop segment is part of a (closed) loop which is either contractible or non-contractible. For the model of critical dense polymers, the fugacity $\beta$ of contractible loops is zero, meaning that a configuration $\sigma$ with one or more such loops is assigned the weight $W(\sigma) = 0$. Other configurations, those without contractible loops and with $n_\gamma$ non-contractible loops, are weighted by 
\be
W(\sigma) = \gamma^{n_{\gamma}} p_1^{\#(\,
\psset{unit=.25cm}
\begin{pspicture}[shift=-0.2](0,0)(1,1)
\facegrid{(0,0)}{(1,1)}
\psarc[linewidth=1.0pt,linecolor=blue](1,0){.5}{90}{180}
\psarc[linewidth=1.0pt,linecolor=blue](0,1){.5}{-90}{0}
\end{pspicture}
\,)}
p_2^{\#(\,
\psset{unit=.25cm}
\begin{pspicture}[shift=-0.2](0,0)(1,1)
\facegrid{(0,0)}{(1,1)}
\psarc[linewidth=1.0pt,linecolor=blue](0,0){.5}{0}{90}
\psarc[linewidth=1.0pt,linecolor=blue](1,1){.5}{180}{270}
\end{pspicture}
\,)}.
\label{eq:Ws}
\ee
Here, $\gamma \in \mathbb R$ is the fugacity of the non-contractible loops, and $p_1, p_2 \in \mathbb R$ are local weights associated with the two types of $1\times 1$ diagrams. More generally, $p_1$ and $p_2$ can be position-dependent and allowed to take different values on the various faces. Finally, the partition function is the sum of the weights over the set of configurations:
\be
Z^{\textrm{cdp}} = \sum_{\sigma} W(\sigma).
\label{eq:Zcdp}
\ee

%%%%%%%%%%%%%%%%%
\subsection{Temperley-Lieb algebra}
%%%%%%%%%%%%%%%%%

The Temperley-Lieb algebra \cite{TL71,Jones} on $n$ sites, $\tl_n(\beta)$, appears naturally in the model of critical dense polymers because its rules for products of connectivities mathematically encode the nonlocal rules that define the loop model. The parameter $\beta \in \mathbb C$ is the fugacity of contractible loops. For critical dense polymers, contractible loops are disallowed, so $\beta=0$. The representation theory of $\tl_n(\beta)$ was studied by Martin~\cite{MartinBook}, Goodman and Wenzl \cite{GW93}, and Westbury \cite{W95}, see also the recent review article by Ridout and Saint-Aubin \cite{RSA1204}.

The elementary objects spanning the algebra $\tl_n(\beta)$ are {\it connectivities}. Let us draw a rectangle with $n$ distinct nodes on the top edge and $n$ more on the bottom one. A connectivity is a pairwise connection of these nodes by non-intersecting loop segments, all drawn inside the rectangle.
For instance, 
\be a_1 =\!
\begin{pspicture}[shift=-0.45](-0.2,-0.5)(3.0,0.5)
\pspolygon[fillstyle=solid,fillcolor=lightlightblue](0,-0.5)(2.8,-0.5)(2.8,0.5)(0,0.5)
\psarc[linecolor=blue,linewidth=1.5pt]{-}(0.4,-0.5){0.2}{0}{180}
\psarc[linecolor=blue,linewidth=1.5pt]{-}(1.6,-0.5){0.2}{0}{180}
\psarc[linecolor=blue,linewidth=1.5pt]{-}(1.2,0.5){0.2}{180}{0}
\psarc[linecolor=blue,linewidth=1.5pt]{-}(2.4,0.5){0.2}{180}{0}
\psbezier[linecolor=blue,linewidth=1.5pt]{-}(0.2,0.5)(0.2,0.0)(1.0,0.0)(1.0,-0.5)
\psbezier[linecolor=blue,linewidth=1.5pt]{-}(0.6,0.5)(0.6,-0.2)(2.2,0.2)(2.2,-0.5)
\psbezier[linecolor=blue,linewidth=1.5pt]{-}(1.8,0.5)(1.8,0)(2.6,0)(2.6,-0.5)
\end{pspicture} 
\ee
is a connectivity in $\tl_7(\beta)$. Two connectivities are considered equal if their nodes have the same connections. The number of distinct connectivities is given by
\be
\dim \tl_n(\beta)  =  \frac{1}{n+1} \begin{pmatrix} 2n \\ n \end{pmatrix}.
\ee
Addition is commutative and the linear combinations of connectivities over $\mathbb C$ are called {\it tangles}. The algebra $\tl_n(\beta)$ is the vector space spanned by the connectivities, endowed with the following product rule. Let $a_1$ and $a_2$ be two connectivities in $\tl_n(\beta)$. The product $a_1a_2$ is obtained by drawing $a_1$ under $a_2$, superposing the top edge of $a_1$ and the bottom edge of $a_2$ in such a way that the $n$ nodes of the two edges match up. The edge common to $a_1$ and $a_2$ is then removed. This produces a bigger rectangle where the nodes from the top and bottom edges are connected, thus giving rise to a new connectivity $a_3$. If closed loops are formed, they are erased and each is replaced by a multiplicative factor of $\beta$. For example,
\be a_1 a_2=\! 
\begin{pspicture}[shift=-0.9](-0.2,-0.5)(3.0,1.5)
\pspolygon[fillstyle=solid,fillcolor=lightlightblue](0,-0.5)(2.8,-0.5)(2.8,0.5)(0,0.5)
\psarc[linecolor=blue,linewidth=1.5pt]{-}(0.4,-0.5){0.2}{0}{180}
\psarc[linecolor=blue,linewidth=1.5pt]{-}(1.6,-0.5){0.2}{0}{180}
\psarc[linecolor=blue,linewidth=1.5pt]{-}(1.2,0.5){0.2}{180}{0}
\psarc[linecolor=blue,linewidth=1.5pt]{-}(2.4,0.5){0.2}{180}{0}
\psbezier[linecolor=blue,linewidth=1.5pt]{-}(0.2,0.5)(0.2,0.0)(1.0,0.0)(1.0,-0.5)
\psbezier[linecolor=blue,linewidth=1.5pt]{-}(0.6,0.5)(0.6,-0.2)(2.2,0.2)(2.2,-0.5)
\psbezier[linecolor=blue,linewidth=1.5pt]{-}(1.8,0.5)(1.8,0)(2.6,0)(2.6,-0.5)
\pspolygon[fillstyle=solid,fillcolor=lightlightblue](0,1.5)(2.8,1.5)(2.8,0.5)(0,0.5)
\psarc[linecolor=blue,linewidth=1.5pt]{-}(1.2,0.5){0.2}{0}{180}
\psarc[linecolor=blue,linewidth=1.5pt]{-}(2.4,0.5){0.2}{0}{180}
\psarc[linecolor=blue,linewidth=1.5pt]{-}(0.4,1.5){0.2}{180}{0}
\psarc[linecolor=blue,linewidth=1.5pt]{-}(2.0,1.5){0.2}{180}{0}
\psbezier[linecolor=blue,linewidth=1.5pt]{-}(0.6,0.5)(0.6,1)(1.8,1)(1.8,0.5)
\psbezier[linecolor=blue,linewidth=1.5pt]{-}(0.2,0.5)(0.2,1.0)(1.0,1.0)(1.0,1.5)
\psbezier[linecolor=blue,linewidth=1.5pt]{-}(1.4,1.5)(1.4,1)(2.6,1)(2.6,1.5)
\end{pspicture}%
\ = \beta^2 
\begin{pspicture}[shift=-0.45](-0.2,-0.5)(3.0,0.5)
\pspolygon[fillstyle=solid,fillcolor=lightlightblue](0,-0.5)(2.8,-0.5)(2.8,0.5)(0,0.5)
\psarc[linecolor=blue,linewidth=1.5pt]{-}(0.4,-0.5){0.2}{0}{180}
\psarc[linecolor=blue,linewidth=1.5pt]{-}(1.6,-0.5){0.2}{0}{180}
\psarc[linecolor=blue,linewidth=1.5pt]{-}(2.4,-0.5){0.2}{0}{180}
\psarc[linecolor=blue,linewidth=1.5pt]{-}(0.4,0.5){0.2}{180}{0}
\psarc[linecolor=blue,linewidth=1.5pt]{-}(2.0,0.5){0.2}{180}{0}
\psline[linecolor=blue,linewidth=1.5pt]{-}(1.0,-0.5)(1.0,0.5)
\psbezier[linecolor=blue,linewidth=1.5pt]{-}(1.4,0.5)(1.4,0)(2.6,0)(2.6,0.5)
\end{pspicture} = \beta^2 a_3 .
\ee
The algebra $\tl_n(\beta)$ is alternatively defined in terms of generators, 
\be
 \tl_n(\beta)=\big\langle I,\,e_j ;\,j=1,\ldots,n-1\big\rangle,\qquad
I =\!
\begin{pspicture}[shift=-0.55](-0.2,-0.65)(2.2,0.45)
\pspolygon[fillstyle=solid,fillcolor=lightlightblue](0,-0.35)(2,-0.35)(2,0.35)(0,0.35)
\rput(1.4,0.0){\small$...$}
\psline[linecolor=blue,linewidth=1.5pt]{-}(0.2,0.35)(0.2,-0.35)\rput(0.2,-0.55){$_1$}
\psline[linecolor=blue,linewidth=1.5pt]{-}(0.6,0.35)(0.6,-0.35)\rput(0.6,-0.55){$_2$}
\psline[linecolor=blue,linewidth=1.5pt]{-}(1.0,0.35)(1.0,-0.35)\rput(1.0,-0.55){$_3$}
\psline[linecolor=blue,linewidth=1.5pt]{-}(1.8,0.35)(1.8,-0.35)\rput(1.8,-0.55){$_n$}
\end{pspicture}  ,
\qquad
 e_j=\!
 \begin{pspicture}[shift=-0.55](-0.2,-0.65)(3.4,0.45)
 \pspolygon[fillstyle=solid,fillcolor=lightlightblue](0,-0.35)(3.2,-0.35)(3.2,0.35)(0,0.35)
\rput(0.6,0.0){\small$...$}
\rput(2.6,0.0){\small$...$}
\psline[linecolor=blue,linewidth=1.5pt]{-}(0.2,0.35)(0.2,-0.35)\rput(0.2,-0.55){$_1$}
\psline[linecolor=blue,linewidth=1.5pt]{-}(1.0,0.35)(1.0,-0.35)
\psline[linecolor=blue,linewidth=1.5pt]{-}(2.2,0.35)(2.2,-0.35)
\psline[linecolor=blue,linewidth=1.5pt]{-}(3.0,0.35)(3.0,-0.35)\rput(3.0,-0.55){$_{n}$}
\psarc[linecolor=blue,linewidth=1.5pt]{-}(1.6,0.35){0.2}{180}{0}\rput(1.35,-0.55){$_j$}
\psarc[linecolor=blue,linewidth=1.5pt]{-}(1.6,-0.35){0.2}{0}{180}\rput(1.85,-0.55){$_{j+1}$}
\end{pspicture},
\ee
satisfying the relations
\be
Ig=gI=g,\qquad e_j^2=\beta e^{}_j, \qquad e_j e_{j\pm1} e_j = e_j, \qquad e_i e_j = e_j e_i \qquad (|i-j|>1),
\ee
where $g$ is any of the generators.

%%%%%%%%%%%%%%%%%%%%%%%%%
\subsection{Standard representations} 
%%%%%%%%%%%%%%%%%%%%%%%%%

Computing physical quantities in the loop model typically requires working with representations of the Temperley-Lieb algebra. A family of representations that appears naturally, the {\it standard representations}, is based on the construction of {\it link states}. Let there be $n$ distinct nodes on a horizontal line. A link state is a diagram where each node is occupied by a loop segment living above the horizontal line. These non-intersecting loop segments either connect nodes pairwise or are vertical loop segments, called {\it defects}, that cannot be overarched. For obvious reasons, the number $d$ of defects is constrained to have the same parity as $n$. The linear span of link states on $n$ nodes with $d$ defects is denoted by $\stan_n^d$. For example,
\be
\begin{pspicture}[shift=-0.15](-0.0,0)(3.2,0.5)
\psline{-}(0,0)(3.2,0)
\psline[linecolor=blue,linewidth=1.5pt]{-}(0.2,0)(0.2,0.5)
\psline[linecolor=blue,linewidth=1.5pt]{-}(0.6,0)(0.6,0.5)
\psarc[linecolor=blue,linewidth=1.5pt]{-}(1.6,0){0.2}{0}{180}
\psarc[linecolor=blue,linewidth=1.5pt]{-}(2.4,0){0.2}{0}{180}
\psbezier[linecolor=blue,linewidth=1.5pt]{-}(1,0)(1,0.7)(3,0.7)(3,0)
\end{pspicture} \ \in \stan_8^2.
\ee

The action of a connectivity $a$ on a link state $w \in \stan_n^d$ closely resembles the rule given for the product of two connectivities. To compute $aw$, one draws $w$ above $a$, erases the common horizontal line segment, reads the new link state from the bottom $n$ nodes and adds a factor of $\beta$ for each contractible loop created (and subsequently erased) in the process. The standard action has an extra rule: If the number of defects has decreased, the result is set to zero. For instance, 
\be
\begin{pspicture}[shift=-0.50](-0,-1)(3.2,0.5)
\pspolygon[fillstyle=solid,fillcolor=lightlightblue](0,-1)(3.2,-1)(3.2,0)(0,0)
\psline[linecolor=blue,linewidth=1.5pt]{-}(0.2,0)(0.2,0.5)
\psline[linecolor=blue,linewidth=1.5pt]{-}(0.6,0)(0.6,0.5)
\psarc[linecolor=blue,linewidth=1.5pt]{-}(1.6,0){0.2}{0}{180}
\psarc[linecolor=blue,linewidth=1.5pt]{-}(2.4,0){0.2}{0}{180}
\psbezier[linecolor=blue,linewidth=1.5pt]{-}(1,0)(1,0.7)(3,0.7)(3,0)
\psline[linecolor=blue,linewidth=1.5pt]{-}(0.2,0)(0.2,-1)
\psline[linecolor=blue,linewidth=1.5pt]{-}(3,0)(3,-1)
\psarc[linecolor=blue,linewidth=1.5pt]{-}(0.8,-1){0.2}{0}{180}
\psarc[linecolor=blue,linewidth=1.5pt]{-}(2.4,-1){0.2}{0}{180}
\psarc[linecolor=blue,linewidth=1.5pt]{-}(2.0,0){0.2}{180}{0}
\psbezier[linecolor=blue,linewidth=1.5pt]{-}(1.4,0)(1.4,-0.5)(2.6,-0.5)(2.6,0)
\psbezier[linecolor=blue,linewidth=1.5pt]{-}(1,0)(1,-0.5)(1.8,-0.5)(1.8,-1)
\psbezier[linecolor=blue,linewidth=1.5pt]{-}(0.6,0)(0.6,-0.5)(1.4,-0.5)(1.4,-1)
\end{pspicture} \ = \beta \
\begin{pspicture}[shift=-0.00](-0.0,0)(3.2,0.5)
\psline{-}(0,0)(3.2,0)
\psline[linecolor=blue,linewidth=1.5pt]{-}(0.2,0)(0.2,0.5)
\psline[linecolor=blue,linewidth=1.5pt]{-}(1.4,0)(1.4,0.5)
\psarc[linecolor=blue,linewidth=1.5pt]{-}(0.8,0){0.2}{0}{180}
\psarc[linecolor=blue,linewidth=1.5pt]{-}(2.4,0){0.2}{0}{180}
\psbezier[linecolor=blue,linewidth=1.5pt]{-}(1.8,0)(1.8,0.5)(3.0,0.5)(3.0,0)
\end{pspicture}\ \ , \qquad \qquad
\begin{pspicture}[shift=-0.50](-0,-1)(3.2,0.5)
\pspolygon[fillstyle=solid,fillcolor=lightlightblue](0,-1)(3.2,-1)(3.2,0)(0,0)
\psline[linecolor=blue,linewidth=1.5pt]{-}(0.2,0)(0.2,0.5)
\psline[linecolor=blue,linewidth=1.5pt]{-}(0.6,0)(0.6,0.5)
\psarc[linecolor=blue,linewidth=1.5pt]{-}(1.6,0){0.2}{0}{180}
\psarc[linecolor=blue,linewidth=1.5pt]{-}(2.4,0){0.2}{0}{180}
\psbezier[linecolor=blue,linewidth=1.5pt]{-}(1,0)(1,0.7)(3,0.7)(3,0)
\psarc[linecolor=blue,linewidth=1.5pt]{-}(1.2,-1){0.2}{0}{180}
\psarc[linecolor=blue,linewidth=1.5pt]{-}(2.4,-1){0.2}{0}{180}
\psarc[linecolor=blue,linewidth=1.5pt]{-}(0.4,0){0.2}{180}{0}
\psarc[linecolor=blue,linewidth=1.5pt]{-}(2.0,0){0.2}{180}{0}
\psbezier[linecolor=blue,linewidth=1.5pt]{-}(0.6,-1)(0.6,-0.5)(1.8,-0.5)(1.8,-1)
\psbezier[linecolor=blue,linewidth=1.5pt]{-}(0.2,-1)(0.2,-0.5)(1.0,-0.5)(1.0,0)
\psbezier[linecolor=blue,linewidth=1.5pt]{-}(1.4,0)(1.4,-0.5)(2.6,-0.5)(2.6,0)
\psline[linecolor=blue,linewidth=1.5pt]{-}(3,0)(3,-1)
\end{pspicture} \ = 0.
\ee
For each $0 \le d \le n$ with $d = n \bmod 2$, this defines a representation of $\tl_n(\beta)$, denoted by $\rho_d$, whose dimension is 
\be
{\rm dim} \,\rho_d = {\rm dim} \, \stan_n^d = {n \choose \frac{n-d}2} - {n \choose \frac{n-d-2}2}.
\ee

%%%%%%%%%%%%%%%%%%%%%
\subsection{Gluing operator}\label{sec:gluing}
%%%%%%%%%%%%%%%%%%%%%

From here onwards, we focus exclusively on the case $\beta = 0$. Forgetting for the moment about the vertical periodicity imposed on the lattice, configurations of the loop model can be viewed as connectivities of $\tl_{n}(\beta = 0)$. The example in \cref{fig:cdpconf} corresponds to
\be
a =
\begin{pspicture}[shift=-0.45](-0.2,-0.5)(3.8,0.5)
\pspolygon[fillstyle=solid,fillcolor=lightlightblue](0,-0.5)(3.6,-0.5)(3.6,0.5)(0,0.5)
\psline[linecolor=blue,linewidth=1.5pt]{-}(0.2,0.5)(0.2,-0.5)
\psarc[linecolor=blue,linewidth=1.5pt]{-}(0.8,-0.5){0.2}{0}{180}
\psarc[linecolor=blue,linewidth=1.5pt]{-}(2.0,-0.5){0.2}{0}{180}
\psarc[linecolor=blue,linewidth=1.5pt]{-}(2.8,-0.5){0.2}{0}{180}
\psarc[linecolor=blue,linewidth=1.5pt]{-}(1.2,0.5){0.2}{180}{0}
\psarc[linecolor=blue,linewidth=1.5pt]{-}(2.4,0.5){0.2}{180}{0}
\psarc[linecolor=blue,linewidth=1.5pt]{-}(3.2,0.5){0.2}{180}{0}
\psbezier[linecolor=blue,linewidth=1.5pt]{-}(0.6,0.5)(0.6,0)(1.4,0)(1.4,-0.5)
\psbezier[linecolor=blue,linewidth=1.5pt]{-}(1.8,0.5)(1.8,0)(3.4,0)(3.4,-0.5)
\end{pspicture}.
\label{eq:a}
\ee
By splitting the loop configuration into the three $2 \times 9$ configurations corresponding to the connectivities
\be
\psset{unit=0.95}
a_1 =
\begin{pspicture}[shift=-0.45](-0.2,-0.5)(3.8,0.5)
\pspolygon[fillstyle=solid,fillcolor=lightlightblue](0,-0.5)(3.6,-0.5)(3.6,0.5)(0,0.5)
\psline[linecolor=blue,linewidth=1.5pt]{-}(0.2,0.5)(0.2,-0.5)
\psarc[linecolor=blue,linewidth=1.5pt]{-}(0.8,0.5){0.2}{180}{0}
\psarc[linecolor=blue,linewidth=1.5pt]{-}(3.2,0.5){0.2}{180}{0}
\psarc[linecolor=blue,linewidth=1.5pt]{-}(2.0,-0.5){0.2}{0}{180}
\psarc[linecolor=blue,linewidth=1.5pt]{-}(2.8,-0.5){0.2}{0}{180}
\psbezier[linecolor=blue,linewidth=1.5pt]{-}(0.6,-0.5)(0.6,0)(1.4,0)(1.4,0.5)
\psbezier[linecolor=blue,linewidth=1.5pt]{-}(1.0,-0.5)(1.0,0)(1.8,0)(1.8,0.5)
\psbezier[linecolor=blue,linewidth=1.5pt]{-}(1.4,-0.5)(1.4,0)(2.2,0)(2.2,0.5)
\psbezier[linecolor=blue,linewidth=1.5pt]{-}(2.6,0.5)(2.6,0)(3.4,0)(3.4,-0.5)
\end{pspicture}, \quad 
a_2 =
\begin{pspicture}[shift=-0.45](-0.2,-0.5)(3.8,0.5)
\pspolygon[fillstyle=solid,fillcolor=lightlightblue](0,-0.5)(3.6,-0.5)(3.6,0.5)(0,0.5)
\psline[linecolor=blue,linewidth=1.5pt]{-}(0.2,0.5)(0.2,-0.5)
\psline[linecolor=blue,linewidth=1.5pt]{-}(0.6,0.5)(0.6,-0.5)
\psline[linecolor=blue,linewidth=1.5pt]{-}(2.6,0.5)(2.6,-0.5)
\psline[linecolor=blue,linewidth=1.5pt]{-}(3.0,0.5)(3.0,-0.5)
\psline[linecolor=blue,linewidth=1.5pt]{-}(3.4,0.5)(3.4,-0.5)
\psarc[linecolor=blue,linewidth=1.5pt]{-}(1.6,-0.5){0.2}{0}{180}
\psarc[linecolor=blue,linewidth=1.5pt]{-}(1.2,0.5){0.2}{180}{0}
\psarc[linecolor=blue,linewidth=1.5pt]{-}(2.0,0.5){0.2}{180}{0}
\psbezier[linecolor=blue,linewidth=1.5pt]{-}(1,-0.5)(1,0)(2.2,0)(2.2,-0.5)
\end{pspicture}, \quad 
a_3 =
\begin{pspicture}[shift=-0.45](-0.2,-0.5)(3.8,0.5)
\pspolygon[fillstyle=solid,fillcolor=lightlightblue](0,-0.5)(3.6,-0.5)(3.6,0.5)(0,0.5)
\psline[linecolor=blue,linewidth=1.5pt]{-}(0.2,0.5)(0.2,-0.5)
\psline[linecolor=blue,linewidth=1.5pt]{-}(0.6,0.5)(0.6,-0.5)
\psline[linecolor=blue,linewidth=1.5pt]{-}(3.0,0.5)(3.0,-0.5)
\psline[linecolor=blue,linewidth=1.5pt]{-}(3.4,0.5)(3.4,-0.5)
\psarc[linecolor=blue,linewidth=1.5pt]{-}(1.6,-0.5){0.2}{0}{180}
\psarc[linecolor=blue,linewidth=1.5pt]{-}(2.4,-0.5){0.2}{0}{180}
\psarc[linecolor=blue,linewidth=1.5pt]{-}(1.2,0.5){0.2}{180}{0}
\psarc[linecolor=blue,linewidth=1.5pt]{-}(2.4,0.5){0.2}{180}{0}
\psbezier[linecolor=blue,linewidth=1.5pt]{-}(1,-0.5)(1,0)(1.8,0)(1.8,0.5)
\end{pspicture}, 
\ee
the connectivity $a$ can be expressed as the product $a = a_1 a_2 a_3$.

The partition function $Z^{\textrm{cdp}}$ can be computed using the linear operator 
\be 
\mathcal F\;:\; \tl_n(0) \rightarrow \mathbb C.
\label{eq:glue}
\ee
For each connectivity $a$, $\mathcal F(a)$ outputs the product of the fugacities of the loops produced by folding $a$ around the cylinder and gluing together its top and bottom edges. We therefore call $\mathcal F$ the {\it gluing operator}. If the folding produces contractible loops, then $\mathcal F(a) = 0$. If only non-contractible loops are formed, then $\mathcal F(a) = \gamma^{n_\gamma}$, with $\gamma$ and $n_\gamma$ defined below \eqref{eq:Ws}. For example, for \eqref{eq:a}, $\mathcal F(a) = \gamma^3$. 

One known realisation of $\mathcal F$ is in terms of traces in the standard representations \cite{RJ06,MDSA11}:
\be
\mathcal F(a) = \hspace{-0.3cm}\sum_{\substack{d = 0\\[0.05cm] d = n \bmod 2}}^n \hspace{-0.3cm} U_d(\tfrac \gamma 2) \,\textrm{Tr}\, \rho_d(a),
\label{eq:F}
\ee
where $U_k(x)$ is the $k$-th Chebyshev polynomial of the second kind. The linearity of the realisation \eqref{eq:F} of $\mathcal F$ is inherited from the same property of the matrix trace.

%%%%%%%%%%%%%%%%%%
\subsection{Transfer tangle}\label{sec:Du}
%%%%%%%%%%%%%%%%%%

It is possible to encode all the connectivities needed for the calculation of $Z^{\textrm{cdp}}$ in a single $\tl_n(0)$ tangle, the {\em double-row transfer tangle} $\Db(u,\xi)$. It is defined as
\be 
\Db (u,\xi) = \frac 1 {\sin 2u} \ 
\psset{unit=0.9}
\begin{pspicture}[shift=-0.9](-0.5,-0.0)(5.5,2.0)
\facegrid{(0,0)}{(5,2)}
\psarc[linewidth=0.025]{-}(0,0){0.16}{0}{90}
\psarc[linewidth=0.025]{-}(1,1){0.16}{90}{180}
\psarc[linewidth=0.025]{-}(1,0){0.16}{0}{90}
\psarc[linewidth=0.025]{-}(2,1){0.16}{90}{180}
\psarc[linewidth=0.025]{-}(4,0){0.16}{0}{90}
\psarc[linewidth=0.025]{-}(5,1){0.16}{90}{180}
\rput(2.5,0.5){$\ldots$}
\rput(2.5,1.5){$\ldots$}
\rput(3.5,0.5){$\ldots$}
\rput(3.5,1.5){$\ldots$}
\psarc[linewidth=1.5pt,linecolor=blue]{-}(0,1){0.5}{90}{-90}
\psarc[linewidth=1.5pt,linecolor=blue]{-}(5,1){0.5}{-90}{90}
\rput(0.5,.5){\scriptsize$u\!+\!\xi_1$}
\rput(0.5,1.5){\scriptsize$u\!-\!\xi_1$}
\rput(1.5,.5){\scriptsize$u\!+\!\xi_2$}
\rput(1.5,1.5){\scriptsize$u\!-\!\xi_2$}
\rput(4.5,.5){\scriptsize$u\!+\!\xi_n$}
\rput(4.5,1.5){\scriptsize$u\!-\!\xi_n$}
\end{pspicture} \ ,
\label{eq:Du}
\ee  
where each square tile is called a {\it face operator} and is formally given by
\be
\psset{unit=.7cm}
\begin{pspicture}[shift=-.4](1,1)
\facegrid{(0,0)}{(1,1)}
\psarc[linewidth=0.025]{-}(0,0){0.16}{0}{90}
\rput(.5,.5){\small $u$}
\end{pspicture}
\ =\ 
\cos u\ \, \begin{pspicture}[shift=-.4](1,1)
\facegrid{(0,0)}{(1,1)}
\rput[bl](0,0){\loopa}
\end{pspicture}
\;+\,\sin u\ \,
\begin{pspicture}[shift=-.4](1,1)
\facegrid{(0,0)}{(1,1)}
\rput[bl](0,0){\loopb}
\end{pspicture} \ .
\label{eq:1x1}
\ee
The small quarter-circle in the lower-left corner of 
$\,
\psset{unit=.4cm}
\begin{pspicture}[shift=-.2](0,0)(1,1)
\facegrid{(0,0)}{(1,1)}
\psarc[linewidth=0.025]{-}(0,0){0.25}{0}{90}
\rput(.5,.5){\scriptsize $u$}
\end{pspicture}\,
$
is a marker indicating the orientation of the diagrams in the decomposition. A face operator with its marker in another corner is obtained by the corresponding rotation of the diagrams on the right-hand side of \eqref{eq:1x1}. The parameter $u \in \mathbb C$ is the {\it spectral parameter} while $\xi_1, \xi_2, \dots, \xi_n \in \mathbb C$ are {\it inhomogeneities}. We often use the shorthand notation $\xi = (\xi_1, \xi_2, \dots, \xi_n)$, as in the left-hand side of \eqref{eq:Du}. We note that the normalisation of $\Db(u,\xi)$ ensures that the 0th order term in its Taylor series expansion in $u$ is non-zero. For $\xi = 0$, the resulting term is a multiple of the identity connectivity.

As a $\tl_n(0)$-tangle, $\Db(u,\xi)$ is defined in a diagrammatic way. Each face operator has two contributions from which one obtains a linear combination of $4^n$ diagrams. Each of these takes the form of a  $2\times n$ loop configuration from which a connectivity can be read off. Some of these configurations contain contractible loops and are discarded because $\beta = 0$. $\Db(u,\xi)$ is defined to be the linear combination of the remaining ones, each weighted by the corresponding powers of $p_1$ and $p_2$, here parametrised by the position-dependent functions $\cos(u \pm \xi_i)$ and $\sin(u \pm \xi_i)$. Two transfer tangles with identical homogeneities commute \cite{BPO96,PRZ06}, 
\be 
[\Db(u,\xi), \Db(v,\xi)] = 0, \qquad u,v \in \mathbb C.
\label{eq:Ducomm}
\ee
The parameter $u$ in $\Db(u,\xi)$ thus explores a family of commuting matrices --- accordingly, it is usually referred to as {\it spectral}. In stark contrast, $\Db(u,\xi)$ and $\Db(u,\xi')$ do not commute in general if $\xi \neq \xi'$, and the inhomogeneities are therefore not spectral parameters. The multi-parameter definition of $\Db(u,\xi)$ provides a considerable freedom in choosing the position-dependent weights
$p_1$ and $p_2$ in \eqref{eq:Ws} while ensuring that the model is integrable. 

The tangle $\big(\Db(u,\xi)\big)^{m/2}$ is a linear combination of connectivities, each weighted by appropriate powers of local weights such as $p_1$ and $p_2$ in (\ref{eq:Ws}).
They include those produced from the loop configurations contributing to $Z^{\textrm{cdp}}$, but also some which are mapped to zero by the gluing operator. Applying $\mathcal F$ to this linear combination removes the unwanted connectivities and correctly assigns $\gamma^{n_\gamma}$ to the surviving ones, so that
\be
Z^{\textrm{cdp}} = \mathcal F \big(\Db(u,\xi)^{m/2}\big).
\ee
From \eqref{eq:F}, the calculation of $Z^{\textrm{cdp}}$ amounts to finding the eigenvalues of the {\it transfer matrices} $\rho_d\big(\Db(u,\xi)\big)$ for each $d$.

%%%%%%%%%%%%%%%%%%%
\subsection{Spectra}\label{sec:cdpspec}
%%%%%%%%%%%%%%%%%%%

The technique used to find closed forms for the eigenvalues of $\rho_d\big(\Db(u,\xi)\big)$ mimics a method originally used to obtain the spectrum of the Ising model transfer matrix \cite{BaxterBook,OPW96}. The idea is to show that two transfer matrices, with their spectral parameters related in a specific way, are inverses to one another, up to a scalar function of these spectral parameters. The eigenvalues are then computed from this scalar function. 

For critical dense polymers, an inversion identity indeed exists. In fact, it is satisfied at the level of the algebra $\tl_n(0)$, in the sense that the product of two specifically related transfer tangles gives a scalar multiple of the identity connectivity. This was first shown in \cite{PR07} for $\Db(u,\xi)$ for $\xi = 0$ and later generalised to other geometries and boundary conditions \cite{PRV10,PRV12,PRT14}. The ideas used for $\Db(u,\xi)$ at $\xi = 0$ easily extend to the general case where $\xi$ is free. The result is
\begin{subequations}\label{eq:Geninv}
\be
\Db(u,\xi)\Db(u+\tfrac \pi 2,\xi) = I\, \bigg( \frac{f_1(u,\xi) - f_2(u,\xi)}{\cos^2 u - \sin^2 u}\bigg)^2
\label{eq:geninv}
\ee
with
\be 
f_1(u,\xi) = \prod_{j=1}^n \cos(u-\xi_j)\cos(u+\xi_j), \qquad f_2(u,\xi) = \prod_{j=1}^n \sin(u-\xi_j)\sin(u+\xi_j).
\label{eq:geninv2}
\ee
\end{subequations}
In any representation, each (parameter-dependent) eigenvalue $\textrm{Eig}\big(\Db(u,\xi)\big)$ of $\Db(u,\xi)$ then satisfies the relation
\be
\textrm{Eig}\big(\Db(u,\xi)\big)\textrm{Eig}\big(\Db(u+\tfrac \pi 2,\xi)\big) = \bigg( \frac{f_1(u,\xi) - f_2(u,\xi)}{\cos^2 u - \sin^2 u}\bigg)^2.
\label{eq:EE1}
\ee

Factorising the right-hand side of \eqref{eq:EE1} is not possible in the general case, but it can be done for special choices of $\xi$. For $\xi=0$, this factorisation yields the following expressions for the eigenvalues, 
\be
\textrm{Eig}\big(\Db(u,0)\big) = \hspace{-0.1cm}\prod_{j=1}^{\lfloor \tfrac {n-1} 2 \rfloor} \Big(1 + \epsilon_j \sin2u\, \sin t_j\Big) \Big(1 + \mu_j \sin 2u \,\sin t_j\Big), \ \quad 
t_j = \left\{\begin{array}{c l}
\frac{j \pi}{n} & n\, \textrm{even,} \\[0.2cm]
\frac{(2j-1) \pi}{2n} & n\, \textrm{odd,}
\end{array}\right.
\label{eq:PReigs}
\ee
whose forms are simplified here compared to those in \cite{PR07}. Different eigenvalues correspond to different choices of $\epsilon_j, \mu_j \in \{ +1, -1 \}$. With $\nu_k=\frac{1-\epsilon_k}{2}$ and $\tau_k=\frac{1-\mu_k}{2}$, these can be recast into
\be
\textrm{Eig}\big(\Db(u,0)\big) = \hspace{-0.3cm}\prod_{\substack{k = 1 \\[0.05cm] k = n\bmod 2}}^{n-2}\hspace{-0.3cm} \Big(1-\sin^2 2u\, \sin^2 q_k\Big) \bigg[\frac{1+\sin2u\, \sin q_k}{1-\sin2u\,\sin q_k}\bigg]^{1-\delta_k}\hspace{-0.1cm}, \quad \delta_k = \nu_k + \tau_k, \quad  q_k = \frac{\pi k}{2n},
\label{eq:speccdp}
\ee
where the eigenvalues are now specified by a selection of $\nu_k, \tau_k \in \{0, 1\}$. Eigenvalues of $\Db(u,0)$ are denoted by $\lambda$. For $\sin 2u >0$, the largest one, $\lambda_0$, is obtained by setting $\nu_k = \tau_k = 0$ for all $k$, for both $n$ even and odd.\footnote{For $\sin 2u <0$, the maximal eigenvalue instead corresponds to $\nu_k = \tau_k = 1$ for all $k$.} Because $\Db(u,0)$ covers two rows of $n$ tiles each, we include, by convention, a factor of $\tfrac 12$ in the definition of the energies:
\begin{subequations}
\begin{align}
E_0^{\textrm{cdp}} &= -\tfrac 12 \log \lambda_0 = -\hspace{-0.3cm}\sum^{n-2}_{\substack{k=1 \\[0.05cm] k = n \bmod 2}}\hspace{-0.2cm} \log(1+\sin2u\, \sin q_k), \label{eq:E0cdp}\\
E^{\textrm{cdp}} - E_0^{\textrm{cdp}} &= -\tfrac 12 \log \Big(\frac{\lambda}{\lambda_0}\Big) = \hspace{-0.3cm}\sum^{n-2}_{\substack{k=1 \\[0.05cm] k = n \bmod 2}}\hspace{-0.2cm} \delta_k\, \log \sqrt \frac{1+\sin 2u\,\sin q_k}{1-\sin 2u\, \sin q_k}.
\end{align}
\end{subequations}

In a given standard representation $\rho_d$, there are restrictions on the possible values that the  quantum numbers $\nu_k$ and $\tau_k$ can take (otherwise, the number of eigenvalues would exceed the dimension of $\rho_d(\Db(u,0))$). One therefore needs to state which pairs of vectors $(\nu,\tau)$ are retained to form the eigenvalues in a given standard representation. The appropriate admissibility criterion was conjectured in~\cite{PR07} and later proven in~\cite{MD11}, and is more complicated than the one for the dimer model. This criterion states that a pair $(\nu,\tau)$ is $d$-admissible, that is, defines through \eqref{eq:speccdp} an eigenvalue of $\rho_d(\Db(u,0))$, if its components satisfy the two following constraints,
\be
 \sum_{\substack {k \ge K \\[0.05cm] k = n \bmod 2}} \hspace{-0.2cm}(\nu_k - \tau_k) \ge 0\quad \  (K \in\mathbb{N}), \qquad  \sum_{\substack {k = 1 \\[0.05cm] k = n \bmod 2}}^{n-2} \hspace{-0.2cm}(\nu_k - \tau_k) = \left\{ 
 \begin{array}{cl}
 \tfrac d 2 \ \textrm{or} \ \tfrac{d-2}2 & n \,\textrm{even,} \\[0.15cm]
  \tfrac {d-1} 2  & n \,\textrm{odd.} \\
 \end{array}\right.
\label{adm}
\ee
Although the values $\nu_k,\tau_k$ are severely constrained, all the possible eigenvalues \eqref{eq:speccdp} appear in at least one standard representation, that is, every $\delta$ with $\delta_k \in \{0,1,2\}$ corresponds to a $d$-admissible pair $(\nu,\tau)$ for at least one value of $d$.

%%%%%%%%%%%%%%%%%%%%%
\subsection{Partition functions}\label{sec:partfcdp}
%%%%%%%%%%%%%%%%%%%%%%

Equipped with the closed-form expressions for the eigenvalues and their degeneracy criteria, one can calculate the cylinder partition functions restricted to each $\rho_d$ as well as the full partition function:
\be
Z^{\textrm{cdp}}_d = \textrm{Tr} \, \big(\rho_d\big(\Db(u,0)^{m/2}\big)\big) = \hspace{-0.15cm} \sum_{d\textrm{-admissible}\, \lambda} \hspace{-0.3cm} {\rm mult}_\lambda\,\lambda^{m/2}, \qquad Z^{\textrm{cdp}} = \hspace{-0.3cm} \sum_{\substack{d=0 \\ d = n \bmod 2}}^n \hspace{-0.3cm}U_d(\tfrac \gamma 2) \, Z^{\textrm{cdp}}_d. 
\label{eq:Zscdp}
\ee
We now study the behaviour of the partition function as $m,n \rightarrow \infty$ and $\frac mn \rightarrow \delta$, with $\delta\in \mathbb R$. The continuum scaling limit will be discussed further in \cref{sec:dimH}. By performing an Euler-Maclaurin expansion of $E_0^{\textrm{cdp}}$ in \eqref{eq:E0cdp}, the bulk and boundary free energies are found to be
\be
f_{\textrm{bulk}} = -\frac1\pi \int_{0}^{\pi/2} \log \big(1+\sin 2u\, \sin t\big) {\rm d}t, \qquad \quad f_{\textrm{bdy}} = \frac 12\log \big(1+\sin 2u \big),
\ee
while the speed of sound and effective central charge are 
\be 
\vartheta = \sin 2u, \qquad c_{\textrm{eff}} 
= \left\{\begin{array}{cl}
-2 & n \, \textrm{even,} \\[0.1cm]
1 & n \, \textrm{odd.}
\end{array}\right.
\ee

The conformal partition functions are obtained by renormalising the partition functions to remove the non-universal terms and by taking the scaling limit:
\be 
e^{mnf_{\textrm{bulk}}+mf_{\textrm{bdy}}} \: Z^{\textrm{cdp}}_d \; \xrightarrow{m,n \rightarrow \infty} \; \tilde Z^{\textrm{cdp}}_d(q), \qquad  e^{mnf_{\textrm{bulk}}+mf_{\textrm{bdy}}} \: Z^{\textrm{cdp}} \;  \xrightarrow{m,n \rightarrow \infty} \; \tilde Z^{\textrm{cdp}}(q).
\ee
For each $d$, $\tilde Z^{\textrm{cdp}}_d(q)$ produces the so-called {\it Kac character} $\chit_{1,d+1}(q)$ \cite{PRZ06,PR07},
\be
\tilde Z^{\textrm{cdp}}_d(q) = \chit_{1,d+1}(q)=  \frac{q^{\frac{(d-1)^2}8}}{\eta(q)}(1-q^{d+1}), \qquad q = e^{-\pi \delta \sin 2u}.
\ee
From \eqref{eq:Zscdp}, this gives an explicit expression for the full conformal partition function $\tilde Z^{\textrm{cdp}}(q)$, for all $\gamma$. Because $U_d(1) = d+1$, the result simplifies significantly for $\gamma = 2$:
\be
\tilde Z^{\textrm{cdp}}(q)\big|_{\gamma = 2} = \sum_d \,(d+1)\,\tilde Z^{\textrm{cdp}}_d(q) = \left\{\begin{array}{ll}
\displaystyle \frac {2\,\theta_2(q)}{\eta(q)}& n\, \textrm{even,} \\[0.5cm]
\displaystyle \frac {2\,\theta_3(q)}{\eta(q)}& n\, \textrm{odd.}
\end{array}\right. 
\label{eq:Zcdpfinal}
\ee
The similarity between \eqref{eq:Zdimfinal} and \eqref{eq:Zcdpfinal} is striking and is discussed in \cref{sec:dimersandloops}.

%%%%%%%%%%%%%%%%%%%%
\section{Dimers, loops and integrability}
\label{sec:dimersandloops}
%%%%%%%%%%%%%%%%%%%%

One of our goals is to study the conformal properties of the dimer model. In this section, we initiate this investigation by establishing a relationship between the dimer model and the critical dense polymer model, as finite-size lattice models.

%%%%%
\subsection{Comparison of spectra}\label{sec:spectra}
%%%%%

A priori, the models of dimers and critical dense polymers are quite different. The objects defining the former, the dimers, cover only two neighbouring sites and are in this sense local objects, while the polymer model is based on nonlocal entities, the loop segments, that can extend over long distances but not form contractible loops. Each model can be solved by analytically computing the eigenvalues of a transfer matrix, although the techniques employed to find their closed forms are different. Another distinguishing feature is that the loop transfer tangles $\Db(u,0)$ and their matrix representatives form a commuting family, while this is not the case for the dimer transfer matrices $T(\alpha)$. 

Nevertheless, one sees a formal but striking similarity in the finite-size spectra \eqref{eq:convL} and \eqref{eq:speccdp} of $T^2(\alpha)$ and $\Db(u,0)$ provided the system sizes are related as
\be
n=N+1.
\label{eq:Nn}
\ee
In what follows, we will assume this relation. As observed in \cite{RR12}, the similarity of the spectra strengthens when we look at the finite-size corrections of the energies. In both models, the ground-state energy is of the form
\be
E_0 = -\hspace{-0.3cm}\sum_{k=1 \atop k = n \bmod 2}^{n-2} \hspace{-0.1cm} \omega(q_k), \qquad q_k = \frac{\pi k}{2n}\,, 
\label{eq:E0}
\ee
but with model-dependent functions $\omega(t)$,
\be
\omega^{\rm dim}(t) = {\rm arcsinh}(\alpha \sin{t}), \qquad \omega^{\rm cdp}(t) = \log{(1 + \zeta \sin{t})},
\label{eq:w(t)}
\ee
each depending on an extra parameter, $\alpha$ or $\zeta$, where 
\be 
\zeta = \sin 2u. 
\ee

In both models, the $\frac1n$ expansion of $E_0$ is computed using the Euler-MacLaurin formula and expressed in terms of the series expansion of the corresponding $\omega(t)$,
\be
\omega(t) = \sum_{p=1}^\infty \: {\lambda_p \over p!} \, t^p\,.
\ee
The ground-state energy $E_0$ then has an expansion of the form \eqref{eq:Elargen} in the variable $n$, with
\begin{subequations}
\begin{alignat}{3}
& f_{\rm bulk} = -{1 \over \pi} \int_0^{\pi \over 2} \omega(t)\,{\rm d}t , \qquad && f_{\rm bdy} = \tfrac 12\,\omega(\textstyle{\pi \over 2}) , \\[.1cm] 
& a_p^{(0)} = {\pi^{2p-1} {\rm B}_{2p}(r) \over (2p)!} \, \lambda_{2p-1}, \qquad && r = \tfrac n2 \bmod 1, \quad r \in \{0,\textstyle{1 \over 2}\},
\end{alignat}
\end{subequations}
where B$_{2p}(z)$ are Bernoulli polynomials. Useful values are 
\be
{\rm B}_2(0) = \frac 16, \quad {\rm B}_4(0) = -\frac 1{30}, \quad {\rm B}_6(0) = \frac1{42}, \;\cdots \quad {\rm and} \quad {\rm B}_{2p}(\tfrac12) =  \Big({1 \over 2^{2p-1}}-1\Big){\rm B}_{2p}(0),
\ee
where the B$_{2p}(0)$ are the Bernoulli numbers. It is noted that, although the series expansion of $\omega(t)$ may involve both even and odd powers, the large $n$ expansion of $E_0$ depends only on the coefficients of the odd powers. For $p=1$, setting $\vartheta=\lambda_1$ and comparing with \eqref{eq:Elargen} yields the value of the effective central charge,
\be
c_{\rm eff} = -12 \, {\rm B}_2(r) = \left\{\begin{array}{cl} -2 & n \,{\rm even,} \\ 1 & n \,{\rm odd.}\end{array}\right.
\label{ceff}
\ee

The analysis of the energy gaps is similar. In both models, they are given by
\be\label{eq:DE}
\Delta E = E - E_0 = \sum_{k=1 \atop k = n \bmod 2}^{n-2} \hspace{-0.3cm} \tfrac12 \, (\nu_k + \tau_k) \,[\omega(q_k) - \omega(-q_k)]\,.
\ee
The odd combination $\omega(t) - \omega(-t)$ in this expression implies that the large $n$ expansion of the energy gaps, as in the case of $E_0$, only depends on the odd-indexed coefficients $\lambda_{2p-1}$. In fact, for the dimer model, $\omega(t)$ is an odd function, so the antisymmetric combination $\frac{\omega(t) - \omega(-t)}2$ coincides with $\omega(t)$. As the finite energy excitations involve only a finite number of non-zero $\nu_k$ and $\tau_k$, it is not difficult to see that the $\frac1n$ expansion for the energy gaps yields
\be
a_p - a_p^{(0)} = \Big({\pi \over 2}\Big)^{2p-1} \, {\lambda_{2p-1} \over (2p-1)!} \sum_{k \ge 1 \atop k= n \bmod 2}  \hspace{-0.3cm}k^{2p-1}\,(\nu_k + \tau_k).
\ee

Putting it all together, we find that, in both models, the energy levels are unequivocally specified in terms of a set of admissible numbers $\{\nu_k,\tau_k\}$ and an appropriate function, $\omega^{\rm dim}$ or $\omega^{\rm cdp}$. Albeit different, the finite-size corrections in the two models are structurally identical to all orders in terms of $n$ and the corresponding function $\omega$. Explicitly, the expansion coefficients read
\be
a_p = {\pi^{2p-1} \over (2p-1)!} \, \lambda_{2p-1} \Bigg[{{\rm B}_{2p}(r) \over 2p} + {1 \over 2^{2p-1}}\hspace{-0.3cm} \sum_{k \ge 1 \atop k= n \bmod 2}\hspace{-0.3cm}  k^{2p-1}\,(\nu_k + \tau_k)  \Bigg], 
\label{ap}
\ee
and the only model dependence is in the factors $\lambda_{2p-1}$. These factors are odd polynomials of degree $2p-1$ in their respective parameter, but are distinct, as the first few values demonstrate:
\begin{subequations}
\begin{alignat}{3}
&\lambda_{1}^{\rm dim}(\alpha) = \alpha, \qquad &&\lambda_{3}^{\rm dim}(\alpha) = -\alpha(\alpha^2 + 1),\qquad &&\lambda_{5}^{\rm dim}(\alpha) = \alpha(9\alpha^4 + 10\alpha^2 + 1), \\
&\lambda_1^{\rm cdp}(\zeta)=\zeta, && \lambda_3^{\rm cdp}(\zeta) = \zeta(2\zeta^2 - 1), && \lambda_5^{\rm cdp}(\zeta) = \zeta(24\zeta^4 - 20 \zeta^2 + 1).
\end{alignat}
\end{subequations}
We see that the finite-size correction coefficients $a_p$ are not universal. However, the ratios $a_p/\lambda_{2p-1}$, or equivalently the ratios $a_p/a_p^{(0)}$, are believed to be universal. 
The universal nature of the similar ratios for the Ising model and critical dense polymers was discussed in~\cite{IH01} and~\cite{IRH12}, respectively.

Up until now, the analysis of the finite-size corrections has been performed without any assumptions regarding the central charge of the underlying conformal field theory. In the remainder of this section, we will work under the hypothesis that the dimer model, like critical dense polymers, 
can be described by a conformal field theory with $c=-2$. We will return to this issue in \cref{sec:CFT}.

With this assumption, it follows that the set of distinct conformal weights in the scaling limit, computed from the coefficients $a_1$, 
\be
\Delta = {c - c_{\rm eff} \over 24} + {1 \over 2} \hspace{-0.2cm}\sum_{k \ge 1 \atop k= n \bmod 2} \hspace{-0.2cm} k \, (\nu_k + \tau_k),
\ee
is exactly the same in the two models. The two conformal spectra, however, are {\em not} identical, because the eigenvalue degeneracies are different. For instance, for fixed $d$ and $v$, two pairs $(\nu_k,\tau_k)$ and $(\nu'_k,\tau'_k)$ may be such that $\nu_k + \tau_k = \nu'_k+\tau'_k$ for all $k$, with one being both $d$- and $v$-admissible while the other is only $v$-admissible. In fact, the admissibility criteria \eqref{eq:vadm} and \eqref{adm} show that the vectors $(\nu, \tau)$ admissible in $\stan_n^d$ are a subset of those admissible in $E_{n-1}^v$, $v = \frac{d-1}2$.

This can also be seen from their conformal partition functions, when the two models are defined on a cylinder of length $n=N+1$ and perimeter $m=M$. The conformal partition functions of both models can be written in terms of the $c=-2$ irreducible Virasoro characters
\be
\ch_{r,s}(q) =q^{\frac{1-c}{24}+\Delta_{r,s}} \frac{(1-q^{rs})}{\eta(q)}, \qquad (r \in 
\mathbb Z_{\ge 0},\ s = 1,2),
\ee
where $r,s$ are Kac labels and the conformal weight is 
\be
\Delta_{r,s} = {(2r-s)^2-1 \over 8}.
\label{KacD}
\ee
For the two models considered here, the explicit forms of the cylinder partition functions strongly depend on the parity of $n$. They read, for $n$ even,
\begin{subequations}\label{eq:char}
\begin{alignat}{3}
&\tilde Z_v^{\rm dim}(q) = \sum_{r=0}^\infty  \, \ch_{|v| + r+\frac12,1}(q) \quad (v \in \mathbb Z + \tfrac 12), \qquad &&\tilde Z^{\rm cdp}_d(q) = \ch_{\frac d 2,1}(q) + \ch_{\frac {d+2} 2,1}(q) \quad (d \in 2 \mathbb Z),
\label{eq:chareven}\\
\intertext{whereas, for $n$ odd,}
&\tilde Z_v^{\rm dim}(q) = \sum_{r=0}^\infty  \, \ch_{|v| + 2r+1,2}(q) \quad (v \in \mathbb Z), \qquad &&\tilde Z^{\rm cdp}_d(q) = \ch_{\frac{d+1}2,2}(q) \quad (d \in 2 \mathbb Z+1),
\label{eq:charodd}
\end{alignat}
\end{subequations}
where we note that $\ch_{0,1}(q) = 0$.

\cref{sec:dimtrees,sec:dimloops,sec:dimerrep} 
provide a deeper understanding of the similarities between the spectra, by building a direct connection between the two models. We also emphasise at this point that the conformal partition functions are mere generating functions for the conformal spectra; they contain very little information about the {\em structure} of the Virasoro representations making up the spectra. We will come back to this in \cref{sec:structure}.

%%%%%%%%%%%%%%%%%%%%
\subsection{Dimers and webs}\label{sec:dimtrees}
%%%%%%%%%%%%%%%%%%%%

\begin{figure}[h!]
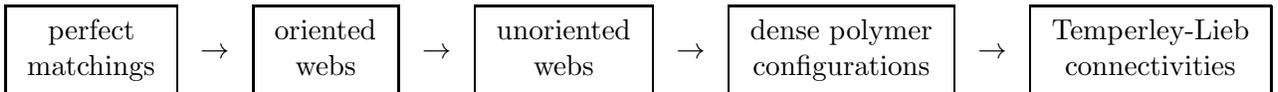
 
\begin{equation*}
\boxed{\begin{array}{c} \textrm{perfect} \\ \textrm{matchings} \end{array}} \hspace{0.2cm} \rightarrow \hspace{0.2cm}
\boxed{\begin{array}{c} \textrm{oriented} \\ \textrm{webs} \end{array}} \hspace{0.2cm}\rightarrow\hspace{0.2cm}
\boxed{\begin{array}{c} \textrm{unoriented} \\ \textrm{webs}  \end{array}}\hspace{0.2cm}\rightarrow\hspace{0.2cm}
\boxed{\begin{array}{c} \textrm{dense polymer} \\ \textrm{configurations} \end{array}}\hspace{0.2cm}\rightarrow\hspace{0.2cm}
\boxed{\begin{array}{c} \textrm{Temperley-Lieb} \\ \textrm{connectivities} \end{array}}
\end{equation*}
\caption{The organisational diagram of maps from perfect matchings to $\tl_n(0)$ connectivities.}
\label{fig:org}
\end{figure}

Starting with Temperley's correspondence~\cite{T74} between dimers and spanning trees, we now establish a direct connection between the dimer model and critical dense polymers. While the authors of~\cite{BPPR14} have used the transfer matrix formulation of the dimer model to probe combinatorial aspects of spanning trees
and spanning webs (see below), 
here we exploit Temperley's correspondence to understand the similarity of spectra described in \cref{sec:spectra} and ultimately to study the conformal properties of the dimer model. Along the way, it will yield a solution of the dimer model entirely in terms of a loop model. The organisational diagram of maps used in this section and the next is displayed in \cref{fig:org}.

The first arrow in this diagram is Temperley's correspondence first discussed in \cite{T74}. The construction is applicable to lattices with a variety of boundary conditions, but here we focus on the horizontal $M \times N$ cylinder defined in \cref{sec:dimersstatmodel}. 

\begin{figure}[h!]
\psset{unit=.5cm}
\begin{center}
\begin{pspicture}(-0.2,-0.7)(8.2,6.25)
\multiput(0,0)(0,1){6}{\psline[linecolor=gray]{-}(0,0)(8,0)}\multiput(0,0)(1,0){9}{\psline[linecolor=gray]{-}(0,-0.5)(0,5.5)}
\rput(0,0){\dimerhblue}
\rput(0,1){\dimerhred}
\rput(1,2){\dimerhblue}
\rput(1,3){\dimerhred}
\rput(2,4){\dimerhblue}
\rput(3,5){\dimerhred}
\rput(2,1){\dimerhred}
\rput(3,0){\dimerhblue}
\rput(7,5){\dimerhred}
\rput(6,3){\dimerhred}
\rput(6,4){\dimerhblue}
\rput(7,2){\dimerhblue}
\rput(0,2){\dimervblue}
\rput(0,4){\dimervblue}
\rput(1,4){\dimervred}
\rput(3,2){\dimervred}
\rput(4,1){\dimervblue}
\rput(4,3){\dimervblue}
\rput(5,3){\dimervred}
\rput(5,1){\dimervred}
\rput(8,3){\dimervblue}
\rput(6,1){\dimervblue}
\rput(7,0){\dimervred}
\rput(8,0){\dimervblue}
\rput(2,0){\halfdimerdownblue}
\rput(2,5){\halfdimerupblue}
\rput(5,0){\halfdimerdownred}
\rput(5,5){\halfdimerupred}
\rput(6,0){\halfdimerdownblue}
\rput(6,5){\halfdimerupblue}
\multiput(0,0)(0,1){6}{\multiput(0,0)(1,0){9}{\pscircle[linewidth=0.025,fillstyle=solid,fillcolor=gray](0,0){0.08}}}
\rput(4,-1.2){(a)}
\end{pspicture} 
\qquad
\begin{pspicture}(-1,-0.7)(9,6.25)
\multiput(0,0)(0,1){6}{\psline[linecolor=gray]{-}(0,0)(8,0)}\multiput(0,0)(1,0){9}{\psline[linecolor=gray]{-}(0,-0.5)(0,5.5)}
\rput(0,0){\rarrowblue\psline[linecolor=darkgreen,linewidth=1.2pt,arrowscale=1.3,arrowinset=0.1]{-}(0,-0.5)(0,0)}
\rput(0,6){\psline[linecolor=darkgreen,linewidth=1.2pt,arrowscale=1.3,arrowinset=0.1]{-}(0,-0.5)(0,-2)
\psline[linecolor=darkgreen,linewidth=1.2pt,arrowscale=1.3,arrowinset=0.1]{->}(0,-2)(0,-0.65)}
\rput(2,0){\psline[linecolor=darkgreen,linewidth=1.2pt,arrowscale=1.3,arrowinset=0.1]{-}(0,-0.5)(0,0)}
\rput(2,6){\psline[linecolor=darkgreen,linewidth=1.2pt,arrowscale=1.3,arrowinset=0.1]{-}(0,-0.5)(0,-2)
\psline[linecolor=darkgreen,linewidth=1.2pt,arrowscale=1.3,arrowinset=0.1]{->}(0,-0.5)(0,-1.35)}
\rput(4,0){\larrowblue}
\rput(6,0){\psline[linecolor=darkgreen,linewidth=1.2pt,arrowscale=1.3,arrowinset=0.1]{-}(0,-0.5)(0,0)}
\rput(6,6){\psline[linecolor=darkgreen,linewidth=1.2pt,arrowscale=1.3,arrowinset=0.1]{-}(0,-0.5)(0,-2)
\psline[linecolor=darkgreen,linewidth=1.2pt,arrowscale=1.3,arrowinset=0.1]{->}(0,-0.5)(0,-1.35)}
\rput(8,0){\uarrowblue}
\rput(0,2){\uarrowblue}
\rput(2,2){\larrowblue}
\rput(4,2){\darrowblue}
\rput(6,2){\darrowblue}
\rput(8,2){\larrowblue}
\rput(2,4){\rarrowblue}
\rput(4,4){\darrowblue}
\rput(6,4){\rarrowblue}
\rput(8,4){\darrowblue}
\rput(1,1){\larrowred}
\rput(3,1){\larrowred}
\rput(5,1){\uarrowred}
\rput(7,1){\psline[linecolor=red,linewidth=1.2pt,arrowscale=1.3,arrowinset=0.1]{->}(0,0)(0,-1.35)
\psline[linecolor=red,linewidth=1.2pt,arrowscale=1.3,arrowinset=0.1]{-}(0,-0.5)(0,-1.5)}
\rput(1,3){\rarrowred}
\rput(3,3){\darrowred}
\rput(5,3){\uarrowred}
\rput(7,3){\larrowred}
\rput(1,5){\darrowred}
\rput(3,5){\rarrowred}
\rput(5,5){\psline[linecolor=red,linewidth=1.2pt,arrowscale=1.3,arrowinset=0.1]{-}(0,0)(0,0.5)} 
\rput(5,-1){\psline[linecolor=red,linewidth=1.2pt,arrowscale=1.3,arrowinset=0.1]{->}(0,0.5)(0,1.35)
\psline[linecolor=red,linewidth=1.2pt,arrowscale=1.3,arrowinset=0.1]{-}(0,0.5)(0,2)} 
\rput(7,5){\rarrowred}
\rput(7,5){\psline[linecolor=red,linewidth=1.2pt,arrowscale=1.3,arrowinset=0.1]{-}(0,0.5)(0,0)}
\multiput(0,0)(0,2){3}{\multiput(0,0)(2,0){5}{\pscircle[linewidth=0.025,fillstyle=solid,fillcolor=darkgreen](0,0){0.08}}}
\multiput(1,1)(0,2){3}{\multiput(0,0)(2,0){4}{\pscircle[linewidth=0.025,fillstyle=solid,fillcolor=red](0,0){0.08}}}
\multiput(-1,1)(0,2){3}{\pscircle[linewidth=0.025,fillstyle=solid,fillcolor=white](0,0){0.08}}
\multiput(9,1)(0,2){3}{\pscircle[linewidth=0.025,fillstyle=solid,fillcolor=white](0,0){0.08}}
\rput(4,-1.2){(b)}
\end{pspicture} 
\qquad
\begin{pspicture}[shift=-1](-1,-1.7)(9,6.25)
\multiput(0,-2)(2,0){5}{\pspolygon[fillstyle=solid,fillcolor=lightlightblue,linewidth=0.75pt](0,0)(1,1)(0,2)(-1,1)}
\multiput(1,-1)(2,0){4}{\pspolygon[fillstyle=solid,fillcolor=lightlightblue,linewidth=0.75pt](0,0)(1,1)(0,2)(-1,1)}
\multiput(0,0)(2,0){5}{\pspolygon[fillstyle=solid,fillcolor=lightlightblue,linewidth=0.75pt](0,0)(1,1)(0,2)(-1,1)}
\multiput(1,1)(2,0){4}{\pspolygon[fillstyle=solid,fillcolor=lightlightblue,linewidth=0.75pt](0,0)(1,1)(0,2)(-1,1)}
\multiput(0,2)(2,0){5}{\pspolygon[fillstyle=solid,fillcolor=lightlightblue,linewidth=0.75pt](0,0)(1,1)(0,2)(-1,1)}
\multiput(1,3)(2,0){4}{\pspolygon[fillstyle=solid,fillcolor=lightlightblue,linewidth=0.75pt](0,0)(1,1)(0,2)(-1,1)}
\multiput(0,4)(2,0){5}{\pspolygon[fillstyle=solid,fillcolor=lightlightblue,linewidth=0.75pt](0,0)(1,1)(0,2)(-1,1)}
\multiput(1,5)(2,0){4}{\pspolygon[fillstyle=solid,fillcolor=lightlightblue,linewidth=0.75pt](0,0)(1,1)(0,2)(-1,1)}
\multiput(0,0)(0,2){4}{\pspolygon[fillstyle=solid,fillcolor=lightlightblue,linewidth=0.75pt](0,0)(-1,1)(-1,-1)}
\multiput(8,0)(0,2){4}{\pspolygon[fillstyle=solid,fillcolor=lightlightblue,linewidth=0.75pt](0,0)(1,1)(1,-1)}
\rput(0,0){\rnoarrowblue\psline[linecolor=darkgreen,linewidth=1.2pt,arrowscale=1.0,arrowinset=0.1]{-}(0,-0.5)(0,0)}
\rput(0,6){\psline[linecolor=darkgreen,linewidth=1.2pt,arrowscale=1.0,arrowinset=0.1]{-}(0,-0.5)(0,-2)
\psline[linecolor=darkgreen,linewidth=1.2pt,arrowscale=1.0,arrowinset=0.1]{-}(0,-2)(0,-0.75)}
\rput(2,0){\psline[linecolor=darkgreen,linewidth=1.2pt,arrowscale=1.0,arrowinset=0.1]{-}(0,-0.5)(0,0)}
\rput(2,6){\psline[linecolor=darkgreen,linewidth=1.2pt,arrowscale=1.0,arrowinset=0.1]{-}(0,-0.5)(0,-2)
\psline[linecolor=darkgreen,linewidth=1.2pt,arrowscale=1.0,arrowinset=0.1]{-}(0,-0.5)(0,-1.25)}
\rput(4,0){\lnoarrowblue}
\rput(6,0){\psline[linecolor=darkgreen,linewidth=1.2pt,arrowscale=1.0,arrowinset=0.1]{-}(0,-0.5)(0,0)}
\rput(6,6){\psline[linecolor=darkgreen,linewidth=1.2pt,arrowscale=1.0,arrowinset=0.1]{-}(0,-0.5)(0,-2)
\psline[linecolor=darkgreen,linewidth=1.2pt,arrowscale=1.0,arrowinset=0.1]{-}(0,-0.5)(0,-1.25)}
\rput(8,0){\unoarrowblue}
\rput(0,2){\unoarrowblue}
\rput(2,2){\lnoarrowblue}
\rput(4,2){\dnoarrowblue}
\rput(6,2){\dnoarrowblue}
\rput(8,2){\lnoarrowblue}
\rput(2,4){\rnoarrowblue}
\rput(4,4){\dnoarrowblue}
\rput(6,4){\rnoarrowblue}
\rput(8,4){\dnoarrowblue}
\rput(1,1){\lnoarrowred}
\rput(3,1){\lnoarrowred}
\rput(5,1){\unoarrowred}
\rput(7,1){\psline[linecolor=red,linewidth=1.2pt,arrowscale=1.0,arrowinset=0.1]{-}(0,0)(0,-1.25)
\psline[linecolor=red,linewidth=1.2pt,arrowscale=1.0,arrowinset=0.1]{-}(0,-0.5)(0,-1.5)}
\rput(1,3){\rnoarrowred}
\rput(3,3){\dnoarrowred}
\rput(5,3){\unoarrowred}
\rput(7,3){\lnoarrowred}
\rput(1,5){\dnoarrowred}
\rput(3,5){\rnoarrowred}
\rput(5,5){\psline[linecolor=red,linewidth=1.2pt,arrowscale=1.0,arrowinset=0.1]{-}(0,0)(0,0.5)} 
\rput(5,-1){\psline[linecolor=red,linewidth=1.2pt,arrowscale=1.0,arrowinset=0.1]{-}(0,0.5)(0,1.25)
\psline[linecolor=red,linewidth=1.2pt,arrowscale=1.0,arrowinset=0.1]{-}(0,0.5)(0,2)} 
\rput(7,5){\rnoarrowred}
\rput(7,5){\psline[linecolor=red,linewidth=1.2pt,arrowscale=1.0,arrowinset=0.1]{-}(0,0.5)(0,0)}
\rput(0,-2){\loopd}\rput(2,-2){\loopd}\rput(4,-2){\loopc}\rput(6,-2){\loopd}\rput(8,-2){\loopc}
\rput(0,0){\loopc}\rput(2,0){\loopc}\rput(4,0){\loopd}\rput(6,0){\loopd}\rput(8,0){\loopd}
\rput(0,2){\loopd}\rput(2,2){\loopc}\rput(4,2){\loopd}\rput(6,2){\loopc}\rput(8,2){\loopd}
\rput(0,4){\loopd}\rput(2,4){\loopd}\rput(4,4){\loopc}\rput(6,4){\loopd}\rput(8,4){\loopc}
\rput(1,-1){\loopc}\rput(3,-1){\loopc}\rput(5,-1){\loopd}\rput(7,-1){\loopd}
\rput(1,1){\loopc}\rput(3,1){\loopd}\rput(5,1){\loopd}\rput(7,1){\loopc}
\rput(1,3){\loopd}\rput(3,3){\loopc}\rput(5,3){\loopd}\rput(7,3){\loopc}
\rput(1,5){\loopc}\rput(3,5){\loopc}\rput(5,5){\loopd}\rput(7,5){\loopd}
\multiput(0,0)(0,2){4}{\psarc[linewidth=1.5pt,linecolor=blue](0,0){.707}{135}{-135}}
\multiput(8,0)(0,2){4}{\psarc[linewidth=1.5pt,linecolor=blue](0,0){.707}{-45}{45}}
\multiput(0,0)(0,2){3}{\multiput(0,0)(2,0){5}{\pscircle[linewidth=0.025,fillstyle=solid,fillcolor=darkgreen](0,0){0.08}}}
\multiput(1,1)(0,2){3}{\multiput(0,0)(2,0){4}{\pscircle[linewidth=0.025,fillstyle=solid,fillcolor=red](0,0){0.08}}}
\multiput(-1,1)(0,2){3}{\pscircle[linewidth=0.025,fillstyle=solid,fillcolor=white](0,0){0.08}}
\multiput(9,1)(0,2){3}{\pscircle[linewidth=0.025,fillstyle=solid,fillcolor=white](0,0){0.08}}
\psframe[fillstyle=solid,linecolor=white,linewidth=0pt](-1.1,-2.1)(9.1,-0.5)
\psframe[fillstyle=solid,linecolor=white,linewidth=0pt](-1.1,5.5)(9.1,7.1)
\rput(4,-1.2){(c)}
\end{pspicture} 
\caption{The results of mapping a given perfect matching to an oriented web, an unoriented web and a loop configuration.}
\label{fig:spanningtrees}
\end{center}
\end{figure}
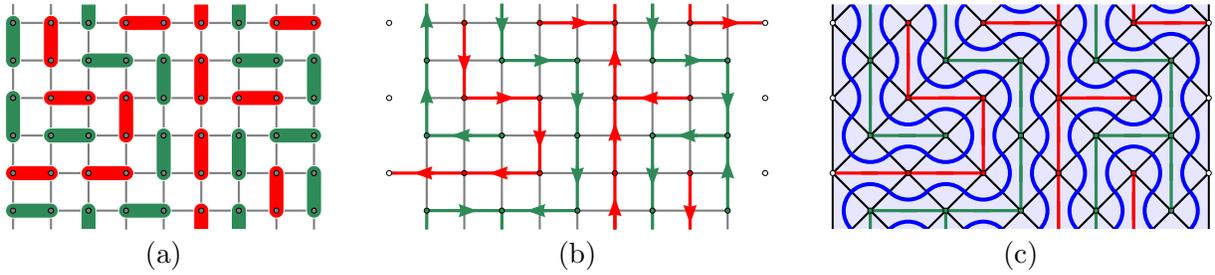

Starting with the $M\times N$ square lattice, one can define a sublattice made of the sites $(i,j)$ with $i = j \bmod 2$. In any perfect matching, each dimer covers one site of the sublattice. Instead of characterising dimers with regard to their horizontal or vertical orientations (as in \cref{fig:dimerconf}), we now distinguish them according to the parity of $i$ (and $j$) of the site $(i,j)$ of the sublattice that it covers. In the example of \cref{fig:spanningtrees}~(a), dimers on odd and even sublattice sites are, respectively, coloured in green and red. Requiring that the sublattice be periodic in the vertical direction constrains $M$ to be even, thus justifying this choice already made in \cref{sec:dimersintro}.

For a given perfect matching, Temperley's correspondence is constructed as follows: For each site $(i,j)$ of the sublattice, one draws an arrow of the appropriate colour, starting from $(i,j)$ and pointing to $(i-2, j)$, $(i+2, j)$, $(i, j-2)$ or $(i, j+2)$, following the orientation of the dimer occupying $(i,j)$. Here, $j$ is taken modulo $M$, so that arrows starting at $(i,1)$ and $(i,2)$ can, respectively, point to $(i,M-1)$ and $(i,M)$, and vice versa. For a site $(i,j)$ of the sublattice with $i = 2$ or $N-1$, it may happen that the arrow produced by this prescription points to a site outside of the lattice, in position $(0,j)$ or $(N+1,j)$. These external sites are referred to as {\it roots} and drawn using white circles in \cref{fig:spanningtrees}~(b). For $N$ odd, the roots on both sides of the cylinder are reached by red arrows, whereas for $N$ even, the roots on the left and right are reached by red and green arrows, respectively. 

In the diagram that emerges, called an {\it oriented web}, the sites of the sublattice are connected by arrows that either (i) flow to a root, producing {\it a rooted tree}, (ii) are part of {\it a unicycle}, that is, a chain of arrows that wraps around the cylinder, forming a non-contractible cycle, or (iii) flow to a unicycle. Red and green structures are intertwined and space-filling \cite{BPPR14}, but non-intersecting. For obvious reasons, the number $L$ of unicycles of an oriented web is constrained to have the same parity as $N$. Other important features are that there are no free ends with arrows pointing towards them, except at the roots, and that there are no contractible cycles of arrows. Indeed, such formations would cover an odd number of sites of the original $M \times N$ dimer lattice and thus never appear from Temperley's correspondence, simply because any chain of dimers covers an even number of sites. 

Starting from an oriented web, it is not hard to reconstruct the unique perfect matching in the pre-image. Temperley's correspondence is therefore {\em bijective}. Moreover, the number of horizontal dimers $h$ is equal to the number of horizontal arrows in the oriented web. Temperley's correspondence thus provides 
an alternative way to calculate $Z^{\textrm{dim}}$:
\be
 Z^{\textrm{dim}} = \hspace{-0.1cm}\sum_{\substack{\textrm{oriented}\\[0.05cm]\textrm{webs}}}\hspace{-0.1cm} \alpha^h.
\label{eq:Zoriented}
\ee

By removing the orientations of the arrows, one obtains an {\it unoriented web}, composed of line segments called {\it branches}, all of equal length (twice the lattice spacing). This is the second map in the organisational diagram of \cref{fig:org}. The concepts of trees and unicycles carry over to unoriented webs in the obvious way. In the example of \cref{fig:spanningtrees}, the unoriented web appears on panel~(c), where a loop configuration relevant to the discussion of \cref{sec:dimloops} has also been superimposed. 
The map from oriented webs to unoriented webs is clearly
not one-to-one, but it is surjective. Indeed, given an unoriented web with $L\ge1$ unicycles, there are more than one oriented web in the pre-image. Other oriented webs mapped to the same unoriented one are obtained by reversing the orientation of all arrows in one or more of the unicycles. At the level of the dimers, this is equivalent to shifting by one position the dimers in the selected unicycles, while leaving the other dimers unchanged. We conclude that an unoriented web with no unicycles has a unique pre-image, and, because each unicycle can be oriented in two ways, an unoriented web with $L$ unicycles has a pre-image of dimension $2^L$. 

The partition function of the dimer model can then be expressed in terms of sums of unoriented webs.  The number $h$ of horizontal dimers is now given by the number of horizontal branches. Following the previous discussion, to reproduce the correct counting of the oriented webs in \eqref{eq:Zoriented}, 
each unoriented web must be weighted by an extra factor of $2^L$. This yields
\be
 Z^{\textrm{dim}} = \hspace{-0.2cm}\sum_{\substack{\textrm{unoriented}\\[0.05cm]\textrm{webs}}} \hspace{-0.2cm} \alpha^h  \, 2^L.
\label{eq:Zunoriented}
\ee

%%%%%
\subsection{Webs and loops}\label{sec:dimloops}
%%%%%

In the previous section, perfect matchings were mapped to unoriented webs made of branches that never form contractible cycles. This is reminiscent of the rule imposed on loop configurations in the critical dense polymer model, where contractible loops are disallowed. We now establish a direct link between unoriented webs and loop configurations. This is the third map in \cref{fig:org}.

The first step is to superimpose, on the unoriented web, a new square grid, tilted at a $45^\circ$ angle, that draws edges between each site $(i,j)$ of the even sublattice and the site's four neighbours $(i\pm1,j\pm1)$. Extra edges are added to connect each site $(1,j)$ with $j$ odd  to the two roots in positions $(0,j\pm1)$, and similarly for $(N,j)$ and $(N+1,j\pm1)$ with $j = N \bmod 2$, while taking the vertical periodicity into account. The result is a grid of $\tfrac{MN}2$ diamond-shaped tiles covering the cylinder. Each of these tiles is crossed by a branch that connects two opposite corners. To establish the relation with the loop model, we decorate each tile with two quarter circles using the rule
\be
\psset{unit=.5cm}
\begin{pspicture}[shift=-.8](-1,-1)(1,1)
\pspolygon[fillstyle=solid,fillcolor=lightlightblue](-1,0)(0,-1)(1,0)(0,1)
\rput(-1,0){\psline[linecolor=purple,linewidth=1.2pt,arrowscale=1.0,arrowinset=0.1]{-}(0,0)(2,0)}
\end{pspicture}
\ \rightarrow\ 
\begin{pspicture}[shift=-.8](-1,-1)(1,1)
\pspolygon[fillstyle=solid,fillcolor=lightlightblue](-1,0)(0,-1)(1,0)(0,1)
\rput(-1,0){\psline[linecolor=purple,linewidth=1.2pt,arrowscale=1.0,arrowinset=0.1]{-}(0,0)(2,0)}
\rput(0,-1){\loopc}
\end{pspicture}\ , \qquad \quad
\begin{pspicture}[shift=-.8](-1,-1)(1,1)
\pspolygon[fillstyle=solid,fillcolor=lightlightblue](-1,0)(0,-1)(1,0)(0,1)
\rput(0,-1){\psline[linecolor=purple,linewidth=1.2pt]{-}(0,0)(0,2)}
\end{pspicture}
\ \rightarrow\ 
\begin{pspicture}[shift=-.8](-1,-1)(1,1)
\pspolygon[fillstyle=solid,fillcolor=lightlightblue](-1,0)(0,-1)(1,0)(0,1)
\rput(0,-1){\psline[linecolor=purple,linewidth=1.2pt]{-}(0,0)(0,2)}
\rput(0,-1){\loopd}
\end{pspicture}\ ,
\ee
for branches of either colour. By adding the boundary triangles 
\be
\psset{unit=.5cm}
\begin{pspicture}[shift=-.8](0,-1)(1,1)
\pspolygon[fillstyle=solid,fillcolor=lightlightblue](0,-1)(1,0)(0,1)
\psarc[linewidth=1.5pt,linecolor=blue](1,0){.707}{135}{-135}
\end{pspicture} \qquad \textrm{and}\qquad
\begin{pspicture}[shift=-.8](-1,-1)(0,1)
\pspolygon[fillstyle=solid,fillcolor=lightlightblue](0,-1)(-1,0)(0,1)
\psarc[linewidth=1.5pt,linecolor=blue](-1,0){.707}{-45}{45}
\end{pspicture}
\ee
on the left and right edges of the tilted grid, we obtain a diagram where the quarter circles form loop segments that trace the contours of the structures of the unoriented web. This is illustrated in \cref{fig:spanningtrees}~(c). 
Because contractible cycles are disallowed and each tree is attached to a root, the loop segments never form contractible loops. Instead, they produce a loop configuration of critical dense polymers on the cylinder, but on a {\em tilted} lattice compared to the one described in \cref{sec:Polymers}.

Starting from the left, the unicycles in any unoriented web appear alternately in green and red, separated by single non-contractible loops. For any configuration with at least one unicycle, the loop configuration it maps to contains two non-contractible loops that act as outer contours for the leftmost and rightmost unicycles. The number of non-contractible loops is thus $n_\gamma = L+1$. This also applies in the case $L=0$ (and $N$ even) where there are no unicycles and only a single non-contractible loop separating the left (red) and right (green) trees.

This map from unoriented webs to loop configurations is readily seen to be bijective, and as the branches do not provide any information not already encoded in the contours, they are now erased. The partition function for dimers can thus be computed using weights $\hat W(\sigma)$ for loop configurations $\sigma$ on the tilted lattice: $\hat W(\sigma) = 0$ if $\sigma$ contains contractible loops, and
\be 
\hat W(\sigma) = \gamma^{n_{\gamma}} p_1^{\#(
\psset{unit=.177cm}
\begin{pspicture}[shift=-.5](-1,-1)(1,1)
\pspolygon[fillstyle=solid,fillcolor=lightlightblue](-1,0)(0,-1)(1,0)(0,1)
\rput(0,-1){\loopc}
\end{pspicture}
)}
p_2^{\#(
\psset{unit=.177cm}
\begin{pspicture}[shift=-.5](-1,-1)(1,1)
\pspolygon[fillstyle=solid,fillcolor=lightlightblue](-1,0)(0,-1)(1,0)(0,1)
\rput(0,-1){\loopd}
\end{pspicture}
)}, \qquad (\gamma = 2),
\ee
otherwise. Because 
$
\#(
\psset{unit=.177cm}
\begin{pspicture}[shift=-.5](-1,-1)(1,1)
\pspolygon[fillstyle=solid,fillcolor=lightlightblue](-1,0)(0,-1)(1,0)(0,1)
\rput(0,-1){\loopc}
\end{pspicture}
) + 
\#(
\psset{unit=.177cm}
\begin{pspicture}[shift=-.5](-1,-1)(1,1)
\pspolygon[fillstyle=solid,fillcolor=lightlightblue](-1,0)(0,-1)(1,0)(0,1)
\rput(0,-1){\loopd}
\end{pspicture}
)
= \frac{MN}2$,
$\hat W(\sigma)$ only depends on the ratio $p_1/p_2$ up to an overall factor. 
For fixed $\alpha$, the values of $p_1$ and $p_2$ for the dimer model are given by
\be 
 p_1 = \frac\alpha\kappa, \qquad p_2 = \frac1\kappa,
\label{eq:p1p2}
\ee
for some value of the otherwise free parameter $\kappa$. From \eqref{eq:Zunoriented}, we find
\be
 Z^{\textrm{dim}} = \kappa^{MN/2} \sum_\sigma 2^{n_{\gamma}-1} 
p_1^{\#(
\psset{unit=.177cm}
\begin{pspicture}[shift=-.5](-1,-1)(1,1)
\pspolygon[fillstyle=solid,fillcolor=lightlightblue](-1,0)(0,-1)(1,0)(0,1)
\rput(0,-1){\loopc}
\end{pspicture}
)}
p_2^{\#(
\psset{unit=.177cm}
\begin{pspicture}[shift=-.5](-1,-1)(1,1)
\pspolygon[fillstyle=solid,fillcolor=lightlightblue](-1,0)(0,-1)(1,0)(0,1)
\rput(0,-1){\loopd}
\end{pspicture}
)} = \tfrac12 \, \kappa^{MN/2} \, \hat Z^{\textrm{cdp}} \Big|_{\gamma = 2}, 
\ee
where $\hat Z^{\textrm{cdp}}$ is the critical dense polymer partition function on the tilted lattice.\footnote{We note that the factor of $\tfrac12$ is due to the difference of $1$ between $n_\gamma$ and $L$. It is ultimately responsible for the relative factor of $2$ between \eqref{eq:Zdimfinal} and \eqref{eq:Zcdpfinal}.}

From \cref{sec:gluing,sec:Du}, $Z^{\textrm{dim}}$ can be computed using a transfer tangle in $\tl_n(0)$ and its matrix representatives in the standard representations $\rho_d$. Omitting the folding on the cylinder, the loop configuration produced from each unoriented web can be viewed as a Temperley-Lieb connectivity. This is the last map in \cref{fig:org}. In \cref{fig:spanningtrees}, for instance, the perfect matching on the cylinder with $N = 9$ is mapped to a loop configuration corresponding to a connectivity in $\tl_{n=10}(0)$:
\be
\begin{pspicture}[shift=-0.45](-0.0,-0.5)(4.0,0.5)
\pspolygon[fillstyle=solid,fillcolor=lightlightblue](0,-0.5)(4.0,-0.5)(4.0,0.5)(0,0.5)
\psline[linecolor=blue,linewidth=1.5pt]{-}(2.6,0.5)(2.6,-0.5)
\psarc[linecolor=blue,linewidth=1.5pt]{-}(0.4,0.5){-0.2}{0}{180}
\psarc[linecolor=blue,linewidth=1.5pt]{-}(2.0,0.5){-0.2}{0}{180}
\psarc[linecolor=blue,linewidth=1.5pt]{-}(3.6,0.5){-0.2}{0}{180}
\psarc[linecolor=blue,linewidth=1.5pt]{-}(0.8,-0.5){0.2}{0}{180}
\psarc[linecolor=blue,linewidth=1.5pt]{-}(1.6,-0.5){0.2}{0}{180}
\psarc[linecolor=blue,linewidth=1.5pt]{-}(3.2,-0.5){0.2}{0}{180}
\psbezier[linecolor=blue,linewidth=1.5pt]{-}(0.2,-0.5)(0.2,0)(1.0,0)(1.0,0.5)
\psbezier[linecolor=blue,linewidth=1.5pt]{-}(2.2,-0.5)(2.2,0)(1.4,0)(1.4,0.5)
\psbezier[linecolor=blue,linewidth=1.5pt]{-}(3.0,0.5)(3.0,0)(3.8,0)(3.8,-0.5)
\end{pspicture}\ .
\ee
It is now clear why similarities in the spectra were found in \cref{sec:spectra}, with the lattice widths of the two models related as $n = N+1$: The loop segments draw the contours of the web and intersect a given slice of the cylinder $n = N+1$ times.

The form of the transfer tangle on the tilted lattice\footnote{Note that one can instead use transfer tangles that are reflected vertically compared to \eqref{eq:tiltedDv}. Then, in \cref{sec:dimerrep}, arriving at a relation akin to \eqref{tau} would require using the $\tl_n(0)$ representation $\tau^\star$ defined by $\tau^\star(e_j) = \big(\tau(e_j)\big)^T$.} depends on the parity of $n$:
\be
\psset{unit=.5cm}
\funkyDb(v) = \left\{\begin{array}{c l}
\begin{pspicture}[shift=-2.2](-1.2,-2)(7,2.4)
\multiput(1,0)(2,0){3}{\pspolygon[fillstyle=solid,fillcolor=lightlightblue,linewidth=0.75pt](0,0)(1,1)(0,2)(-1,1)}
\multiput(0,-1)(2,0){4}{\pspolygon[fillstyle=solid,fillcolor=lightlightblue,linewidth=0.75pt](0,0)(1,1)(0,2)(-1,1)}
\rput(0,1){\pspolygon[fillstyle=solid,fillcolor=lightlightblue,linewidth=0.75pt](0,0)(-1,1)(-1,-1)
\psarc[linewidth=1.5pt,linecolor=blue](0,0){.707}{135}{-135}}
\rput(6,1){\pspolygon[fillstyle=solid,fillcolor=lightlightblue,linewidth=0.75pt](0,0)(1,1)(1,-1)
\psarc[linewidth=1.5pt,linecolor=blue](0,0){.707}{-45}{45}}
\rput(0,-1){\psarc[linewidth=0.025]{-}(0,0){0.22}{45}{135}\rput(0,1){\scriptsize $v$}}
\rput(2,-1){\psarc[linewidth=0.025]{-}(0,0){0.22}{45}{135}\rput(0,1){\scriptsize $v$}}
\rput(4,-1){\rput(0,1){\scriptsize $\ldots$}}
\rput(6,-1){\psarc[linewidth=0.025]{-}(0,0){0.22}{45}{135}\rput(0,1){\scriptsize $v$}}
\rput(1,0){\psarc[linewidth=0.025]{-}(0,0){0.22}{45}{135}\rput(0,1){\scriptsize $v$}}
\rput(3,0){\psarc[linewidth=0.025]{-}(0,0){0.22}{45}{135}\rput(0,1){\scriptsize $v$}}
\rput(5,0){\rput(0,1){\scriptsize $\ldots$}}
\end{pspicture} & \quad n \, \textrm{even,}\\
\begin{pspicture}[shift=-1.7](-1.2,-1.2)(6,2)
\multiput(1,0)(2,0){3}{\pspolygon[fillstyle=solid,fillcolor=lightlightblue,linewidth=0.75pt](0,0)(1,1)(0,2)(-1,1)}
\multiput(0,-1)(2,0){3}{\pspolygon[fillstyle=solid,fillcolor=lightlightblue,linewidth=0.75pt](0,0)(1,1)(0,2)(-1,1)}
\rput(0,1){\pspolygon[fillstyle=solid,fillcolor=lightlightblue,linewidth=0.75pt](0,0)(-1,1)(-1,-1)
\psarc[linewidth=1.5pt,linecolor=blue](0,0){.707}{135}{-135}}
\rput(5,0){\pspolygon[fillstyle=solid,fillcolor=lightlightblue,linewidth=0.75pt](0,0)(1,1)(1,-1)
\psarc[linewidth=1.5pt,linecolor=blue](0,0){.707}{-45}{45}}
\rput(0,-1){\psarc[linewidth=0.025]{-}(0,0){0.22}{45}{135}\rput(0,1){\scriptsize $v$}}
\rput(2,-1){\psarc[linewidth=0.025]{-}(0,0){0.22}{45}{135}\rput(0,1){\scriptsize $v$}}
\rput(4,-1){\rput(0,1){\scriptsize $\ldots$}}
\rput(1,0){\psarc[linewidth=0.025]{-}(0,0){0.22}{45}{135}\rput(0,1){\scriptsize $v$}}
\rput(3,0){\rput(0,1){\scriptsize $\ldots$}}
\rput(5,0){\psarc[linewidth=0.025]{-}(0,0){0.22}{45}{135}\rput(0,1){\scriptsize $v$}}
\end{pspicture} & \quad n \, \textrm{odd.} 
\end{array} \right.
\label{eq:tiltedDv}
\ee
We call it the {\it tilted transfer tangle}. It decomposes as a linear combination of connectivities weighted by powers of $p_1 = \sin v$ and $p_2 = \cos v$, where we set
\be
\cos v = \frac{1}{\sqrt{1+\alpha^2}}, \qquad \sin v = \frac{\alpha}{\sqrt{1+\alpha^2}},
\label{eq:alphav}
\ee
corresponding to \eqref{eq:p1p2} with $\kappa = \sqrt{1+\alpha^2}$. 

Some connectivities contributing to $(\funkyDb(v))^{M/2}$ contain loop segments that will form contractible loops when the lattice is folded on the cylinder. The gluing operator \eqref{eq:glue} sends these to zero and maps the surviving connectivities to their correct weight, $2^{n_\gamma}$ in this case. This yields 
\be
 Z^{\textrm{dim}} = \tfrac 12\,  (1+\alpha^2)^{MN/2} \, \mathcal F\big((\funkyDb(v))^{M/2}\big) \Big|_{\gamma = 2}.
\ee 
From \eqref{eq:F}, the partition function for dimers can thus be calculated using the transfer matrices of critical dense polymers on the tilted lattice, and we arrive at
\be
 Z^{\textrm{dim}} = \tfrac 12\,  (1+\alpha^2)^{MN/2}  \hspace{-0.3cm}\sum_{\substack{d = 0\\[0.05cm] d = n \bmod 2}}^n 
  \hspace{-0.3cm} (d+1) \,\textrm{Tr}\, \rho_d\big( (\funkyDb(v))^{M/2}\big).
\label{che}
\ee

The rest of the evaluation of $Z^{\textrm{dim}}$ using the tilted transfer tangle is similar to the one of $Z^{\textrm{cdp}}$ on the regular (untilted) lattice. One first establishes an inversion identity for $\funkyDb(v)$ and then uses it to compute the eigenvalues in terms of quantum numbers $\nu_k$ and $\tau_k$. This is done in \cref{sec:integ}. As argued at the end of that section, the admissibility criterion for a vector $(\nu, \tau)$ to appear in the spectrum of $\rho_d(\funkyDb(v))$ turns out to be identical to the similar criterion for $\rho_d(\Db(u,0))$. Partition functions restricted to a given $d$ are subsequently obtained by taking the trace of $(\funkyDb(v))^{M/2}$ in $\rho_d$, and the steps follow those used for the untilted lattice. The conformal, $d$-specific partition functions are also the same up to a redefinition of the nome $q$. By weighting them by $d+1$ and computing their sum, one finally arrives at \eqref{eq:Zdimfinal}.

%%%%
\subsection{Dimer representation of the Temperley-Lieb algebra}\label{sec:dimerrep}
%%%%

\cref{sec:dimtrees,sec:dimloops} established a connection between the dimer model and the Temperley-Lieb algebra through the series of maps in \cref{fig:org}. This did not involve Lieb's transfer matrix, but as shown in \cite{MDRR14}, $T^2(\alpha)$ directly ties the first and last items of \cref{fig:org} via the so-called {\it dimer representation} of $\tl_n(0)$,
\be 
\tau\;: \; \tl_n(0) \rightarrow  \textrm{End}\big((\mathbb C^{2})^{\otimes \, n-1}\big), \qquad \tau(I) = \mathbb I, \qquad \tau(e_j) = \sigma_{j-1}^- \sigma_{j}^++ \sigma_{j}^+ \sigma_{j+1}^-, 
\ee  
using the convention $\sigma^\pm_0 = \sigma^\pm_{n} = 0$. We note that $\funkyDb(v)$ admits a simple expression in terms of the generators of $\tl_n(0)$:
\be 
\funkyDb(v) = \Big( \prod_{j=1}^{\lfloor \frac{n}2 \rfloor} \, \boldsymbol X_{2j-1}(v)\Big) \Big(\prod_{j=1}^{\lfloor \frac{n-1}2 \rfloor}\hspace{-0.1cm} \boldsymbol X_{2j}(v) \Big) , \qquad \boldsymbol X_j(v) = I \, \cos v  +  e_j \, \sin v \,.
\label{eq:DvTL}
\ee
Here onward, we use the variable $n$ (instead of $N+1$) in computations in the dimer model. The transfer matrix squared $T^2(\alpha)$ is, up to a constant, the matrix representative of $\funkyDb(v)$ in the dimer representation. Indeed, by rearranging the terms of the first exponential in \eqref{eq:newT}, we find
\be
 T^2(\alpha) = \prod_{j=1}^{\lfloor \frac{n}2 \rfloor} 
  \Big(\mathbb I + \alpha\, (\sigma^-_{2j-2}\,\sigma^+_{2j-1}+\sigma^+_{2j-1}\,\sigma^-_{2j})\Big) 
  \prod_{j=1}^{\lfloor\! \frac{n-1}2 \!\rfloor}
  \Big(\mathbb I+\alpha\, (\sigma^-_{2j-1}\,\sigma^+_{2j}+\sigma^+_{2j}\,\sigma^-_{2j+1})\Big)
\label{eq:T2TL}
\ee
as well as
\be
\tau(\boldsymbol X_j(v)) = \frac{\mathbb I + \alpha \, \tau(e_j)}{(1+\alpha^2)^{1/2}},\qquad \tau(\funkyDb(v)) = \frac{T^2(\alpha)}{(1+\alpha^2)^{(n-1)/2}}, \qquad (\alpha = \tan v).
\label{tau}
\ee
 
%%%%%%%
\subsection{Lattice integrability and integrals of motion}
\label{sec:integ}
%%%%%%%

Based on examples for small $n$, it is not hard to see that the commutator $[\funkyDb(u), \funkyDb(v)]$ is non-zero in general. This reflects the non-commutativity of $T^2(\alpha)$ and $T^2(\alpha')$ in the dimer model. As we now show, the description of $T^2(\alpha)$ in terms of $\tl_n(0)$ reveals the integrability of the dimer model on the lattice.

Following \cite{YB95,DJS09}, we fix the spectral parameter and inhomogeneities to have the specific values
\be
 u = \tfrac v 2, \qquad \xi  = \xi_v= (-\tfrac v2, \tfrac v2, -\tfrac v2, \tfrac v2, \dots),
\ee
in which case $\Db(u,\xi)$ becomes
\be
\Db (\tfrac v 2,\xi_v) = \frac 1 {\sin v} \ 
\psset{unit=0.8}
\begin{pspicture}[shift=-0.9](-0.5,-0.0)(6.5,2.0)
\facegrid{(0,0)}{(6,2)}
\psarc[linewidth=0.025]{-}(1,1){0.16}{90}{180}
\psarc[linewidth=0.025]{-}(1,0){0.16}{0}{90}
\psarc[linewidth=0.025]{-}(3,1){0.16}{90}{180}
\psarc[linewidth=0.025]{-}(3,0){0.16}{0}{90}
\rput(4.5,0.5){$\ldots$}
\rput(4.5,1.5){$\ldots$}
\rput(5.5,0.5){$\ldots$}
\rput(5.5,1.5){$\ldots$}
\psarc[linewidth=1.5pt,linecolor=blue]{-}(0,1){0.5}{90}{-90}
\psarc[linewidth=1.5pt,linecolor=blue]{-}(6,1){0.5}{-90}{90}
\rput(0.5,1.5){\small$v$}
\rput(1.5,0.5){\small$v$}
\rput(2.5,1.5){\small$v$}
\rput(3.5,0.5){\small$v$}
\rput(0,0){\loopa}
\rput(1,1){\loopb}
\rput(2,0){\loopa}
\rput(3,1){\loopb}
\end{pspicture} \ .
\label{eq:Du2}
\ee   
By using the diagrammatic identity
\be
\psset{unit=0.8}
\begin{pspicture}[shift=-0.7](-0.6,-0.5)(1.1,1.0)
\facegrid{(0,0)}{(1,1)}
\psarc[linewidth=0.025]{-}(1,0){0.16}{90}{180}
\psarc[linewidth=1.5pt,linecolor=blue]{-}(0,0){0.5}{90}{360}
\rput(0.5,.5){\small$v$}
\end{pspicture} \ = \ \sin v \
\begin{pspicture}[shift=-0.7](-0.6,-0.5)(1.1,1.0)
\facegrid{(0,0)}{(1,1)}
\rput(0,0){\loopa}
\psarc[linewidth=1.5pt,linecolor=blue]{-}(0,0){0.5}{90}{360}
\end{pspicture}
\ee
and deforming the diagram in \eqref{eq:Du2} in such a way that each of the face operators 
$\,
\psset{unit=.4cm}
\begin{pspicture}[shift=-.2](0,0)(1,1)
\facegrid{(0,0)}{(1,1)}
\psarc[linewidth=0.025]{-}(0,0){0.25}{0}{90}
\rput(.5,.5){\scriptsize $v$}
\end{pspicture}\,
$
and 
$\,
\psset{unit=.4cm}
\begin{pspicture}[shift=-.2](0,0)(1,1)
\facegrid{(0,0)}{(1,1)}
\psarc[linewidth=0.025]{-}(1,0){0.25}{90}{180}
\rput(.5,.5){\scriptsize $v$}
\end{pspicture}\,
$
is oriented as
$\,
\psset{unit=.35cm}
\begin{pspicture}[shift=-.2](0,-0.7)(1.4,0.7)
\rput{-45}(0,0){\facegrid{(0,0)}{(1,1)}
\psarc[linewidth=0.025]{-}(1,0){0.25}{90}{180}
\rput{45}(.5,.5){\scriptsize $v$}}
\end{pspicture}\,
$, we see that
\be 
 \Db(\tfrac v 2, \xi_v) =  \funkyDb(v).
 \label{eq:DDv}
\ee

It is now clear from \eqref{eq:DDv} and the discussion below \eqref{eq:Ducomm} why $\funkyDb(u)$ and $\funkyDb(v)$ do not commute: They simply correspond to different choices of the inhomogeneities, $\xi_u$ and $\xi_v$. However, for any fixed $v\in \mathbb C$, we can construct a one-parameter family of transfer tangles,
\be
\Db (u,\xi_v) = \frac 1 {\sin 2u} \ 
\psset{unit=0.9}
\begin{pspicture}[shift=-0.9](-0.5,-0.0)(6.5,2.0)
\facegrid{(0,0)}{(6,2)}
\psarc[linewidth=0.025]{-}(0,0){0.16}{0}{90}
\psarc[linewidth=0.025]{-}(1,1){0.16}{90}{180}
\psarc[linewidth=0.025]{-}(1,0){0.16}{0}{90}
\psarc[linewidth=0.025]{-}(2,1){0.16}{90}{180}
\psarc[linewidth=0.025]{-}(2,0){0.16}{0}{90}
\psarc[linewidth=0.025]{-}(3,1){0.16}{90}{180}
\psarc[linewidth=0.025]{-}(3,0){0.16}{0}{90}
\psarc[linewidth=0.025]{-}(4,1){0.16}{90}{180}
\rput(4.5,0.5){$\ldots$}
\rput(4.5,1.5){$\ldots$}
\rput(5.5,0.5){$\ldots$}
\rput(5.5,1.5){$\ldots$}
\psarc[linewidth=1.5pt,linecolor=blue]{-}(0,1){0.5}{90}{-90}
\psarc[linewidth=1.5pt,linecolor=blue]{-}(6,1){0.5}{-90}{90}
\rput(0.5,.5){\scriptsize$u\!-\!\frac v2$}
\rput(0.5,1.5){\scriptsize$u\!+\!\frac v2$}
\rput(1.5,.5){\scriptsize$u\!+\!\frac v2$}
\rput(1.5,1.5){\scriptsize$u\!-\!\frac v2$}
\rput(2.5,.5){\scriptsize$u\!-\!\frac v2$}
\rput(2.5,1.5){\scriptsize$u\!+\!\frac v2$}
\rput(3.5,.5){\scriptsize$u\!+\!\frac v2$}
\rput(3.5,1.5){\scriptsize$u\!-\!\frac v2$}
\end{pspicture}   \qquad (u \in \mathbb C),
\label{eq:Duv}
\ee 
which {\em does} commute with $\funkyDb(v)$, see \eqref{eq:Ducomm}: 
\begin{equation} 
[\funkyDb(v),\Db(u,\xi_v)]=[\Db(\tfrac v 2,\xi_v),\Db(u,\xi_v)] = 0.
\end{equation}
Using \eqref{tau}, it follows that $\tau\big(\Db(u,\xi_v)\big)$ 
constitutes a one-parameter family of matrices commuting with $T^2(\alpha)$. Tangles that are independent of $u$ and commute with $\Db(\tfrac v 2,\xi_v)$ are obtained from the coefficients in the expansion of $\Db(u, \xi_v)$ in powers of $\sin 2u$,
\be
\Db(u,\xi_v) = \sum_{m=0}^{n-1} \Jb_m\,(\sin 2u)^m,
\label{eq:lattIOM}
\ee
and are here called {\it lattice integrals of motion}. For $n=3$, for example, $\Jb_0, \Jb_1$ and $\Jb_2$ are readily read off from
\begin{alignat}{2}
\Db(u,\xi_v) = \tfrac 18( 5 + 3 \cos 2v)\, I + \sin 2u \Big[ \cos v\,(e_1 + e_2) &+ \tfrac12 \sin v\,(e_1e_2-e_2e_1)\Big] \label{eq:iom3}\\ &+ (\sin2u)^2 \Big[-\tfrac14 I+\tfrac12(e_1e_2+e_2e_1)\Big]. \nonumber
\end{alignat}
The matrix realisation of $\Jb_m=\Jb_m(v)$ in the representation $\tau$ commutes with $T^2(\alpha)$ for $\alpha = \tan v$: 
\be
[\tau(\Jb_m(v)),T^2(\tan v)]=0,\qquad m=0,1,\dots,n-1.
\ee
As indicated, the lattice integrals of motion generally depend on $v$, or equivalently on $\alpha$. 

In the end, lattice integrability is built into the dimer model, albeit in a somewhat peculiar way. 
On the finite lattice, different values of $\alpha$ belong to different integrable families, simply because such families do not have the same ($\alpha$-dependent) lattice integrals of motion. In \cref{sec:inv,sec:integrals}, we will compute the eigenvalues of $\Db(u,\xi_v)$ and those of the lattice integrals of motion and compare them with the spectra of the {\em conformal} integrals of motion for $c=-2$. We will find that the $\alpha$-dependence nicely factorises and that the conformal integrals of motion are universal and independent of $\alpha$, as expected.

%%%%%
\subsection{Inversion identities}
\label{sec:inv}
%%%%%

From \eqref{eq:Geninv}, the transfer tangle $\Db(u,\xi_v)$ satisfies the following inversion identity:
\be
 \Db(u,\xi_v)\Db(u+\tfrac \pi 2,\xi_v) = I \left(\frac{\big(\cos(u-\tfrac v2)\cos(u+\tfrac v2)\big)^n 
  - \big(\sin(u-\tfrac v2)\sin(u+\tfrac v2)\big)^n}{\cos(u-\tfrac v2)\cos(u+\tfrac v2) - \sin(u-\tfrac v2)\sin(u+\tfrac v2)}\right)^2.
\ee
The function on the right-hand side factorises, yielding the following functional relations for the eigenvalues:
\be
\textrm {Eig}\big(\Db(u,\xi_v)\big) \textrm {Eig}\big(\Db(u+\tfrac \pi 2,\xi_v)\big) = \left\{\begin{array}{ll}
\displaystyle
\frac1{2^{2n-2}}\prod_{j=1}^{\frac{n-1}2} \bigg[ \frac{\cos^2 v}{\sin^2 \big(\frac{(2j-1)\pi}{2n}\big)} + \sin^2 v - \sin^2 2u\bigg]^2
& n \, \textrm{odd,} \\[0.6cm]
\displaystyle
\frac{n^2 \cos^2 v}{2^{2n-2}}\prod_{j=1}^{\frac {n-2}2} \bigg[ \frac{\cos^2 v}{\sin^2 \big(\frac{j\pi}{n}\big)} + \sin^2 v - \sin^2 2u\bigg]^2
& n \, \textrm{even.}\end{array}\right. 
\ee
The general solution is
\be
 \textrm {Eig}\big(\Db(u,\xi_v)\big) = (\cos v)^{n-1}\hspace{-0.1cm} \prod_{j=1}^{\lfloor \frac{n-1}2\rfloor}\Big[ \sqrt{1+\alpha^2 \sin^2 t_j}
  + \epsilon_j \sin t_j\frac {\sin 2u}{\cos v}\Big] \Big[ \sqrt{1+\alpha^2 \sin^2 t_j}+ \mu_j \sin t_j \frac {\sin 2u}{\cos v}\Big]
\label{eq:EigDuv}
\ee
where $\epsilon_j, \mu_j \in \{+1,-1\}$ and $\alpha = \tan v$, while $t_j$ is defined in \eqref{eq:PReigs}. This expression can be rearranged to resemble \eqref{eq:speccdp} and \eqref{eq:convL},
\be
 \textrm{Eig}\big(\Db(u,\xi_v)\big) =  (\cos v)^{n-1} \hspace{-0.4cm}  \prod_{\substack{k = 1 
\\[0.05cm] k = n\bmod 2}}^{n-2}\hspace{-0.3cm} \bigg[1+\Big(\alpha^2 - \frac{\sin^22u}{\cos^2v}\Big)\, \sin^2 q_k\bigg] 
 \bigg[\frac{\sqrt{1+\alpha^2 \sin^2 q_k}+\frac{\sin 2u}{\cos v}\,\sin q_k}{\sqrt{1+\alpha^2 \sin^2 q_k}- \frac{\sin 2u}{\cos v}\,\sin q_k}\bigg]^{1-\delta_k} 
\label{eq:genspectra}
\ee
where $\nu_k = \frac{1-\epsilon_k}{2}$, $\tau_k = \frac{1-\mu_k}{2}$ and $\delta_k = \nu_k+\tau_k \in \{0,1,2\}$, while $q_k$ is defined in \eqref{eq:speccdp}. 

Remarkably, the transfer tangle $\Db(u,\xi_v)$ provides a unified transfer matrix description of the dimer model and critical dense polymers.
For $\alpha = v = 0$, it becomes the transfer tangle $\Db(u,0)$ discussed in \cref{sec:cdpspec}, so \eqref{eq:genspectra} yields the eigenvalues \eqref{eq:speccdp} under this specialisation. For $u = \frac v 2$, $\Db(u,\xi_v)$ becomes the tilted transfer tangle $\funkyDb(v)$, and \eqref{eq:genspectra} indeed reduces to the spectra \eqref{eq:convL} of $T^2(\alpha)$ up to an overall factor of $\kappa^{n-1}=(\cos v)^{n-1}$, as it should according to \eqref{tau}.

The eigenvalues \eqref{eq:genspectra} are continuous functions of the parameters $u$ and $v$, so in any given representation, the admissibility criterion for the allowed $(\nu, \tau)$ vectors is independent of $u$ and $v$. This allows us to deduce the following criterion for $\funkyDb(v)$ in $\rho_d$: The constraints on $(\nu,\tau)$ are the same as those for $\Db(u,0)$ given in \eqref{adm}. As discussed at the end of \cref{sec:dimloops}, the traces of $\rho_d\big((\funkyDb(v))^{M/2}\big)$ in \eqref{che} can then be computed explicitly, reproducing the expressions \eqref{eq:Zdimfinal} for the dimer partition functions.

Finally, the analysis of finite-size corrections in \cref{sec:spectra} can be extended to $\Db(u,\xi_v)$ simply by replacing $\omega(t)$ in \eqref{eq:E0} and \eqref{eq:DE} by
\be
\omega(t) = \log \big( \sqrt{1+\alpha^2 \sin^2 t} + \sqrt{1+\alpha^2}\, \sin 2u\, \sin t  \big).
\label{eq:omega!}
\ee
Indeed, this function reproduces both $\omega^{\rm dim}(t)$ and $\omega^{\rm cdp}(t)$ in \eqref{eq:w(t)} under the respective specialisations. We will return to this in \cref{sec:integrals}.

%%%%%%%%%%%%%%%%%%%%
\section{Conformal data}
\label{sec:CFT}
%%%%%%%%%%%%%%%%%%%%

A key objective of this section is to refine the arguments tying the dimer model to a conformal field theory with central charge $c=-2$. \cref{sec:dimH} discusses the continuum scaling limit of the dimer model, defines the dimer Hamiltonian and identifies the first integral of motion $L_0-\frac c{24}$. In \cref{sec:higher}, we construct expressions for the higher Virasoro modes using their lattice approximate realisations as Temperley-Lieb tangles. \cref{sec:structure} then discusses the relation with the fermionic $bc$ ghost system with $c=-2$. It culminates with \cref{thm:structurethm}, which asserts that, as modules over the Virasoro algebra, the dimer Fock spaces decompose into direct sums of Feigin-Fuchs modules and exhibit reducible yet indecomposable structures. \cref{sec:integrals} then examines the conformal integrals of motion for Feigin-Fuchs modules at $c=-2$ and finds complete consistency with the eigenvalue analysis of the lattice integrals of motion. Finally, in \cref{sec:higherc=1}, we show that the expression for $L_0-\frac c{24}$ found in \cref{sec:dimH} is also compatible with a $c = 1$ realisation of the Virasoro algebra. Accordingly, we explicitly construct Virasoro modes for two non-interacting fermions and find in \cref{thm:structurethm1} that, under their action, the dimer Fock spaces are completely reducible Virasoro modules. However, the transfer tangle $\Db(u,\xi_v)$ is found {\it not} to be a generating function for the $c=1$ integrals of motion.

%%%%%
\subsection{Scaling limit and dimer Hamiltonian}\label{sec:dimH}
%%%%%

In \cref{sec:diago}, we used a Jordan-Wigner transformation to express $T^2(\alpha)$ as a product of the mutually commuting operators $T_q$, each written in terms of the lattice fermions $\Zeta_q$ and $\Zetad_{q}$. From \eqref{eq:Li}, the logarithm of $T^2(\alpha)$ can be written as 
\be\label{eq:logT2}
-\tfrac 12\log T^2(\alpha) = \sum_{q=1}^{\lfloor \frac n2 \rfloor} \arcsinh \big(\alpha \cos \tfrac{\pi q} n\big) \big( - \Zetad_q \Zeta_q + \Zetad_{n-q}\Zeta_{n-q}\big).
\ee
For $n$ odd, it has a single ground state,
\begin{subequations}\label{eq:gs}
\be 
|\Omega \rangle = \Zetad_{\frac {n-1}2}\Zetad_{\frac {n-3}2} \cdots \Zetad_2 \Zetad_1 |{\sf d} \rangle,
\label{eq:gsodd}
\ee 
whereas it has two for $n$ even, namely
\be
|\Omega \rangle = \Zetad_{\frac {n-2}2}\Zetad_{\frac {n-4}2} \cdots \Zetad_2 \Zetad_1 |{\sf d} \rangle, \qquad |\Omega' \rangle = \Zetad_{\frac n2}\Zetad_{\frac {n-2}2} \cdots  \Zetad_2 \Zetad_1 |{\sf d} \rangle = \Zetad_{\frac n2} |\Omega \rangle.
\label{eq:gseven}
\ee
\end{subequations}

Eigenvectors of $T^2(\alpha)$ are naturally built from the reference state $|{\sf d} \rangle$ in the form $\Zetad_{q_1} \Zetad_{q_2} \dots \Zetad_{q_r}|{\sf d}\rangle$. To study the scaling limit, however, it is convenient to use a basis for the fermions that builds excitations on the ground state(s) instead. We thus proceed by introducing
\be
\PPhi_q = \Zetad_{\frac n2 -q} \quad (0 \le q < \tfrac n2), \qquad \PPsi_q = \Zeta_{\frac n2+q} \quad (0 < q < \tfrac n2) \label{eq:phipsi},
\ee
which merely amounts to a relabelling. Here, the mode labels $q$ of $\PPhi_q$ and $\PPsi_q$, respectively, take integer and half-integer values for $n$ even and odd. The non-trivial anticommutation relations are 
\be
\{\PPhi_p, \PPhid_q\}= \delta_{p,q}, \qquad \{\PPsi_p, \PPsid_q\} = \delta_{p,q}.
\label{eq:comms}
\ee
From \eqref{eq:logT2}, we have
\be
-\tfrac 12\log T^2(\alpha) = \hspace{-0.1cm}\sideset{}{'}\sum_{0 \le \,q\, \le \frac{n-2}2} \hspace{-0.1cm} \arcsinh \big(\alpha \sin \tfrac{\pi q} n\big) \big( \PPhid_q \PPhi_q + \PPsid_{q}\PPsi_{q} - \mathbb I\big), 
\label{eq:noinspiration}
\ee
where the primed sum indicates that the index $q$ runs over half-integer values for $n$ odd and integer ones for $n$ even. For $q>0$, the modes $\PPhi_q$ and $\PPsi_q$ annihilate the ground states \eqref{eq:gs}, so the ground-state energy can be directly read off:
\be
E_0 = -\hspace{-0.15cm}\sideset{}{'}\sum_{0 \le \,q\, \le \frac{n-2}2}\hspace{-0.1cm}  \arcsinh \big(\alpha \sin \tfrac{\pi q} n\big).
\ee
As will be discussed in \cref{sec:higherc=1}, $\PPhi_q$ and $\PPsi_q$ can be seen as non-interacting 
fermions. 

We now turn to the {\it continuum scaling limit}, or {\it scaling limit} for short, of the dimer model. Roughly speaking, this limit retains the states that are the lowest excitations over the ground state and discards those whose energy differences with the ground-state energy do not scale the right way as $n \rightarrow \infty$, see \eqref{eq:limit}.

This can be defined more precisely. The dimer model actually constitutes a family of models characterised by their lattice width and height. The transfer matrices act on the vector spaces $\ctwotimes {n-1}$ whose dimensions increase with $n$. Because some properties of $T^2(\alpha)$ are specific to $n$ odd and even, it is necessary to study the two parities of $n$ separately. One can then define the sequences of eigenstates 
\be
\PPhid_{p_1}\PPhid_{p_2}\cdots\PPhid_{p_i} \PPsid_{q_1} \PPsid_{q_2} \cdots \PPsid_{q_j} | \Omega\rangle,\label{eq:seqs}
\ee
of $T^2(\alpha)$ in $\ctwotimes {n-1}$, made of one state for each $n$ (in steps of $2$, so the parity remains fixed), and study the behaviour of these sequences of states as $n \rightarrow \infty$. In \eqref{eq:seqs}, it is understood that $i$ and $j$ do not vary with $n$, but that $p_k$ and $q_\ell$ may. Here, focus is on sequences of eigenstates of $T^2(\alpha)$, whereas in \cref{sec:higher} we will discuss sequences similar to \eqref{eq:seqs}, but based on eigenstates of the {\it dimer Hamiltonian} $H$ defined below. The operator whose eigenstates are used to define the scaling limit, that is, $T^2(\alpha)$ or $H$, will be referred to as the {\it reference operator}.

A sequence \eqref{eq:seqs} will be said to be {\it conformal} if the following two conditions are satisfied:
\begin{itemize}
\item[(i)] The energies $E(n)$ of the sequence satisfy
\be
\lim_{n \rightarrow \infty}n \,\big(E(n)- E_0(n)\big) = \kappa \qquad {\rm for\ some} \qquad 
\kappa < \infty, 
\label{eq:limit}
\ee
where $E_0(n)$ is the ground-state energy for given $n$;
\item[(ii)] The limits $\lim_{n \rightarrow \infty} p_k(n)$ and $\lim_{n \rightarrow \infty} q_\ell(n)$ exist for all $1 \le k\le i$ and $1 \le \ell \le j$. 
\end{itemize}
As discussed in \cref{sec:spectra}, the eigenvalues of conformal sequences have finite-size corrections of the form \eqref{eq:Elargen}. Each conformal sequence is expected to converge to a state in a Virasoro module, with $\kappa$ in \eqref{eq:limit} equal to $\pi \vartheta(\Delta-\Delta_0)$. The difference in conformal dimensions is thus finite, and the corresponding conformal state is often referred to as {\it a finite excitation}.

A sequence for which at least one of the above conditions fails is not conformal. In particular, if the limit in \eqref{eq:limit} diverges, the energy difference with the ground state is too large and the sequence is said to {\it drift off to infinity}. Such sequences completely decouple in the conformal
field theory.

In this description, the sequences associated with the states $|\Omega \rangle$ and $\PPhid_{2}\PPhid_{5}\PPsid_{3}|\Omega \rangle$ on increasing values of $n$ are two examples of conformal sequences. In contrast, $\PPhid_{n/2} |\Omega \rangle$ is not a conformal sequence (neither condition is met). We say that labels $p_i$ or $q_j$ satisfying condition (ii) {\it lie in the vicinity of zero}. In the $\zeta_q^\dagger$ formulation, $q(n)$ must instead lie in the vicinity of $\frac n2$ for the sequence $\zeta^\dagger_{q(n)}|\Omega \rangle$ to be conformal, meaning that $\lim_{n\rightarrow \infty} (q(n)-\tfrac n2)$ must exist. We shall denote by $\ctwotimes \infty$ the infinite-dimensional vector space generated by the scaling limit of the conformal sequences and refer to it as the {\it dimer Fock space}.

Mimicking the translation from the Schr\"odinger to the Heisenberg picture, the concept of conformal sequences is lifted from the states to the operators: A sequence of operators $\mathcal O(n) \in {\rm End}(\ctwotimes{n-1})$ is conformal if the action of $\mathcal O(n)$ on any conformal sequence is also a conformal sequence or a linear combination thereof. Conformal sequences of operators survive the scaling limit. In \cref{sec:higher}, we shall encounter operators of the form $\mathcal O_1(n) + \mathcal O_2(n)$ where $\mathcal O_1(n)$ is conformal, whereas $\mathcal O_2(n)$ is not, as it maps any conformal sequence to a non-conformal one. In this scenario, only $\mathcal O_1(n)$ survives the scaling limit.

In \eqref{eq:noinspiration}, the operators $\PPhid_q$ and $\PPsid_q$ that are conformal 
are those for which $q$ is near the lower bound of the sum. The large $n$ expansion of $\log T^2(\alpha)$ thus starts out as
\be
-\tfrac 12 \log T^2(\alpha) = n f_{\rm bulk} + f_{\rm bdy} + \frac{\alpha \pi}n \Big(-\frac{c_{\rm eff}}{24}+\sideset{}{'}\sum_{q \ge 0} q \big( \pphid_q \pphi_q + \ppsid_{q}\ppsi_{q} \big)\Big) + o(n^{-2})
\label{eq:T2L0}
\ee
where $c_{\rm eff}$ is given in \eqref{ceff} and
\be 
\pphi_q = \PPhi_q \big|_{\alpha = 0}\,, \qquad \ppsi_q = \PPsi_q \big|_{\alpha = 0}\,.
\label{eq:smallphipsi}
\ee
In the scaling limit, the $\frac 1n$ term is expected to become $L_0 - \tfrac{c}{24}$: 
\be
\frac n{\alpha \pi}\big(-\tfrac 12 \log T^2(\alpha) - n f_{\rm bulk} - f_{\rm bdy}\big) \; \xrightarrow{n \rightarrow \infty}\; L_0 - \frac {c}{24} = -\frac{c_{\textrm{eff}}}{24} + \sideset{}{'}\sum_{q\ge0} q \big( \pphid_q \pphi_q + \ppsid_{q}\ppsi_{q}\big).
\label{eq:L0dim}
\ee
As discussed in the introduction, analysing this $\frac 1n$ term is insufficient to determine the central charge.

For finite $n$, $\log T^2(\alpha)$ and $\log T^2(\alpha')$ do not commute, because the fermions $\Phi_q$ and $\Psi_q$ in \eqref{eq:noinspiration} depend on $\alpha$, and the relations \eqref{eq:comms} generally hold only if both entries of a given anticommutator are evaluated at the same value of $\alpha$. In the scaling limit, however, the $\alpha$ dependence of the $\frac 1n$ term takes the form of an overall factor of $\alpha$, so in this limit, $\log T^2(\alpha)$ and $\log T^2(\alpha')$ commute at order $\frac 1n$. 

The {\it dimer Hamiltonian} is defined as
\be
H = - \frac{\rm d}{{\rm d}\alpha}\log T^2(\alpha) \Big|_{\alpha = 0} = - \sum_{j = 1}^{n-2} \,(\sigma_j^+ \sigma_{j+1}^- + \sigma_j^- \sigma_{j+1}^+)
\label{eq:dimerH}
\ee
and can be written in terms of the {\it Hamiltonian tangle} $\mathcal H$ of the Temperley-Lieb algebra as
\be
H = \tau(\mathcal H), \qquad \mathcal H = - \sum_{j=1}^{n-1} e_j \in \tl_n(0).
\ee
The Hamiltonian tangle is also the leading non-trivial term in the Taylor series expansions of $\Db(u,0)$ and $\funkyDb(v)$:
\be
\Db(u,0) = I - 2 u\, \mathcal H + O(u^2), \qquad \funkyDb(v) = I - v\, \mathcal H + O(v^2).
\label{eq:powerseries}
\ee
Expressing $H$ in terms of the fermions is done using \eqref{eq:noinspiration} and \eqref{eq:dimerH} and yields
\be
H = 2 \hspace{-0.1cm}\sideset{}{'}\sum_{0 \le \,q\, \le \frac{n-2}2} \hspace{-0.1cm}  \sin (\tfrac{\pi q} n) \big( \pphid_q \pphi_q + \ppsid_{q}\ppsi_{q}\big) + 
\left \{\begin{array}{ll}
1-\cot (\tfrac \pi{2n}),  & \quad n \, \textrm{even}, \\[0.2cm]
1-\csc (\tfrac \pi{2n}), & \quad n \, \textrm{odd}.
\end{array}\right. 
\label{eq:Hfinite}
\ee
Its $\frac 1n$ expansion has the form
\be
H = -\frac {2n} \pi + 1 + \frac{2 \pi}{n}\Big(-\frac{c_{\textrm{eff}}}{24} +\sideset{}{'}\sum_{q \ge 0} q\, \big( \pphid_q \pphi_q + \ppsid_{q}\ppsi_{q}\big)\Big)  + o(n^{-2}).
\label{eq:Hexp}
\ee
Up to the multiplicative factor of $2 \pi$, the $\frac 1n$ coefficient is exactly $L_0 - \frac c{24}$ given in \eqref{eq:L0dim}.

%%%%%
\subsection[Virasoro modes at $c = -2$]{Higher Virasoro modes at $\boldsymbol{c = -2}$}\label{sec:higher}
%%%%%

One direct way to extract the central charge of a critical model is to construct the Virasoro modes explicitly. For lattice models with underlying representations of the Temperley-Lieb algebra, this can be achieved via the so-called {\it Virasoro mode approximations}. The original idea is due to Koo and Saleur \cite{KS94} and was more recently applied to the $gl(1|1)$ spin chain \cite{GST12,GRS13}. First, the {\it Hamiltonian densities} are defined as
\be
\mathcal H_k = - \sum_{j=1}^{n-1}\,\cos (\tfrac {\pi k j}n) \, e_j \in \tl_n(0), \qquad k\in
\mathbb{Z}_{\ge 0}.
\ee
For $k = 0$, this is just the Temperley-Lieb Hamiltonian, which in the scaling limit is believed to converge to $L_0 - \frac c{24}$, up to a multiplicative scalar and addition of a constant term.
For $k>0$, in the scaling limit, $\mathcal H_k$ should converge to the linear combination $L_k + L_{-k}$, after an appropriate rescaling. In fact, this is believed to hold for any $\tl_n(0)$ representation, in particular in the dimer representation. Because 
\be
\big[L_k + L_{-k},L_0\big] = k \,(L_k - L_{-k}), 
\label{eq:trick}
\ee
the commutator $[\mathcal H_k, \mathcal H_0]$ should then become $k \,(L_k - L_{-k})$ in the scaling limit, thus allowing one to extract each $L_k$ individually. 

Our objective is to investigate this for the dimer representation $\tau$, so we apply the Jordan-Wigner transformation to
\be 
H_k = \tau(\mathcal H_k)
\ee
and consider the scaling limit. As discussed in \cref{sec:dimH}, this limit is defined in terms of the energies $E(n)$ of a reference operator. In the present context, there are two natural choices for this operator: $-\log T^2(\alpha)$ and $H = H_0$. For finite $n$, $-\log T^2(\alpha)$ and $H$ do not commute and have different eigenbases. For finite $n$, the expression for $H_k$ will of course depend on whether it is written in terms of $\Phi_q,\Psi_q$ or of $\phi_q, \psi_q$. Carrying out both computations, however, we find that, in the scaling limit, the two choices actually produce {\it the same result} at order $\frac 1n$. 
This is due to the fact that, for $q$ in the vicinity of zero, 
\be
\Phi_q,\Psi_q \,\xrightarrow{n \rightarrow \infty}\, \phi_q,\psi_q. 
\ee
In the end, the $\Ldim_k$ are written exclusively in terms of the fermions $\phi_q$ and $\psi_q$, and they are independent of the choice of $-\log T^2(\alpha)$ or $H$ as the reference operator. For $\Ldim_0$, this is already seen from \eqref{eq:L0dim} and \eqref{eq:Hexp}. We now present the steps of the (slightly shorter) calculation that uses $H$ as the reference operator.

By applying \eqref{eq:Ceta} and subsequently \eqref{eq:zetadef}, we find
\be
H_k = - \sum_{j=1}^{n-1}\, \cos (\tfrac {\pi k j}n)\,\tau(e_j) = - \sum_{j=1}^{n-2} \big[\cos\big(\tfrac{\pi k j} n\big) \Cjd_j  \Cj_{j+1} + \cos\big(\tfrac{\pi k (j+1)} n\big) \Cjd_{j+1}  \Cj_{j}\big] = -\! \sum_{p,q=1}^{n-1} c_{p,q}\, \Etad_p \Eta_q
\label{eq:Hkfermions}
\ee
where
\be
c_{p,q} = \cos(\tfrac{\pi q} n) \big( \delta_{p-q,-k \textrm{\,mod\,} 2n} +  \delta_{p-q,k \textrm{\,mod\,} 2n} -  \delta_{p+q,-k \textrm{\,mod\,} 2n} - \delta_{p+q,k \textrm{\,mod\,} 2n}\big).
\label{eq:cpq}
\ee
To take the scaling limit, we restrict $k$ to values much smaller than $n$. Many contributions to $H_k$ involve operators that belong to sequences that are not conformal. For instance, for the $\delta_{p + q,- k + 2n}$ contribution, the operator $\Etad_p \Eta_q = \Etad_{-q-k+2n}\Eta_q$ acts on any conformal sequence by outputting one that is not conformal: If $q$ is in the vicinity of $\frac n2$, $p$ is not. The same applies for the $\delta_{p + q,k}$ contribution. In fact, the last two terms in \eqref{eq:cpq} only yield operator sequences that are not conformal. For the same reason, the contributions $\delta_{p-q,-k \pm 2n}$ and $\delta_{p-q,k \pm 2n}$ of the first two terms in \eqref{eq:cpq} also drop out. The only surviving contributions are 
\be
A_k = - \sum_{p = 1}^{n-1} \Etad_p \Eta_q \cos(\tfrac{\pi q}n) \delta_{p-q,-k}, \qquad B_{k} = - \sum_{p = 1}^{n-1} \Etad_p \Eta_q \cos(\tfrac{\pi q}n) \delta_{p-q,k}.
\ee 
For $k = 0$, the results of the computation are already given in \eqref{eq:Hfinite} and \eqref{eq:Hexp}. Setting $k>0$, we find that, in terms of $\pphi_q, \ppsi_q$, $A_k$ and $B_k$ are expressed as
\be
\begin{array}{l}\displaystyle
A_k = \hspace{-0.1cm}\sideset{}{'}\sum_{0\, \le \,q\, \le \frac{n-2}2-k} \hspace{-0.2cm} \sin (\tfrac{\pi q} n) \, \pphid_q \pphi_{q+k}
+\hspace{-0.1cm}\sideset{}{'}\sum_{0\, \le \,q\, \le \,k} \hspace{-0.1cm} \sin (\tfrac{\pi q} n) \, \pphi_{k-q} \ppsi_{q}
+\hspace{-0.1cm}\sideset{}{'}\sum_{k\, < \,q\, \le \frac{n-2}2} \hspace{-0.1cm} \sin (\tfrac{\pi q} n) \, \ppsid_{q-k} \pphi_{q}, \\[0.75cm] 
\displaystyle
B_k = \hspace{-0.1cm}\sideset{}{'}\sum_{k\, \le \,q\, \le \frac{n-2}2} \hspace{-0.1cm} \sin (\tfrac{\pi q} n) \, \pphid_q \pphi_{q+k}
+\hspace{-0.1cm}\sideset{}{'}\sum_{0\, \le \,q\,<\,k} \hspace{-0.1cm} \sin (\tfrac{\pi q} n) \, \pphid_{q} \ppsid_{k-q}
+\hspace{-0.1cm}\sideset{}{'}\sum_{0\, \le \,q\, \le \frac{n-2}2-k} \hspace{-0.1cm} \sin (\tfrac{\pi q} n) \, \ppsid_{q-k} \pphi_{q},
\end{array} \quad (k >0).
\label{eq:AkBk}
\ee

These expressions for $A_k$ and $B_k$ can be recast in a simpler form after an additional relabelling,
\be
\theta_q^+ = \left\{\begin{array}{ll} \pphi_q & q \ge 0, \\[0.1cm] \ppsid_{-q} & q < 0, \end{array}\right. \qquad \theta_q^- = \left\{\begin{array}{ll} \ppsi_q &  q > 0, \\[0.1cm] \pphid_{-q} & q \le 0, \end{array}\right. \quad \ (q = \tfrac n2 \bmod 1),
\ee
where
\be
\{\theta^\alpha_p,\theta^\beta_q \} = \delta_{\alpha + \beta, 0} \, \delta_{p + q, 0}, \qquad (\alpha,\beta \in \{ +, - \}).\label{eq:thetacomm}
\ee
Indeed, with this relabelling, we find that
\be
\begin{array}{l}\displaystyle
A_k = \hspace{-0.1cm}\sideset{}{'}\sum_{0\, \le \,q\, \le \frac{n-2}2-k} \hspace{-0.2cm} \sin (\tfrac{\pi q} n) \, \theta^-_{-q} \theta^+_{q+k}
+\hspace{-0.1cm}\sideset{}{'}\sum_{0\, \le \,q\, \le \frac{n-2}2} \hspace{-0.1cm} \sin (\tfrac{\pi q} n) \, \theta^+_{-q+k} \theta^-_{q}, \\[0.85cm]
\displaystyle
B_k = \hspace{-0.1cm}\sideset{}{'}\sum_{0\, \le \,q\, \le \frac{n-2}2} \hspace{-0.1cm} \sin (\tfrac{\pi q} n) \, \theta^-_{-q} \theta^+_{q-k}
+\hspace{-0.1cm}\sideset{}{'}\sum_{0\, \le \,q\, \le \frac{n-2}2-k} \hspace{-0.2cm} \sin (\tfrac{\pi q} n) \, \theta^+_{-q-k} \theta^-_{q},
\end{array} \quad (k >0).
\ee
The scaling limit of $H_k$ then reads
\be
\tfrac n \pi  H_k \; \xrightarrow{n \rightarrow \infty}\; \Ldim_k + \Ldim_{-k} =  \sideset{}{'}\sum_{q \ge 0} q \big( \theta^-_{-q}\theta^+_{q+k} + \theta^+_{-q+k} \theta^-_q + \theta^-_{-q}\theta^+_{q-k}+\theta^+_{-q-k} \theta^-_q\big),\qquad (k > 0).
\label{eq:npiHk}
\ee

By using \eqref{eq:L0dim}, \eqref{eq:trick} and \eqref{eq:npiHk}, we can isolate each $\Ldim_k$. In the scaling limit, we find that $A_k$ and $B_k$ become $\Ldim_k$ and $\Ldim_{-k}$, respectively, with
\be 
\Ldim_k = \sideset{}{'}\sum_{q \ge 0} q\, \big(\theta^-_{-q}\theta^+_{q+k} + \theta^+_{-q+k} \theta^-_q \big), \qquad (k \in \mathbb Z \setminus \{0\}).
\label{eq:VirLk}
\ee
Finally, computing the commutator $[\Ldim_k,\Ldim_{-k}] = 2k \Ldim_0 + \frac c{12} (k^3 - k)$, we arrive at
\be
\Ldim_0= \sideset{}{'}\sum_{q \ge 0} q \big(\theta^-_{-q}\theta^+_{q} + \theta^+_{-q} \theta^-_q \big) + \Delta, \qquad \Delta = \left\{\begin{array}{cl} 0 & n \,{\rm even}, \\[0.1cm] -\tfrac18 & n \,{\rm odd}. \end{array}\right.
\label{eq:VirL0}
\ee
and, crucially,
\be
c = -2.
\ee
The weights $\Delta = 0$ and $\Delta=-\frac18$ are readily recognised as the conformal dimensions $\Delta_{1,1}$ and $\Delta_{1,2}$ of the $c=-2$ Kac table, see \eqref{KacD}. The final compact form for the Virasoro modes is
\be
\Ldim_k = \sideset{}{'}\sum_{q \ge 0} q\, \big(\theta^-_{-q}\theta^+_{q+k} + \theta^+_{-q+k} \theta^-_q \big) + \delta_{k,0}\, \Delta,  \qquad (k \in \mathbb Z).
\label{eq:finalLk}
\ee

%%%%%
\subsection{Module decompositions}\label{sec:structure}
%%%%%

This section investigates the structure of the dimer Fock spaces as Virasoro modules (under the action of \eqref{eq:finalLk}),
comparing them with the representation theory of other realisations of the Virasoro algebra at $c=-2$.

From \eqref{eq:finalLk}, we find that the modes of the fields $\theta^-(z)$ and $\theta^+(z)$ satisfy 
\be
[\Ldim_k, \theta^-_q] = - (k+q)\, \theta^-_{k+q}, \qquad [\Ldim_k, \theta^+_q] = - q\, \theta^+_{k+q},
\ee
demonstrating that they are primary fields of conformal weight $0$ and $1$, respectively. 
In fact, the expression \eqref{eq:finalLk} is a known realisation of the Virasoro algebra at $c=-2$ and has previously appeared in a 
fermionic $bc$ ghost system \cite{FMS86}. It was studied in detail by Kausch~\cite{K95}.

As discussed in \cref{sec:Sz}, for finite $n$, the full state space $\ctwotimes{n-1}$ is naturally divided into $T^2(\alpha)$-invariant
subspaces $E_{n-1}^v$ labeled by the eigenvalues $v$ of $S^z$. 
The space $E_{n-1}^v$ is also invariant under the action of the densities $\mathcal H_k$ (since it is left invariant by the Temperley-Lieb generators) 
and is thus expected to become a Virasoro module in the scaling limit. In that limit, $S^z$ is written in terms of the modes $\theta^\pm_q$ as
\be
S^z = -\sideset{}{'}\sum_{q \ge 0} \theta^-_{-q}\theta^+_q + \sideset{}{'}\sum_{q >0} \theta^+_{-q}\theta^-_q + \tfrac12 \delta_{n, 0 \bmod 2}\,.
\label{eq:Szlim}
\ee
Each invariant subspace $E_{n-1}^v$ becomes infinite-dimensional,
\be
E_{n-1}^v  \; \xrightarrow{n \rightarrow \infty}\; E^v = {\rm span}\big\{ \theta^-_{p_1} \cdots \theta^-_{p_i}\theta^+_{q_1} \cdots \theta^+_{q_j}|\Omega\rangle \big\} \quad {\rm with} \quad j-i = \left\{\begin{array}{cl}v & n\, {\rm odd,}\\[0.1cm] v + \tfrac12 & n\, {\rm even,}\end{array}\right.
\ee
where the labels $p_k$ and $q_\ell$ are non-positive and take half-integer and integer values for $n$ odd and even, respectively. A basis of states is obtained by restricting the labels to $p_1 < p_2< \dots < p_i$ and $q_1 < q_2< \dots < q_j$. The (unique) ground state $|\Omega_v \rangle$ of $E^v$ is 
\be
|\Omega_v \rangle =  \left\{\begin{array}{ll} 
 \displaystyle\sideset{}{'}\prod_{0 < q <-v}\theta^-_{-q}|\Omega \rangle & \ v \le 0, \\[0.7cm]  
\displaystyle\sideset{}{'}\prod_{0\le q <v}\theta^+_{-q}|\Omega \rangle & \ v \ge 0,\end{array}\right. 
\label{omegav}
\ee
and has conformal dimension
\be
\Delta_v = \frac{4 v^2 - 1}8.
\label{eq:confdim}
\ee
Here, the indices of the primed products run over half-integers and integers for $n$ odd and even, respectively. Recalling that
\be
\theta^+_q |\Omega\rangle = 0\, \quad (q>0) 
\qquad {\rm and} \qquad \theta^-_q |\Omega\rangle = 0 \quad (q \ge 0),
\ee
and comparing with \eqref{eq:gs}, we see that the ground state 
for $n$ odd is $|\Omega\rangle = |\Omega_0 \rangle$, while for $n$ even, $|\Omega\rangle= |\Omega_{-1/2} \rangle$ and $|\Omega'\rangle = |\Omega_{1/2} \rangle = \theta^+_0 |\Omega\rangle$.

Because the character over $E^v$ is a Verma module character, the number of states at level $\ell$ is equal to the number $p(\ell)$ of integer partitions of $\ell$. These states, however, are not all descendants of $|\Omega_v\rangle$. As the following theorem states, $E^v$ is not a Verma module but instead a Feigin-Fuchs module. The definition of Feigin-Fuchs modules is reviewed in \cref{sec:FFs}. 

\begin{Theoreme} 
\label{thm:structurethm}
The module $E^v$ is isomorphic to the $c=-2$ Feigin-Fuchs module $\mathcal F_{\lambda}$ with $\lambda = \tfrac12-v$.
Thus, its structure is
\begin{itemize}
\item[(i)] for $n$ odd,
\be
E^v \simeq E^{-v}\; : \; \qquad
\begin{pspicture}[shift=-0.25](0,-0.2)(6,0.7)
\multiput(0,0)(1.5,0){4}{\pscircle[linewidth=0.025,fillstyle=solid,fillcolor=black](0,0){0.12}}
\multiput(0.60,-0.1)(1.5,0){4}{$\oplus$}
\rput(0,0.5){\small $\Delta_v$}
\rput(1.5,0.5){\small $\Delta_{v+2}$}
\rput(3,0.5){\small $\Delta_{v+4}$}
\rput(4.5,0.5){\small $\Delta_{v+6}$}
\rput(5.95,0){$\cdots$}
\end{pspicture}
\qquad (v \ge 0),
\label{odd}
\ee
\item[(ii)] for $n$ even,
\be
E^v \; : \; \  \left\{\quad
\begin{array}{ll}
\begin{pspicture}[shift=-0.3](0,-0.2)(9.5,0.7)
\multiput(0,0)(3,0){3}{\pscircle[linewidth=0.025,fillstyle=solid,fillcolor=black](0,0){0.12}\psline[linewidth=0.8pt,arrowscale=1.2,arrowinset=0.4]{<-}(0.3,0)(1.2,0)}
\multiput(1.5,0)(3,0){3}{\pscircle[linewidth=0.025,fillstyle=solid,fillcolor=mygray](0,0){0.12}\psline[linewidth=0.8pt,arrowscale=1.2,arrowinset=0.4]{->}(0.3,0)(1.2,0)}
\rput(0,0.5){\small $\Delta_v$}
\rput(1.5,0.5){\small $\Delta_{v+1}$}
\rput(3,0.5){\small $\Delta_{v+2}$}
\rput(4.5,0.5){\small $\Delta_{v+3}$}
\rput(6.0,0.5){\small $\Delta_{v+4}$}
\rput(7.5,0.5){\small $\Delta_{v+5}$}
\rput(9.25,0){$\cdots$}
\end{pspicture}
\qquad &(v \ge \tfrac 12), \\[0.6cm]
\begin{pspicture}[shift=-0.3](0,-0.2)(9.5,0.7)
\multiput(0,0)(3,0){3}{\pscircle[linewidth=0.025,fillstyle=solid,fillcolor=mygray](0,0){0.12}\psline[linewidth=0.8pt,arrowscale=1.2,arrowinset=0.4]{->}(0.3,0)(1.2,0)}
\multiput(1.5,0)(3,0){3}{\pscircle[linewidth=0.025,fillstyle=solid,fillcolor=black](0,0){0.12}\psline[linewidth=0.8pt,arrowscale=1.2,arrowinset=0.4]{<-}(0.3,0)(1.2,0)}
\rput(0,0.5){\small $\Delta_{-v}$}
\rput(1.5,0.5){\small $\Delta_{-v+1}$}
\rput(3,0.5){\small $\Delta_{-v+2}$}
\rput(4.5,0.5){\small $\Delta_{-v+3}$}
\rput(6.0,0.5){\small $\Delta_{-v+4}$}
\rput(7.5,0.5){\small $\Delta_{-v+5}$}
\rput(9.25,0){$\cdots$}
\end{pspicture}
\qquad &(v \le -\tfrac 12).
\end{array}\right.
\label{eq:struceven}
\ee
\end{itemize}
\end{Theoreme}
Here, each circle represents an irreducible composition factor whose highest weight is indicated just above it.
A proof of \cref{thm:structurethm} is given in \cref{sec:iso}. Notably, the structure of $E^v$, in the form of chains of irreducible modules with alternating arrows, is similar to the structure of $E^v_{n-1}$ as a Temperley-Lieb module, the difference being that the chains in the latter case are of finite length \cite{MDRR14}. We also note that $\Ldim_0$ is diagonalisable on each $E^v$, as are the transfer matrix and Hamiltonian on $E_{n-1}^v$ for finite $n$.

The structures of the Virasoro modules underlying the scaling limit of the standard modules $\stan_{n}^d$ for critical dense polymers are simpler. They contain either one or two composition factors, see \eqref{eq:char}. If there is a single composition factor, that is if $n$ is odd, or if $n$ is even and $d=0$, then the module is irreducible. For $n$ even and $d>0$, there are two composition factors and the scaling limit of $\stan_n^d$ was conjectured in \cite{R11} to be an indecomposable highest-weight module, with extra evidence given in \cite{MDRR15}. Remarkably, this is actually a consequence of \cref{thm:structurethm}, as shown at the end of \cref{sec:iso}.

We conclude this section by recalling the relation between the $c=-2$ realisation of the fermionic $bc$ ghost system and that of symplectic fermions \cite{K00}.
In the chiral case, the symplectic fermions $\chi^+(z)$ and $\chi^-(z)$ are two interacting fields defined as
\be
\chi^\alpha(z) = \sum_{q \in \mathbb Z+r} \chi_q^\alpha z^{-q-1}, \qquad (\alpha \in \{+,-\}),
\label{eq:chialpha}
\ee 
whose modes satisfy the anticommutation relations
\be
\{\chi_p^\alpha, \chi_q^\beta\} = p\, d^{\alpha \beta} \delta_{p+q,0}, \qquad (\alpha, \beta \in \{+,-\}).
\label{eq:commchi}
\ee
Here, $d^{\alpha\beta}$ is an anti-symmetric tensor which, for simplicity, 
we set to
\be
\begin{pmatrix}
d^{++} & d^{+-} \\ 
d^{-+} & d^{--}
\end{pmatrix} = 
\begin{pmatrix}
0 & 1 \\ 
-1 & 0
\end{pmatrix}.
\ee

In the Neveu-Schwarz and Ramond sectors, the value of $r$ in \eqref{eq:chialpha} is, respectively, $0$ and $\frac12$, corresponding to integer and half-integer labels. The Virasoro modes take the form
\be 
\LSF_k = \sideset{}{'}\sum_{q \ge 0} \chi^-_{-q}\chi^+_{q+k} - \sideset{}{'}\sum_{q > 0} \chi^+_{-q+k}\chi^-_{q} + \delta_{k,0}\,\Delta,
 \qquad \Delta = \left\{\begin{array}{cl} 0 & \ \hbox{Neveu-Schwarz,} \\[0.1cm] -\tfrac18 & \ {\rm Ramond,} \end{array}\right.
 \label{eq:modesSF}
\ee
where the primed sums run over integer or half-integer values, depending on the sector. It is noteworthy that the generator $\LSF_0$ admits non-trivial Jordan blocks \cite{K00}.

One can define a map
\be
\chi_q^+ \mapsto \theta_q^+, \qquad \chi_q^- \mapsto - q\, \theta_q^-,
\label{SFmapstoBC}
\ee
which, respectively, maps the Neveu-Schwarz and Ramond sectors to the even and odd sectors of the $bc$ system described in \cref{sec:higher}. 
Indeed, the commutation relations \eqref{eq:commchi} become \eqref{eq:thetacomm},
and the Virasoro modes \eqref{eq:modesSF} are mapped to \eqref{eq:finalLk}.

In the Ramond sector, $q$ is never zero, so \eqref{SFmapstoBC} is bijective, and the two theories have the same representation content. In the Neveu-Schwarz sector, however, the generator $\chi_0^-$ is mapped to zero in \eqref{SFmapstoBC}, so the map is not bijective. As observed in \cite{K95}, in this case, the $bc$ system is a quotient of the symplectic fermion system, and the representation contents are thus different, with the Jordan blocks of $\LSF_0$ not transferred to the $L_0$ mode of the $bc$ system.

%%%%%
\subsection{Conformal integrals of motion}
\label{sec:integrals}
%%%%%

The tangle $\Db(u,\xi_v)$ discussed in Section \ref{sec:integ} is important for two reasons. First, it uniformises the treatment of critical dense polymers and that of the dimer model, in the sense that it reproduces their corresponding double-row transfer tangles upon specialising the parameters to $v=0$ and $u=\frac v2$, respectively (although the modules on which the two tangles act are different). Second, and more important for the integrability of the dimer model, $\Db(u,\xi_v)$ provides a one-parameter family of self-commuting tangles which also commute with the tilted transfer tangle,
\be
[\Db(u,\xi_v),\Db(u',\xi_v)] = 0\,, \qquad [\Db(u,\xi_v),\funkyDb(v)] = 0.
\ee
As discussed in \cref{sec:integ}, the coefficients in the expansion of $\Db(u,\xi_v)$ in powers of $\sin{2u}$ are lattice integrals of motion, namely, operators commuting with the transfer tangle and being in involution. At finite size, there is only a finite number of such linearly independent operators. 
As the simple example $n=3$ in \eqref{eq:iom3} indicates, their explicit computation is complicated already for small $n$. This is also evident from the expressions given in \cite{Ni09}.

At the level of spectra, on the other hand, a complete picture can be obtained from the eigenvalues of the transfer tangle. From Section \ref{sec:inv}, the eigenvalues of $-\tfrac 12\log \Db(u,\xi_v)$ take the following general form, valid in any representation,
\be
{\rm Eig}\big(\!-\!\tfrac 12\log \Db(u,\xi_v)\big) = -\tfrac 12\log{(\cos^{n-1} v)} + \hspace{-0.25cm}\sum_{\substack{k = 1 \\[0.05cm] k = n\bmod 2}}^{n-2}\hspace{-0.25cm} \Big( -\omega(q_k) + \tfrac 12 (\nu_k + \tau_k) \: \big(\omega(q_k) - \omega(-q_k)\big)\Big),
\ee
with the function $\omega(t)$ given in \eqref{eq:omega!}, namely,
\be
\omega(t) = \log \big( \sqrt{1+\alpha^2 \sin^2 t} + \sqrt{1+\alpha^2}\, \sin 2u\, \sin t  \big).
\label{omeg}
\ee

For finite size, the eigenvalues of $-\tfrac 12\log \Db(u,\xi_v)$ depend on $n$. Their expansion in inverse powers of $n$ has the form \eqref{eq:Elargen} and the coefficients have been worked out in Section \ref{sec:spectra} for a generic function $\omega(t)$. We rewrite the expansion as 
\be
{\rm Eig}\big(\!-\!\tfrac 12\log \Db(u,\xi_v)\big) = n f_{\textrm{bulk}} + f_{\textrm{bdy}} + \sum_{p=1}^\infty {2^p\pi^{2p-1} \over (2p)!} \, \frac{\lambda_{2p-1}(\alpha,u)}{n^{2p-1}} \, I_{2p-1}(\nu,\tau),
\label{Eig}
\ee
where the numerical coefficients $I_{2p-1}(\nu,\tau)$ depend on the specific eigenvalue determined by the choice of $\nu_k,\tau_k$, and, from \eqref{ap}, are given by
\be
I_{2p-1}(\nu,\tau) = {1 \over 2^p} \Bigg[{\rm B}_{2p}(r) + 2p \hspace{-0.3cm} \sum_{\substack{k\ge 1 \\[0.05cm] k = n\bmod 2}}\hspace{-0.3cm}  (\nu_k + \tau_k) \, \Big({k \over 2}\Big)^{2p-1}\Bigg].
\label{I2p}
\ee

In the present case, the functions $\lambda_{2p-1}$ of $\alpha$ and $\sin {2u}$ are the odd-power coefficients in the Taylor expansion in $t$ of the function $\omega(t)$ given in \eqref{omeg}. It is not difficult to see that $\lambda_{2p-1}$ is an odd polynomial in $\sin{2u}$ of degree $2p-1$. The first few functions read (with $\alpha=\tan v$)
\begin{subequations} 
\begin{alignat}{2}
\lambda_1(\alpha,u) &= \frac{\sin{2u}}{\cos{v}}, \qquad 
\lambda_3(\alpha,u) = 2 \, \frac{\sin^3{2u}}{\cos^3{v}} - (3\alpha^2+1) \, \frac{\sin{2u}}{\cos{v}},
\\[0.2cm]
\lambda_5(\alpha,u) &= 24\,\frac{\sin^5{2u}}{\cos^5{v}} - 20 (3 \alpha^2+1) \, \frac{\sin^3{2u}}{\cos^3{v}} + (45 \alpha^4 + 30 \alpha^2 + 1) \, \frac{\sin{2u}}{\cos{v}}.
\end{alignat}
\end{subequations}
Setting 
\be
\lambda_{2p-1}(\alpha,u) = \sum_{m=1}^p \, g^{(2p-1)}_{2m-1}(\alpha) \, (\sin{2u})^{2m-1},
\ee
we can reorganise the expansion \eqref{Eig} as a series in $\sin{2u}$: 
\be
{\rm Eig}\big(\!-\!\tfrac 12\log \Db(u,\xi_v)\big) = n f_{\textrm{bulk}} + f_{\textrm{bdy}} + \sum_{m \ge 1} (\sin{2u})^{2m-1} \: \sum_{p \ge m} \: g^{(2p-1)}_{2m-1}(\alpha)\, {2^p\pi^{2p-1} \over (2p)!} \, \frac{I_{2p-1}(\nu,\tau)}{n^{2p-1}}. 
\ee
By exponentiating this expression and comparing with \eqref{eq:lattIOM}, one obtains the $\frac 1n$ expansion of the eigenvalues of the lattice integrals of motion $\Jb_{m}$. This furnishes an infinite triangular linear system relating the $I_{2p-1}$ to the $J_{m}$, from which we can reasonably infer that the quantities $I_{2p-1}$ are themselves eigenvalues of certain integrals of motion. As argued below, for $c=-2$, they reproduce the spectra of the conformal integrals of motion ${\mathbf I}_{2p-1}$.

For the specific case of the dimer model, for which one simply sets $2u=v$ in \eqref{Eig} (and for the critical dense polymer as well, in which case the relevant tangle is $\Db(u,0)$), we therefore recover what appears to be a generic fact for conformally invariant spectra of transfer matrices: The large $n$ eigenvalues involve coefficients $I_{2p-1}$ which are eigenvalues of integrals of motion. This general observation has been first put forward in \cite{BLZ96,BLZ97}, and later checked explicitly in the tricritical Ising model \cite{FG04}, for critical dense polymers \cite{Ni09} and in the Ising model \cite{Ni10,Nigro15}.

The explicit values of $I_{2p-1}$ given in \eqref{I2p} come directly from lattice computations. In the scaling limit, the integrals of motion ${\mathbf I}_{2p-1}$ are universal conformal operators acting in the conformal representations occurring in the model. 
(This makes them much better behaved and, hence, easier to deal with, than their finite-size counterparts $\Jb_{m}$.) Algebraically, the operators ${\mathbf I}_{2p-1}$ form an infinite-dimensional Abelian subalgebra of the enveloping algebra of the Virasoro algebra. Although their explicit form is not known for all values of $p$, they can be constructed as charges associated to normal-ordered polynomials of degree $p$ in the stress-energy tensor and its derivatives \cite{SY88,EY89}. The first few read
\begin{subequations}\label{eq:I135}
\begin{alignat}{4}
{\mathbf I}_1 &= L_0 - \frac{c}{24}, \label{i1}\\
{\mathbf I}_3 &= 2\sum_{n=1}^{\infty} L_{-n}L_n + L_0^2 - \frac{c+2}{12}L_0 + \frac{c(5c+22)}{2\,880},\label{i3}  \\
{\mathbf I}_5 &= \sum_{m,n,p\in\mathbb{Z}} \delta_{m+n+p,0}:\!L_m L_n L_p\!: + \frac{3}{2} \, \sum_{n=1}^\infty L_{1-2n}L_{2n-1} - \frac{c(3c+14)(7c+68)}{290\,304}\nonumber \\
& \hspace{1.5cm} + \sum_{n=1}^\infty\Bigg(\frac{11+c}{6}n^2-\frac{c}{4}-1\Bigg)L_{-n} L_n - \frac{c+4}{8}L_0^2 + \frac{(c+2)(3c+20)}{576}L_0,
\label{i5}
\end{alignat}
\end{subequations}
where the normal ordering means that the modes are ordered from left to right by increasing values of their indices. Because different triplets $(m,n,p)$ can produce the same ordered product, some $L_mL_nL_p$ will appear with (integer) prefactors in ${\mathbf I}_5$. For example, $L_0^3$, $L^{}_{-2}L_1^2$ and $L_{-1}L_0L_1$ appear with coefficients $1, 3$ and $6$, respectively.

Our main goal here is to verify that $\Db(u,\xi_v)$ is a generating function for the integrals of motion for $c=-2$ 
by comparing the eigenvalues of the first few ${\mathbf I}_{2p-1}$ on the Feigin-Fuchs modules with the results \eqref{I2p} coming from the lattice. 

From their explicit expressions above, one is easily convinced that the actual diagonalisation of the integrals of motion quickly becomes impractical beyond the first few levels. When they are restricted to act on a highest-weight state $|\mathsf v_\Delta\rangle$, many more eigenvalues have been computed. For instance, the eigenvalues of ${\mathbf I}_1,{\mathbf I}_3,\ldots,{\mathbf I}_{15}$ for $|\mathsf v_\Delta\rangle$ are given explicitly in \cite{BLZ97} for any value of $\Delta$ and $c$. In the special case of $c=-2$, the eigenvalues of all integrals of motion on a highest-weight state are 
\be
I_{2p-1}(\mathsf v_{\Delta_v}) = {1 \over 2^p} \, {\rm B}_{2p}\Big(|v| + {1 \over 2}\Big), \qquad \Delta_v = \frac{4v^2-1}{8},
\label{highest}
\ee
where $I_{2p-1}(\mathsf v)$ is the eigenvalue of $\mathbf I_{2p-1}$ associated with the eigenvector $|\mathsf v \rangle$.

The rest of this subsection is devoted to the comparison of lattice and conformal eigenvalues of the integrals of motion for the modules $E^v$, whose structures were given in Theorem 1. Our investigation covers all $\mathbf I_{2p-1}$ for the zeroth level, as well as $\mathbf I_3$ and $\mathbf I_5$ ($\mathbf I_1$ is trivial) for the first and second levels. We will assume $c=-2$ in the above expressions for ${\mathbf I}_3$ and ${\mathbf I}_5$, but will discuss the possibility of having $c=1$ in \cref{sec:higherc=1}. 

Let us note that the expansion \eqref{Eig} assumes a definite normalisation for the quantities $I_{2p-1}(\nu, \tau)$, relative to the normalisation of the conformal integrals of motion of which they are claimed to be the eigenvalues. 
Because this normalisation is independent of the vector $(\nu,\tau)$, it is fixed by the ground-state analysis.
The interpretation of the finite-size corrections to $\Db(u,\xi_v)$ as eigenvalues of conformal integrals of motion is subsequently unambiguous.
To avoid confusion, from now on, we will denote by $I_{2p-1}$ the eigenvalues obtained from the action of ${\mathbf I}_{2p-1}$ in \eqref{eq:I135} on the module $E^v$, and by $I_{2p-1}^{\rm latt}$ the quantities \eqref{I2p} obtained from the lattice. These can be easily computed for Fock states constructed from fermionic excitations of the vacuum $|\Omega\rangle$: The vacuum itself corresponds to all $\nu_k,\tau_k=0$, while the action of $\theta^+_{-q}$ or $\theta^-_{-q}$ on the vacuum sets, respectively, $\nu^{}_{2q}=1$ or $\tau^{}_{2q}=1$, for $q>0$. (The creation operator $\theta_0^+$ does not modify the quantum numbers.) The values of $\delta_{k} = \nu_k + \tau_k$ are given for the first four excited states in \cref{fig:deltak}. As before, we set $r=\tfrac n2 \bmod 1$, taken in $\{0,\tfrac12\}$.

\paragraph{Level 0} 
All modules $E^v$, for $v$ integer or half-integer, have a unique highest-weight state $|\pi_0\rangle=|\Omega_v\rangle$, of conformal dimension $\Delta_v$. From the expression given in \eqref{omegav}, the state $|\Omega_v\rangle$ corresponds to $\nu^{}_{2q}=1$ for $0 < q < v$ (and all other zero) and to $\tau^{}_{2q}=1$ for $0 < q < -v$ (and all other zero). In both cases, using the general identity 
\be
\sum_{m=0}^{t-1} (\alpha m + \beta)^\ell = {\alpha^\ell \over \ell+1} \Big({\rm B}_{\ell+1}\big(t+\tfrac \beta\alpha\big) - {\rm B}_{\ell+1}\big(\tfrac \beta\alpha\big)\Big),
\ee
we readily obtain 
\be \label{eq:IB}
I_{2p-1}^{\rm latt}(\pi_0) = {{\rm B}_{2p}(r) \over 2^p}  + {p \over 2^{p-1}} 
\sideset{}{'}\sum_{0<q<|v|}
q^{2p-1} = {{\rm B}_{2p}(r) \over 2^p}  + {p \over 2^{p-1}} \hspace{-0.1cm} \sum_{m=0}^{|v|-r-\frac 12} (m+r)^{2p-1} = {{\rm B}_{2p}\big(|v|+\frac 12\big) \over 2^p},
\ee
which exactly reproduces the result $I_{2p-1}(\mathsf v_{\Delta_v})$ quoted in \eqref{highest}.

\paragraph{Level 1}
For a given $v$, all states in $E^v$ are constructed by keeping the same balance between the number of $\theta^+_p$ and $\theta^-_q$ as for $|\pi_0\rangle$, but by allowing larger values for $p$ and $q$. 
There is a single state in $E^v$ with $L_0$-eigenvalue equal to $\Delta_v+1$, given by 
\be
|\pi_1\rangle = \theta^+_{-(v+\frac 12)} \, \theta^-_{v-\frac 12} \, |\pi_0\rangle \simeq 
\left\{\begin{array}{ll} 
\theta^-_{v-\frac 12} \;\displaystyle\sideset{}{'}\prod_{0 < q < -v-1} \!\! \theta^-_{-q} |\Omega\rangle & \quad v <0,\\[7mm] 
\theta^+_{-\frac12}\theta^-_{-\frac12} |\Omega\rangle & \quad v = 0,\\[5mm]
\theta^+_{-(v+\frac 12)} \;\displaystyle\sideset{}{'}\prod_{0 \le q < v-1}\! \theta^+_{-q} |\Omega\rangle & \quad v > 0,
\end{array} \right.
\ee
where the symbol $\simeq$ means an equality up to a possible multiplicative sign. The quantum numbers $\delta_k$ for $|\pi_1 \rangle$ are given in \cref{fig:deltak}. For $v \neq 0$, the vector $\delta_k=\nu_k+\tau_k$ corresponding to $|\pi_1\rangle$ is obtained from that of $|\pi_0\rangle$ by changing $\delta^{}_{2|v|-1}: 1 \to 0$ and $\delta^{}_{2|v|+1}: 0 \to 1$ for $v \ne 0$. For $v=0$, $|\pi_1\rangle = \theta^+_{-1/2}\theta^-_{-1/2}|\Omega\rangle$ contains two excitations, so the difference with $|\pi_0\rangle$ is that $\delta_1:0\to 2$. We therefore obtain
\be
I_{2p-1}^{\rm latt}(\pi_1) = {{\rm B}_{2p}\big(|v|+\frac 12\big) \over 2^p} + {p \over 2^{p-1}} \Big( \big(|v|+\tfrac 12\big)^{2p-1} - \big(|v|-\tfrac 12\big)^{2p-1}\Big),
\ee
also valid for $v=0$. Specialising to $p=2,3$ yields
\be
I_{3}^{\rm latt}(\pi_1) = \Delta_v^2 + 6 \Delta^{}_v + \tfrac{119}{120},\qquad
I_{5}^{\rm latt}(\pi_1) = \Delta_v^3 + \tfrac{59}{4} \Delta_v^2 + \tfrac{15}{2} \Delta_v^{} + \tfrac{253}{336},\label{35pi1}
\ee
where we have used 
\be
{\rm B}_4(x) = x^4-2 x^3+x^2-\frac{1}{30}, \qquad
{\rm B}_6(x) = x^6-3 x^5+\frac{5 x^4}{2}-\frac{x^2}{2}+\frac{1}{42}.
\ee

On the conformal side, the module structures described in Theorem 1 show that $|\pi_1\rangle$ is a Virasoro descendant of $|\pi_0\rangle$ except if $v=\frac 12$, as confirmed by a direct computation giving $L_{-1} |\pi_0\rangle = (v-\frac 12) |\pi_1\rangle$ and $L_1 |\pi_1\rangle = (v+\frac 12) |\pi_0\rangle$. In all cases, we obtain 
\be L_{-1}L_1 |\pi_1\rangle = 2\Delta_v |\pi_1\rangle, \qquad L_{-1}L_0L_1 |\pi_1\rangle = 2\Delta_v^2 |\pi_1\rangle.\ee 
The calculation of the eigenvalues of ${\mathbf I}_3$ and ${\mathbf I}_5$ on $|\pi_1\rangle$ is straightforward from \eqref{eq:I135} and reproduces the expressions \eqref{35pi1} exactly.

\begin{figure}
\begin{center}
$
\begin{array}{c||c|c|c|c}
k & 2|v|-3 & 2|v|-1 & 2|v|+1 & 2|v|+3\\ \hline\hline &&&&\\[-0.35cm]
|\pi_0\rangle & 1 & 1 & 0 & 0\\[0.2cm]
|\pi_1\rangle & 1 & 0 & 1 & 0\\[0.2cm]
|\pi_2\rangle {\rm \ or \ }|\pi_2'\rangle 
& 1 & 0 & 0 & 1\\[0.2cm]
|\pi_2'\rangle {\rm \ or \ }|\pi_2\rangle 
& 0 & 1 & 1 & 0
\end{array}
$
\end{center}
\captionof{table}{The quantum numbers $\delta_{k}$ of the first four states in $E^v$ for $|v| \ge 3/2$. For these states, $\delta_{k}=1$ for $k<2|v|-3$ and $\delta_{k} = 0$ for $k>2|v|+3$. The cases $v = 0, \pm \frac12, \pm 1$ are special. For $v = 0$, the pairs $(\delta_1,\delta_3)$ corresponding to the first four states are, in order, $(0,0)$, $(2,0)$, $(1,1)$ and $(1,1)$. For $|v| = \frac12$, the pairs $(\delta_2,\delta_4)$ are $(0,0)$, $(1,0)$, $(0,1)$ and $(1,1)$. For $|v| = 1$, the triples $(\delta_1,\delta_3, \delta_5)$ of the first four states are $(1,0,0)$, $(0,1,0)$, $(0,0,1)$ and $(2,1,0)$.}
\label{fig:deltak}
\end{figure}

\paragraph{Level 2}
A basis for the two states $|\pi_2\rangle,|\pi_2'\rangle$ at level 2 in $E^v$ is 
\be
|\pi_2\rangle =   \theta^+_{-(v+\frac 32)} \, \theta^-_{v-\frac 12} \, |\pi_0\rangle, \qquad |\pi'_2\rangle = \theta^-_{v-\frac 32} \, \theta^+_{-(v+\frac 12)} \, |\pi_0\rangle.
\ee
It follows that for $v \le 0$, $|\pi_2\rangle$ differs from $|\pi_0\rangle$ by one new excitation with mode label $k = |v|+\tfrac 12$ and none for $k = |v|-\tfrac 32$, whereas $|\pi_2'\rangle$ has an excitation for $k =|v|+\tfrac 32$ and none for $k = |v|-\tfrac 12$. For $v \ge 0$, the two mode labels for $|\pi_2\rangle$ and $|\pi_2'\rangle$ are identical but interchanged between the creation and annihilation operators. Thus, 
\begin{subequations} \label{eq:Ilevel2}
\begin{alignat}{2}
I_{2p-1}^{\rm latt}(\pi_2 {\rm \ or \ }\pi_2') &= 
{{\rm B}_{2p}\big(|v|+\frac 12\big) \over 2^p} + {p \over 2^{p-1}} \Big( \big(|v|+\tfrac 32\big)^{2p-1} - \big(|v|-\tfrac 12\big)^{2p-1}\Big),\\[0.2cm]
I_{2p-1}^{\rm latt}(\pi_2' {\rm \ or \ }\pi_2)&= 
{{\rm B}_{2p}\big(|v|+\frac 12\big) \over 2^p} + {p \over 2^{p-1}} \Big( \big(|v|+\tfrac 12\big)^{2p-1} - \big(|v|-\tfrac 32\big)^{2p-1}\Big).
\end{alignat}
\end{subequations}
For $p=2,3$, and in terms of $\Delta_v$, we obtain the following two pairs,
\begin{alignat}{2}
I_3^{\rm latt}(\pi_2 {\rm \ or \ } \pi'_2) &= \Delta_v^2 + 12\Delta_v + \tfrac{599}{120} \pm 3\sqrt{8\Delta_v+1}\label{3pi2}\\[-1.0cm]\nonumber 
\intertext{and}\nonumber\\[-1.0cm] 
I_5^{\rm latt}(\pi_2{\rm \ or \ } \pi'_2) &= \Delta_v^3 + \tfrac{119}{4}\Delta_v^2 + 60\Delta_v + \tfrac{4285}{336} \pm 15(\Delta_v+\tfrac 34)\sqrt{8\Delta_v+1}.\label{5pi2}
\end{alignat}

Let us now compare these values with the eigenvalues of ${\mathbf I}_3$ and ${\mathbf I}_5$ on the two states at level 2 in the modules $E^v$. According to \cref{thm:structurethm}, these two states are descendants of $|\pi_0\rangle$ except for $v = 0, \frac12, \frac 32$. It is not difficult to compute the action of the two integrals of motion on the basis $L_{-2}|\pi_0\rangle,\,L_{-1}^2 |\pi_0\rangle$. We find (still with $c=-2$)
\begin{subequations}
\begin{alignat}{2}
{\mathbf I}_3
\Big|_{{\rm level \, }2} 
&= \Big[\Delta_v^2 + 12 \Delta_v + \tfrac{479}{120}\Big] + 
\begin{pmatrix}
- 2 & 12 \Delta_v \\
6 & 4
\end{pmatrix},\label{bfi3}\\[3mm]
{\mathbf I}_5
\Big|_{{\rm level \, }2} 
 &= \Big[\Delta_v^3 + \tfrac{119}{4} \Delta_v^2\Big] + 
\begin{pmatrix}
45 \Delta_v + {505 \over 336} & 60\Delta_v^2 + 45\Delta_v \\[2mm]
30\Delta_v + {45 \over 2} & 75 \Delta_v + {8\,065 \over 336}
\end{pmatrix}.
\label{bfi5}
\end{alignat}
\end{subequations}
In each case, a simple diagonalisation shows that the eigenvalues are precisely the values given in \eqref{3pi2} and \eqref{5pi2}.

For $v=0$, the highest-weight state $|\pi_0\rangle$ has weight $\Delta_0=-\frac 18$. In this case, the eigenvalues in \eqref{eq:Ilevel2} are degenerate (for all $p$). From the module structure  in \eqref{odd}, the state $|\pi_0\rangle$ has a single descendant at level 2, as follows from the identity $L^{}_{-2}|\pi_0\rangle = 2L_{-1}^2 |\pi_0\rangle$. This single descendant is an eigenstate of ${\mathbf I}_3$ and ${\mathbf I}_5$ and accounts for one of the two degenerate eigenvalues, namely
\be
I_3 = \Delta_0^2 + 12\Delta_0 + \tfrac{599}{120} = \tfrac{3\,367}{960}, \qquad I_5 = \Delta_0^3 + \tfrac{119}{4}\Delta_0^2 + 60\Delta_0 + \tfrac{4\,285}{336} = \tfrac{61\,457}{10\,752}.
\ee
The other state at level 2 is not a descendant of $|\pi_0\rangle$. Instead, it is a highest-weight state with weight $\Delta_2=\frac {15}{8}$. As such, it must be an eigenstate of all ${\mathbf I}_{2p-1}$ with eigenvalues given by \eqref{highest}. For $p=2,3$, these eigenvalues are
\be 
I_3(\mathsf v_{\Delta_2}) = \tfrac 14 {\rm B}_{4}\big(\tfrac 52 \big) = \tfrac{3\,367}{960}, 
\qquad I_5(\mathsf v_{\Delta_2}) = \tfrac 18 {\rm B}_{6}\big(\tfrac 52 \big) = \tfrac{61\,457}{10\,752},
\ee
as expected.

For $v=\frac 12$, $\Delta_{1/2}=0$ and the relevant quantities, \eqref{3pi2} and \eqref{5pi2}, are equal to
\be
\{I_3^{\rm latt}(\pi_2),I_3^{\rm latt}(\pi'_2)\} = \{\tfrac{239}{120},\tfrac{959}{120}\}, \qquad 
\{I_5^{\rm latt}(\pi_2),I_5^{\rm latt}(\pi'_2)\} = \{\tfrac{505}{336},\tfrac{8\,065}{336}\}.
\ee
Because $L_{-1}|\pi_0\rangle=0$, the state $|\pi_0\rangle$ has a unique descendant at level 2, namely $L_{-2} |\pi_0\rangle$. The action of ${\mathbf I}_3$ and ${\mathbf I}_5$ on $L_{-2}|\pi_0\rangle$ is most easily obtained from the upper-left entry in the $2\times2$ matrices given in \eqref{bfi3} and \eqref{bfi5}, namely
\be 
I_3=\Delta_v^2 + 12 \Delta_v + \tfrac{479}{120} - 2 = \tfrac{239}{120}, \qquad 
I_5=\Delta_v^3 + \tfrac{119}{4} \Delta_v^2 + 45 \Delta_v + \tfrac{505}{336} = \tfrac{505}{336}.
\ee
The second state at level 2 is the descendant $L_{-1}|\pi_1\rangle$ of the highest-weight state $|\pi_1\rangle$, which was encountered at level 1 and has $\Delta_{3/2}=1$. The eigenvalues of ${\mathbf I}_3$ and ${\mathbf I}_5$ on such a state are given in \eqref{35pi1}. For $\Delta_{3/2}=1$, they become
\be
I_3 = \Delta_v^2 + 6 \Delta^{}_v + \tfrac{119}{120} = \tfrac{959}{120}, \qquad
I_5 = \Delta_v^3 + \tfrac{59}{4} \Delta_v^2 + \tfrac{15}{2} \Delta_v^{} + \tfrac{253}{336} = \tfrac{8\,065}{336},
\ee
once again in full agreement.

For the last exceptional case, $v=\frac32$, \eqref{3pi2} and \eqref{5pi2} give
\be \label{eq:lastexample}
\{I_3^{\rm latt}(\pi_2),I_3^{\rm latt}(\pi'_2)\} = \{\tfrac{3\,239}{120},\tfrac{1\,079}{120}\}, \qquad 
\{I_5^{\rm latt}(\pi_2),I_5^{\rm latt}(\pi'_2)\} = \{\tfrac{61\,237}{336},\tfrac{8\,317}{336}\}.
\ee
A direct computation, using the expressions \eqref{eq:I135} and the basis $\{\theta^+_{-3} \theta^+_0|\Omega\rangle, \theta^+_{-2} \theta^+_{-1}|\Omega\rangle\}$, shows that the first is an eigenvector of 
both $\mathbf I_3$ and $\mathbf I_5$, and that the eigenvalues indeed coincide with \eqref{eq:lastexample}.

We thus find that the $c=-2$ modules in \cref{thm:structurethm} are entirely consistent with the first finite-size correction terms of the eigenvalues of the dimer transfer matrix (squared), related through the eigenvalues of universal integrals of motion acting on these modules. We note that the finite-size corrections for critical dense polymers are given by the {\em same} numbers $I_{2p-1}^{\rm latt}$ at all orders, albeit generally with different degeneracies. The Virasoro modules in critical dense polymers (discussed in \cref{sec:structure} and \cref{sec:iso}) are structurally different \cite{R11,MDRR15} from the ones in the dimer model. We note, however, that the eigenvalues of the integrals of motion only depend on the composition factors present and are independent of the embedding structures. The consistency with $c=-2$ can therefore also be confirmed in this case~\cite{Ni09}.

%%%%%
\subsection[Virasoro modes at $c = 1$]{Virasoro modes at $\boldsymbol{c = 1}$}\label{sec:higherc=1}
%%%%%

In \eqref{eq:L0dim}, we gave an expression for $\Ldim_0-\frac{c}{24}$ in terms of the fermions $\phi_q$ and $\psi_q$ and built a $c=-2$ realisation 
of the other Virasoro modes in \cref{sec:higher}. In this section, we show that $\Ldim_0$ is also consistent with a $c=1$ conformal description, 
and analyse the resulting Virasoro modules. However, we also argue that this $c=1$ picture is irreconcilable with the 
eigenvalues of the integrals of motion $I_{2p-1}^{\rm latt}$ discussed in \cref{sec:integrals}.

Let us start by recalling the fermionic construction of the Virasoro modes for $c=\tfrac 12$. A single fermionic field has a mode expansion of the 
form 
\be 
f(z) = \sideset{}{'}\sum_{q} z^{-q-\frac12} f_q \qquad {\rm with} \qquad \{f_p,f_q\} = \delta_{p+q,0},
\label{eq:1f}
\ee
where the modes have integer or half-integer labels, depending on whether we consider the Ramond or Neveu-Schwarz sector, respectively (the primed sums run over the corresponding values). The Virasoro modes take the form 
\be
\Lf_k = \sideset{}{'}\sum_{q > \frac k2} (q-\tfrac k2) f_{k-q}f_q +\Deltaf\, 
\delta_{k,0}, \qquad \Deltaf = 
\left\{\begin{array}{cl} \frac1{16} & {\rm Ramond}, \\[0.1cm] 0& {\rm Neveu-Schwarz},
\end{array}\right.
\label{eq:Lk1f}
\ee
and satisfy the Virasoro commutation rules for $c=\frac12$. By computing $[\Lf_k,f_q]$, one confirms that $f(z)$ is a primary field of conformal dimension $\Delta = \frac12$.

The fermions used in the previous sections satisfy $\{\theta^+_p,\theta^-_q\} = \delta_{p+q}$ and give rise to two independent fermions $f^{(1)},f^{(2)}$ of the type given in \eqref{eq:1f}, defined by 
\be
f_q^{\textrm{\tiny ($1$)}} = \left\{\begin{array}{cl} 
\theta^+_q & q>0,\\[0.05cm]
\frac1{\sqrt 2}(\theta^+_0 + \theta^-_0) &q=0,\\
\theta^-_{q}& q<0,
\end{array}\right. \qquad 
f_q^{\textrm{\tiny ($2$)}} = \left\{\begin{array}{cll} 
\theta^-_q & q>0,\\[0.05cm]
\frac\ir{\sqrt 2}(\theta^+_0 - \theta^-_0) &q=0,\\
\theta^+_{q} &q<0,
\end{array}\right.
\ee
in terms of which the anticommutation rules read
\be
\{f_p^{\textrm{\tiny ($\alpha$)}},f_q^{\textrm{\tiny ($\beta$)}}\}
= \delta_{\alpha,\beta}\,\delta_{p+q,0}, 
\qquad (\alpha, \beta \in \{1,2\}).
\ee
The Ramond and Neveu-Schwarz sectors thus correspond to the even and odd parities of $n$ of the dimer model. As the fermions $f_q^{\textrm{\tiny ($1$)}}$ and $f_q^{\textrm{\tiny ($2$)}}$ are 
independent, $\{f_p^{\textrm{\tiny (1)}},f_q^{\textrm{\tiny (2)}}\}
= 0$, the sum of their Virasoro modes yields a realisation of the Virasoro algebra with $c=1$ in the form of two copies of \eqref{eq:Lk1f}, 
\be
\Lff_k = \sideset{}{'}\sum_{q > \frac k2} (q-\tfrac k2)\big(f^{\textrm{\tiny ($1$)}}_{k-q}f^{\textrm{\tiny ($1$)}}_q+ f^{\textrm{\tiny ($2$)}}_{k-q}f^{\textrm{\tiny ($2$)}}_q\big) +\Deltaff \, \delta_{k,0}, \qquad \Deltaff = 
\left\{\begin{array}{cl} \frac1{8} & n \ {\rm even,} \\[0.1cm] 0& n \ {\rm odd.} 
\end{array}\right.
\label{eq:Lk2f}
\ee
In addition, the following operator,
\be
\mathcal X = \ir \Big( \sideset{}{'}\sum_{q \ge 0} f^{\textrm{\tiny ($1$)}}_{-q}f^{\textrm{\tiny ($2$)}}_{q} - \sideset{}{'} \sum_{q>0} f^{\textrm{\tiny ($2$)}}_{-q}f^{\textrm{\tiny ($1$)}}_{q}\Big) 
\label{eq:X}
\ee
commutes with all the Virasoro modes:
\be
[\Lff_k, \mathcal X] = 0, \qquad (k \in \mathbb Z).
\ee

Rewriting the modes $\Lff_k$ in terms of the fermions $\theta^\pm$ provides the announced conformal realisation with $c=1$. We also note that for $k=0$, $\Lff_0 - \tfrac c{24} = \Lff_0 - \tfrac 1{24}$ given in \eqref{eq:Lk2f} for $c=1$ exactly reproduces, for the two parities of $n$, the expression of $\Ldim_0 - \tfrac c{24} = \Ldim_0 + \tfrac 1{12}$ found in \eqref{eq:VirL0} for the dimer model and $c=-2$. The other modes, for $k \neq 0$, have different expressions in the two pictures and are expected to lead to Virasoro modules with different structures. Crucially, the modes $\Lff_k$ were written down heuristically from the observation that $\Ldim_0$ appears to be described by two fermions, and not from the Virasoro mode approximations as in \cref{sec:higher}.

Likewise, the operator $\mathcal X$ differs from the operator $S^z$ used before and therefore partitions the dimer Fock space $\ctwotimes \infty$ in a different way. In particular, the states $|\Omega_v\rangle$ defined in \eqref{omegav} are not eigenstates of $\mathcal X$ for $|v| \ge 1$. To find an eigenbasis of $\mathcal X$, we define two new fermions $\hat \theta^\pm$ by setting
\be
f^{\textrm{\tiny ($1$)}}_{q} = \frac{1}{\sqrt 2} (\hat\theta^+_q + \hat\theta^-_q), \qquad 
f^{\textrm{\tiny ($2$)}}_{q} = \frac{\ir}{\sqrt 2} (\hat\theta^+_q - \hat\theta^-_q),  \qquad({\rm all\ }q)
\ee
so that the two sets of modes $\theta^\pm_q,\hat\theta^\pm_q$ are linearly related. One can easily verify that the anticommutators are preserved,
\be
\{\hat\theta^\alpha_p,\hat\theta^\beta_q \} = \delta_{\alpha + \beta, 0}\,\delta_{p + q,0},  \qquad (\alpha,\beta \in \{ +, - \}).
\ee

In terms of $\hat\theta_q^\pm$, the operator $\mathcal X$ is given by
\be
\mathcal X = - \sideset{}{'}\sum_{q \ge 0} \hat \theta^-_{-q} \hat \theta^+_q + \sideset{}{'}\sum_{q>0}\hat \theta^+_{-q} \hat \theta^-_q + \tfrac12 \delta_{n,0 \bmod 2}\,\label{eq:Xtheta}
\ee
and has exactly the same expression as the operator $S^z$ had in terms of $\theta^\pm_q$. Also, the Virasoro modes acquire a rather simple form,
\be
\Lff_k = \sideset{}{'}\sum_{q > \frac k2}(q-\tfrac k2) \big( \hat \theta^-_{k-q}\hat\theta^+_{q} + \hat \theta^+_{k-q}\hat\theta^-_{q} \big) + \Deltaff\, \delta_{k,0}\,.
\label{eq:L2fktheta}
\ee
Because of
\be
[\Lff_k,\hat \theta^\pm_q] = -(\tfrac k2 + q) \,\hat \theta^\pm_{k+q},
\label{eq:Lk2ftheta}
\ee
the fields $\hat \theta^\pm(z)$ are primary, both of dimension $\Delta=\frac12$. 

The states 
\be 
\hat\theta^-_{p_1} \cdots \hat\theta^-_{p_i}\hat\theta^+_{q_1} \cdots \hat\theta^+_{q_j}|\Omega\rangle \label{eq:Xbasis}
\ee 
form a basis of the dimer Fock space $\ctwotimes \infty$, which is the direct sum of the subspaces $\hat E^x$ where $\mathcal X$ acts as $x\cdot \mathbb I$. These subspaces are generated by the states of the form \eqref{eq:Xbasis} with $j-i = x$ for $n$ odd and $j-i = x+\frac 12$ for $n$ even. Each $\hat E^x$ is invariant under the action of the modes $\Lff_k$ and is thus a module over the Virasoro algebra at $c=1$. 
The (unique) ground state $|\hat \Omega_x \rangle$ of $\hat E^x$ is 
\be
|\hat \Omega_x \rangle =  \left\{\begin{array}{ll} 
 \displaystyle\sideset{}{'}\prod_{0 < k<-x}\hat\theta^-_{-k}|\Omega \rangle & \ x \le 0,\\[0.7cm]  
\displaystyle\sideset{}{'}\prod_{0\le k<x}\hat\theta^+_{-k}|\Omega \rangle & \ x \ge 0,\end{array}\right. 
\label{eq:gss}
\ee
and has conformal dimension
\be
\Deltaff_x = \frac{x^2}2.
\label{eq:deltax}
\ee

Comparing with \cref{sec:higher}, we find that $\theta^\pm_q$ and $\hat \theta^\pm_q$ realise identical anticommutator algebras. However, the Virasoro modes \eqref{eq:finalLk} and \eqref{eq:L2fktheta} have different forms. The next theorem is the counterpart of \cref{thm:structurethm} and gives the structure of $\hat E^x$ as a $c=1$ Virasoro module.

\begin{Theoreme}\label{thm:structurethm1} 
The module $\hat E^x$ is an irreducible $c=1$ Virasoro module, except for $n$ odd and $x = 0$, in which case it is an infinite direct sum of irreducible modules. Thus, its structure is
\be
\hat E^x \; : \; \left\{\;\;
\begin{array}{ll}
\begin{pspicture}[shift=-0.25](-0.2,-0.2)(6,0.7)
\multiput(0,0)(1.5,0){4}{\pscircle[linewidth=0.025,fillstyle=solid,fillcolor=black](0,0){0.12}}
\multiput(0.60,-0.1)(1.5,0){4}{$\oplus$}
\rput(0,0.5){\normalfont $\Deltaff_0
$}
\rput(1.5,0.5){\normalfont$\Deltaff_{2}$}
\rput(3,0.5){\normalfont$\Deltaff_{4}$}
\rput(4.5,0.5){\normalfont$\Deltaff_{6}$}
\rput(5.95,0){$\cdots$}
\end{pspicture} & \quad n {\rm \ odd,}\ x = 0, \\[0.5cm]
\begin{pspicture}[shift=-0.25](-0.2,-0.2)(6,0.7)
\pscircle[linewidth=0.025,fillstyle=solid,fillcolor=black](0,0){0.12}
\rput(0,0.5){\normalfont$\Deltaff_x$}
\end{pspicture} & \quad {\rm otherwise.}
\end{array}\right.
\ee
\end{Theoreme}
A proof of the theorem is presented in \cref{sec:iso1}.

As already indicated, a comparison of the lattice and conformal integrals of motion involves relative normalisations. 
For the values $c_0\in\{-2,1\}$ of the central charge $c$, the question is thus whether
the lattice results $I_{2p-1}^{\rm latt}$ match the spectra of $\gamma_{2p-1}^{c=c_0}\,\mathbf I_{2p-1}\big|_{c=c_0}$
for some normalisation coefficients $\gamma_{2p-1}^{c=c_0}$. In Section~\ref{sec:integrals}, 
we presented evidence that the $c=-2$ integrals of motion reproduce the lattice results $I_{2p-1}^{\rm latt}$ 
if $\gamma_{2p-1}^{c=-2}=1$ for all $p$.
Thus, to match $I_{2p-1}^{\rm latt}$ for given $p$, the corresponding 
spectrum of $\gamma_{2p-1}^{c=1}\,\mathbf I_{2p-1}\big|_{c=1}$ following from 
Theorem~\ref{thm:structurethm1} must match the similar 
spectrum of $\mathbf I_{2p-1}\big|_{c=-2}$ following from Theorem~\ref{thm:structurethm}.
Since
\be
 {\rm ch}[\hat E^x](q)=\frac{q^{x^2/2}}{\eta(q)}
 \qquad {\rm and}\qquad 
 {\rm ch}[E^v](q) = \frac{q^{v^2/2}}{\eta(q)}
\ee
are equal for $x=v$,
the first integrals of motion $\mathbf I_{1}\big |_{c=1}$ and $\mathbf I_{1}\big|_{c=-2}$ 
have matching spectra (meaning that $\gamma_1^{c=1}=1$),
in accordance with the observation that the effective central charge is the same in the two cases.

However, the spectrum of $\mathbf I_{3}\big|_{c=1}$
does {\em not} match the lattice results. To appreciate this, for $n$ odd, 
it suffices to compare the $\mathbf I_{3}$-eigenvalues on the $\mathbf I_{1}$-eigenspaces of
eigenvalues $-\frac{1}{24}$ and $\frac{11}{24}$.
For both central charges, these eigenspaces are spanned by highest-weight vectors.
For $c=1$, the corresponding conformal weights are $0$ and $\frac{1}{2}$, respectively, while for $c=-2$, 
the weights are $-\frac{1}{8}$ and $\frac{3}{8}$, respectively.
Having matching spectra thus requires that
\be
\gamma_3^{c=1}I_3(\mathsf v_0)\big|_{c=1}=I_3(\mathsf v_{-1/8})\big|_{c=-2},\qquad 
\gamma_3^{c=1}I_3(\mathsf v_{1/2})\big|_{c=1}=I_3(\mathsf v_{3/8})\big|_{c=-2},
\ee
that is,
\be
 \tfrac{3}{320}\,\gamma_3^{c=1}=\tfrac{7}{960},\qquad
 \tfrac{43}{320}\,\gamma_3^{c=1} =\tfrac{127}{960},
\ee
to which there is no solution for $\gamma_3^{c=1}$. 
For $n$ even, it suffices to consider the spectrum of $\mathbf I_{3}$ on the four-dimensional 
$\mathbf I_{1}$-eigenspace of eigenvalue $\frac{13}{12}$. For both central charges, $\mathbf I_{3}$ is diagonalisable 
on that subspace. 
For $c=-2$, all four eigenvalues are $\frac{119}{120}$, whereas for $c=1$, there are two distinct eigenvalues, namely
$\frac{159}{160}$ and $\frac{239}{160}$, each appearing twice.

These results for $c=1$ are thus incompatible with the corresponding lattice result.
We conclude that $\Db(u,\xi_v)$, of which $T^2(\alpha)$ is a specific matrix realisation, is a generating function for the integrals of motion for $c=-2$, but not for $c=1$.

%%%%%%%%%%%%%%%%%%%%
\section{Discussion}
\label{sec:Discussion}
%%%%%%%%%%%%%%%%%%%%

Are dimers on the square lattice described by a $c=1$ or a $c=-2$ conformal field theory? Previously, the evidence supporting the $c=-2$ description was unsatisfactory. This paper is an attempt to remedy this. First, we reinforced the connection between the dimer model and critical dense polymers, by mapping the former to the latter through a series of maps. The dimer partition function was then computed using the loop model and its double-row transfer tangles. Second, we used the description of the transfer matrix 
squared in terms of the Temperley-Lieb algebra $\tl_n(0)$ and its dimer representations to understand the lattice integrability of the dimer model and construct expressions for all the Virasoro modes in the scaling limit. The resulting expressions were recognised as
producing a $c=-2$ realisation of the Virasoro algebra previously encountered in fermionic $bc$ ghost systems.
As modules over the Virasoro algebra, the dimer Fock spaces were found to be of Feigin-Fuchs type. For one parity of the lattice width $n$, the corresponding modules are reducible yet indecomposable, while for the other parity, they are completely reducible modules. Third, we found that the lattice integrals of motion generated by the transfer matrices have eigenvalues that precisely match those of the conformal integrals of motion at $c=-2$. Altogether, we believe that this constitutes compelling evidence for the soundness of the $c=-2$ description of the dimer model.

We also investigated a possible $c=1$ description of the dimer model, compatible with the expression for $\Ldim_0-\frac{c}{24}$ obtained from the dimer transfer matrix. These $c=1$ modes take the form of the sum of two copies of $c=\frac12$ modes, and their action on the dimer Fock space turns out to be completely reducible. This description is unitary. However, the corresponding conformal integrals of motion are found not to reproduce the eigenvalues of their lattice counterparts, as the $c=-2$ ones did.

From the point of view of Lieb's transfer matrix approach, the $c=-2$ description therefore seems to be favoured. This conclusion is built on the premise that the $\frac1n$ expansion of the logarithm of the transfer tangles/matrices precisely yields the conformal integrals of motion, as has been observed to be true for other models \cite{FG04,Ni09,Ni10,Nigro15}. Of course, allowing combinations of the $\mathbf I_{2p-1}$ still gives commuting elements of the universal enveloping algebra of the Virasoro algebra, and it is not clear at this point if this could cure the $c=1$ interpretation of the integrals of motion. 

\cref{tab:c.comparison} draws a comparison between the two conformal descriptions. Their only common feature is $\Ldim_0-\frac c{24}$, which, perhaps surprisingly, can be expressed in terms of two different sets of primary fields. We are not aware of other physical models for which such a duality exists. It thus seems that the central charge depends on the questions one hopes to answer. 
As discussed earlier, Lieb's transfer matrix approach provides a natural framework to study properties of spanning trees, consistent with the $c=-2$ description. On the other hand, the $c=1$ description has been found to be suitable for questions involving monomers or heights. 
To investigate this further, a sensible next step is to search for a transfer matrix formulation of the  dimer model, which leads to a $c=1$ description, and investigate the $\frac1n$ expansion of its eigenvalues. Also, we find that the operator
\be
\mathcal X^{\textrm{\tiny $(n)$}}= \ir \!\sum_{q=1}^{\lfloor\frac{n-1}2\rfloor} ( \Etad_q \Etad_{n-q} + \Eta_q\Eta_{n-q}) - \ir \,\delta_{n,0 \bmod 2}\, (\Etad_{\frac n2}\Eta_{\frac n2}-\tfrac12) \; \in \; {\rm End}(\ctwotimes{n-1})
\ee
commutes with $T^2(\alpha)$ at finite size and becomes $\mathcal X$ in the scaling limit. As opposed to $S^z$, $\mathcal X^{\textrm{\tiny $(n)$}}$ appears not to have a simple interpretation when expressed in terms of $C_j, C_j^\dagger$, or $\sigma^+_j, \sigma^-_j$, so it again seems that the $c=-2$ description is more natural for Lieb's transfer matrix approach. Instead, $\mathcal X^{\textrm{\tiny $(n)$}}$ may have a more natural interpretation in a $c=1$ transfer matrix formalism.

\begin{figure}
\begin{center}
\begin{tabular}{l||c|c}
& $c = -2$ & $c = 1$\\\hline\hline&&\\[-0.3cm]
Primary fields & $\theta^+,\theta^-$ & $\hat \theta^+,\hat \theta^-$ \\[0.2cm]
Conserved lattice operator & $S^z$ & $\mathcal X^{\textrm{\tiny $(n)$}}$ \\[0.2cm]
Virasoro realisation & \begin{tabular}{c}fermionic $bc$ ghosts\\(non-unitary) 
\end{tabular} & \begin{tabular}{c}two non-interacting 
fermions\\(unitary)\end{tabular} \\[0.5cm]
Module structures & \begin{tabular}{c} completely reducible ($n$ odd) \\ reducible yet indecomposable ($n$ even) \end{tabular} & completely reducible
\end{tabular}
\captionof{table}{Comparison of the $c=-2$ and $c=1$ descriptions.}   
\label{tab:c.comparison}
\end{center}
\end{figure}

The results of this paper raise several new questions for future work. In addition to the $c=1$ issues above, one may ask whether the Temperley-Lieb connection and the conformal description of the dimer model extend to other lattices and/or boundary conditions. For instance, it is not clear how these extend if we allow monomers on the boundary, define the model on the hexagonal or the triangular lattice, or fold the lattice on a torus. The torus problem addresses the bulk part of the system and is expected to involve non-chiral representations. An understanding of this will yield valuable new insight into the dimer problem and its description as a conformal field theory. We also note that the parity of the cylinder circumference $M$ is necessarily even for the series of maps from perfect matchings to polymer configurations to be well defined. A detailed analysis of the integrability and conformal properties of the dimer model
for $M$ odd remains an open problem.

\subsection*{Acknowledgments}

AMD was supported by the National Sciences and Engineering Research Council of Canada as a postdoctoral fellow. 
JR was supported by the Australian Research Council under the Future Fellowship scheme, project number 
FT100100774. 
PR is Senior Research Associate of the Belgian Fonds National de la Recherche Scientifique (FNRS). AMD and PR acknowledge the support of the Belgian Interuniversity Attraction Poles Program P7/18 through the network DYGEST (Dynamics, Geometry and Statistical Physics). The authors thank Nicolas Allegra, Christian Hagendorf, Paul Pearce and David Ridout for useful discussions.

\bigskip

\bigskip

\appendix

%%%%%%%%%%%%%%%%%%%%%%%%%%%%%%
%
\section{Feigin-Fuchs modules}
\label{sec:FFs}
%
%%%%%%%%%%%%%%%%%%%%%%%%%%%%%%

The Feigin-Fuchs modules are a special class of modules over the Virasoro algebra.
Their structure was first described by Feigin and Fuchs \cite{FF82}, see also
the book by Iohara and Koga \cite{IK11} and the summary by Ridout and Wood \cite[Appendix A]{RW14}.

One realisation of the Feigin-Fuchs modules is built on the Heisenberg algebra $\mathsf H$. 
This algebra is spanned by modes $a_k, k \in \mathbb Z$,
and a central element $\mathbf 1$ satisfying the relations
\be
[a_k, a_\ell] = k \, \delta_{k+\ell,0} \,\mathbf 1, \qquad [a_k, \mathbf 1] = 0.
\ee
A one-parameter family of highest-weight representations $\mathcal F_\lambda$ over $\mathsf H$, with $\lambda \in \mathbb C$, is constructed from the action of the universal enveloping algebra of $\mathsf H$ on the highest-weight state $|\lambda \rangle$, with the rules
\be
a_k |\lambda \rangle = 0 \qquad (k >0), \qquad a_0 |\lambda \rangle = \lambda\, |\lambda \rangle, \qquad \mathbf 1 |\lambda \rangle = |\lambda \rangle.
\ee
A basis of states is given by
\be
\big\{ a_{-k_1}^{n_1} a_{-k_2}^{n_2} \cdots a_{-k_r}^{n_r} |\lambda \rangle \ ; \ r \in \mathbb Z_{\ge 0}, \ k_1> k_2 > \dots > k_r >0, \ n_1, n_2, \dots, n_r \in \mathbb Z_{>0}\big\}.
\label{eq:Hbasis}
\ee
It is not hard to see that $\mathcal F_\lambda$ is irreducible as a module over $\mathsf H$. Indeed, for each element $| \mathsf w\rangle$ of the basis \eqref{eq:Hbasis}, one can construct the word
\be 
m_{\mathsf w} = a_{k_r}^{n_r} \cdots a_{k_2}^{n_2}  a_{k_1}^{n_1}\qquad {\rm such\ that} \qquad
m_{\mathsf w} |{\mathsf w} ' \rangle = \kappa_{\mathsf w}\, \delta_{\mathsf w,\mathsf w'}\, |\lambda \rangle, \qquad \kappa_{\mathsf w} = \prod_{i=1}^r n_i! \,k_i^{n_i} ,
\label{eq:mw}
\ee
for any $|\mathsf w ' \rangle$ of the same basis. One can then get from any state $|\mathsf v\rangle = \sum_{\mathsf w} \alpha_{\mathsf w} |\mathsf w \rangle$ to any other state $|\mathsf v'\rangle$ by 
(i) selecting a $|\mathsf w\rangle$ for which $\alpha_{\mathsf w} \neq 0$, (ii) applying $\kappa_{\mathsf w}^{-1}m_{\mathsf w}$ on $|\mathsf v\rangle$ to arrive at $|\lambda \rangle$, and (iii) descending from $|\lambda\rangle$ to $|\mathsf v' \rangle$.

The universal enveloping algebra of $\mathsf H$ carries a one-parameter realisation of the Virasoro algebra,
\be
\LFF_k = \frac12 \sum_{q \in \mathbb Z} a_q a_{k-q} - \frac12 (k+1)\, Q\, a_k \quad \ (k \neq 0), \qquad \LFF_0 = \frac12 a_0^2 + \sum_{q=1}^\infty a_{-q}a_q - \frac 12 Q\, a_0, \quad \ Q \in \mathbb C,
\ee
for which the central charge is $c = 1-3 Q^2$. As a module over the Virasoro algebra, the Fock space $\mathcal F_\lambda$ is referred to as a Feigin-Fuchs module. Its highest-weight state $|\lambda\rangle$ satisfies $\LFF_0|\lambda\rangle=\Delta_\lambda|\lambda\rangle$ with
\be
\DeltaFF_\lambda = \frac 12 \lambda\,(\lambda - Q).
\ee
The character of $\mathcal F_\lambda$ is the same as that of the Verma module $\mathcal V_{\DeltaFF_\lambda}$ of highest weight $\Delta_\lambda$, 
\be  
{\rm ch}[\mathcal F_{\lambda}](q) = \frac{q^{\DeltaFF_\lambda-\frac c{24}}}{\prod_{j=1}^\infty (1-q^{j})}.
\ee
This implies that $\mathcal F_{\lambda}$ and $\mathcal V_{\DeltaFF_\lambda}$ have identical composition factors, but not necessarily the same structure. Because $\DeltaFF_\lambda = \DeltaFF_{Q-\lambda}$, this also applies to $\mathcal F_{Q-\lambda}$. 
However, ${\rm ch}[\mathcal F_{\lambda}](q) = {\rm ch}[\mathcal F_{Q-\lambda}](q)$ does not imply that $\mathcal F_{\lambda}$ and $\mathcal F_{Q-\lambda}$ are isomorphic. Setting
\be
Q = \sqrt{\frac {2}{pp'}}(p'-p),\qquad \lambda = \lambda_{r,s} = - \sqrt{\frac {p'}{2p}}\, (r-1) +  \sqrt{\frac {p\phantom{'}}{2p'}} \,(s-1)
\ee
yields the central charge and conformal dimensions
\be
c = 1-6 \frac{(p'-p)^2}{p p'}, \qquad \DeltaFF = \Delta_{r,s} = \frac{(p'r - ps)^2 - (p-p')^2}{4 pp'}.
\ee
For 
\be 
p,p' \in \mathbb Z_{>0}, \qquad p\le p', \qquad {\rm gcd}(p,p') = 1,
\ee 
the Feigin-Fuchs module $\mathcal F_\lambda$ is irreducible unless $\Delta_\lambda$ is a conformal dimension in the Kac table. The special values correspond to $\lambda = \lambda_{r,s}$ (and thus $\Delta_\lambda = \Delta_{r,s}$) with $r$ and $s$ integers, and in this case $\mathcal F_{r,s} \equiv \mathcal F_{\lambda_{r,s}}$ is a reducible Virasoro module. Because $\lambda_{r+p,s+p'} = \lambda_{r,s}$ and, hence, $\mathcal F_{r+p,s+p'} \simeq \mathcal F_{r,s}$, we can restrict $r$ and $s$ to be positive integers. 

The module structure of $\mathcal F_{r,s}$ depends on whether $(r,s)$ is (i) a {\em corner entry} of the Kac table, meaning that $r$ is a multiple of $p$ and $s$ is a multiple of $p'$, (ii) a {\em boundary entry} of the Kac table, meaning that either $r$ is a multiple of $p$ or $s$ is a multiple of $p'$, but not both, or (iii) an {\em interior entry} of the Kac table, meaning that $r$ is not a multiple of $p$ and $s$ is not a multiple of $p'$. If $p=1$, there are no interior entries, and if $p= p'=1$, there are no boundary entries either.

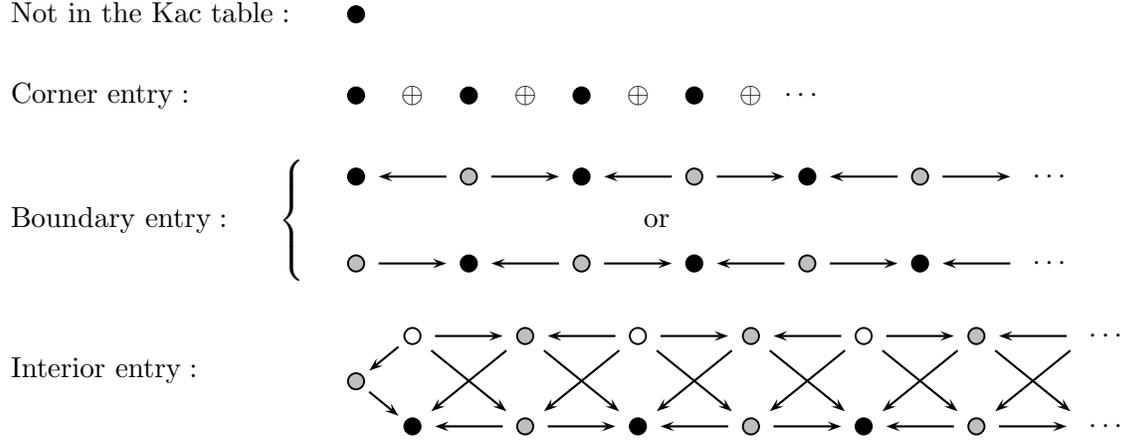
\begin{figure}[h!]
\begin{center}
$
\begin{array}{lll}
{\rm Not\  in\ the\ Kac\ table:} & &
\begin{array}{l}
\begin{pspicture}[shift=-0.1](0,-0.2)(6,0.2)
\multiput(0,0)(1.5,0){1}{\pscircle[linewidth=0.025,fillstyle=solid,fillcolor=black](0,0){0.12}}
\end{pspicture}
\end{array}
\\[0.6cm]
{\rm Corner\ entry:} &&  
\begin{array}{l}
\begin{pspicture}[shift=-0.1](0,-0.2)(6,0.2)
\multiput(0,0)(1.5,0){4}{\pscircle[linewidth=0.025,fillstyle=solid,fillcolor=black](0,0){0.12}}
\multiput(0.60,-0.1)(1.5,0){4}{$\oplus$}
\rput(5.95,0){$\cdots$}
\end{pspicture}
\end{array}
\\[0.6cm]
{\rm Boundary\ entry:} & \hspace{-0.5cm}\left \{ \begin{array}{l}\begin{pspicture}[shift=-0.1](0,-0.2)(0.1,0.2) \end{pspicture} \\[0.6cm]\begin{pspicture}[shift=-0.1](0,-0.2)(0.1,0.2)\end{pspicture}\end{array}\right. \hspace{-0.8cm}
&
\begin{array}{l}
\begin{pspicture}[shift=-0.1](0,-0.2)(9.5,0.2)
\multiput(0,0)(3,0){3}{\pscircle[linewidth=0.025,fillstyle=solid,fillcolor=black](0,0){0.12}\psline[linewidth=0.8pt,arrowscale=1.2,arrowinset=0.4]{<-}(0.3,0)(1.2,0)}
\multiput(1.5,0)(3,0){3}{\pscircle[linewidth=0.025,fillstyle=solid,fillcolor=mygray](0,0){0.12}\psline[linewidth=0.8pt,arrowscale=1.2,arrowinset=0.4]{->}(0.3,0)(1.2,0)}
\rput(9.25,0){$\cdots$}
\end{pspicture}
\\
\begin{pspicture}[shift=-0.1](0,-0.02)(9.5,0.02) \rput(4,0.07){or}\end{pspicture}\\[0.2cm]
\begin{pspicture}[shift=-0.1](0,-0.2)(9.5,0.2)
\multiput(0,0)(3,0){3}{\pscircle[linewidth=0.025,fillstyle=solid,fillcolor=mygray](0,0){0.12}\psline[linewidth=0.8pt,arrowscale=1.2,arrowinset=0.4]{->}(0.3,0)(1.2,0)}
\multiput(1.5,0)(3,0){3}{\pscircle[linewidth=0.025,fillstyle=solid,fillcolor=black](0,0){0.12}\psline[linewidth=0.8pt,arrowscale=1.2,arrowinset=0.4]{<-}(0.3,0)(1.2,0)}
\rput(9.25,0){$\cdots$}
\end{pspicture}
\end{array}
\\[0.8cm]
{\rm Interior\ entry:} & &
\begin{array}{l}
\begin{pspicture}[shift=-0.85](-0.75,-0.2)(9.5,1.7)
\pscircle[linewidth=0.025,fillstyle=solid,fillcolor=mygray](-0.75,0.6){0.12}
\multiput(0,0)(3,0){3}{\pscircle[linewidth=0.025,fillstyle=solid,fillcolor=black](0,0){0.12}\psline[linewidth=0.8pt,arrowscale=1.2,arrowinset=0.4]{<-}(0.3,0)(1.2,0)}
\multiput(1.5,0)(3,0){3}{\pscircle[linewidth=0.025,fillstyle=solid,fillcolor=mygray](0,0){0.12}\psline[linewidth=0.8pt,arrowscale=1.2,arrowinset=0.4]{->}(0.3,0)(1.2,0)}
\multiput(0,1.2)(3,0){3}{\pscircle[linewidth=0.025,fillstyle=solid,fillcolor=white](0,0){0.12}\psline[linewidth=0.8pt,arrowscale=1.2,arrowinset=0.4]{->}(0.3,0)(1.2,0)}
\multiput(1.5,1.2)(3,0){3}{\pscircle[linewidth=0.025,fillstyle=solid,fillcolor=mygray](0,0){0.12}\psline[linewidth=0.8pt,arrowscale=1.2,arrowinset=0.4]{<-}(0.3,0)(1.2,0)}
\multiput(0,0)(1.5,0){6}{\psline[linewidth=0.8pt,arrowscale=1.2,arrowinset=0.4]{<-}(0.25,0.2)(1.25,1.01)\psline[linewidth=0.8pt,arrowscale=1.2,arrowinset=0.4]{<-}(1.25,0.2)(0.25,1.01)}
\psline[linewidth=0.8pt,arrowscale=1.2,arrowinset=0.4]{<-}(-0.185,0.14)(-0.575,0.46)
\psline[linewidth=0.8pt,arrowscale=1.2,arrowinset=0.4]{->}(-0.185,1.06)(-0.575,0.74)
\rput(9.25,0){$\cdots$}\rput(9.25,1.2){$\cdots$}
\end{pspicture}
\end{array}
\end{array}
$
\caption{Module structures of Feigin-Fuchs modules. The conformal dimensions of the composition factors of $\mathcal F_\lambda$ are the same as those of the corresponding Verma module $\mathcal V_{\Delta_\lambda}$, and grow towards the right.
} 
\label{fig:struc}
\end{center}
\end{figure}

The different module structures are displayed in \cref{fig:struc}. There, the composition factors in the {\it socle} --- the maximal completely reducible submodule --- are represented by black circles. Quotienting by the socle gives a new module. The {\it second socle layer} is the socle of this quotient module. Its composition factors are represented by gray circles in \cref{fig:struc}. Repeating the construction defines the {\it third socle layer}, whose composition factors are represented by white circles. Feigin-Fuchs modules never have more than three socle layers. The arrows in \cref{fig:struc} describe the action of the Virasoro algebra. If an arrow points, say, from a composition factor 
$\begin{pspicture}[shift=-0.1](-0.2,-0.2)(0.2,0.2)\pscircle[linewidth=0.025,fillstyle=solid,fillcolor=mygray](0,0){0.12}\end{pspicture}$
to a factor 
$\begin{pspicture}[shift=-0.1](-0.2,-0.2)(0.2,0.2)\pscircle[linewidth=0.025,fillstyle=solid,fillcolor=black](0,0){0.12}\end{pspicture}$, vectors in 
$\begin{pspicture}[shift=-0.1](-0.2,-0.2)(0.2,0.2)\pscircle[linewidth=0.025,fillstyle=solid,fillcolor=black](0,0){0.12}\end{pspicture}$
can be reached from vectors in 
$\begin{pspicture}[shift=-0.1](-0.2,-0.2)(0.2,0.2)\pscircle[linewidth=0.025,fillstyle=solid,fillcolor=mygray](0,0){0.12}\end{pspicture}$ by the action of the Virasoro algebra, whereas no vector in 
$\begin{pspicture}[shift=-0.1](-0.2,-0.2)(0.2,0.2)\pscircle[linewidth=0.025,fillstyle=solid,fillcolor=mygray](0,0){0.12}\end{pspicture}$
 can be reached from 
 $\begin{pspicture}[shift=-0.1](-0.2,-0.2)(0.2,0.2)\pscircle[linewidth=0.025,fillstyle=solid,fillcolor=black](0,0){0.12}\end{pspicture}$.

For corner entries, $\mathcal F_{r,s}$ is completely reducible and is the direct sum of the composition factors of $\mathcal V_{\Delta_{r,s}}$. For the other entries of the Kac table, $\mathcal F_{r,s}$ is indecomposable. For boundary entries, $\mathcal F_{r,s}$ takes the form of a chain of composition factors that alternate between the socle and the second socle layer. As drawn in \cref{fig:struc}, there are two possibilities, and $\mathcal F_{r,s}$ corresponds to the one where the composition factor of conformal weight $\Delta_{r,s}+rs$ is in the second socle layer. For interior entries, the structure is that of a braid, with some arrows pointing in the opposite direction compared to the similar arrows in the corresponding braided Verma module. The composition factor with weight $\Delta_{r,s} + rs$ belongs to the third socle layer. Only the top three cases of \cref{fig:struc} will come up in \cref{sec:iso,sec:iso1}.

%%%%%%%%%%%%%%%%%%%%%%%%%%%%%%
%
\section{Module structures at $\boldsymbol{c = -2}$}
\label{sec:iso}
%
%%%%%%%%%%%%%%%%%%%%%%%%%%%%%%

The main objective of this appendix is to give a proof of \cref{thm:structurethm}. 
Our proof builds a family of isomorphisms between the Virasoro dimer modules $E^v$ and Feigin-Fuchs modules $\mathcal F_\lambda$ at $c=-2$, for which
\be
(p,p') = (1,2), \qquad Q=1, \qquad \lambda = \lambda_{r,s} = \frac{s-2r+1}2, \qquad \Delta_{r,s} = \frac{(2r-s)^2-1}8.\label{eq:c=-2stuff}
\ee
The Kac table for $c=-2$ consists of corner entries (for $s$ even) and boundary entries (for $s$ odd). 
The module structure of $\mathcal F_{r,s}$ is discussed in \cref{sec:FFs}. For $\mathcal F_{1,s}$ in particular, for $s$ even, this structure is
\be
\mathcal F_{1,s} \simeq \mathcal F_{1,4-s}\;:\; \qquad
\begin{pspicture}[shift=-0.25](0,-0.2)(6.4,0.7)
\multiput(0,0)(1.5,0){4}{\pscircle[linewidth=0.025,fillstyle=solid,fillcolor=black](0,0){0.12}}
\multiput(0.60,-0.1)(1.5,0){4}{$\oplus$}
\rput(0,0.5){$\Delta_{1,s}$}
\rput(1.5,0.5){$\Delta_{1,s+4}$}
\rput(3,0.5){$\Delta_{1,s+8}$}
\rput(4.5,0.5){$\Delta_{1,s+12}$}
\rput(5.95,0){$\cdots$}
\end{pspicture}
\quad (s\ge 2,\, {\rm even}),
\label{eq:Feven}
\ee
whereas for $s$ odd, 
\be
\mathcal F_{1,s} \;:\; \left\{\quad\;\;
\begin{array}{ll}
\begin{pspicture}[shift=-0.3](0,-0.2)(9.5,0.7)
\multiput(0,0)(3,0){3}{\pscircle[linewidth=0.025,fillstyle=solid,fillcolor=black](0,0){0.12}\psline[linewidth=0.8pt,arrowscale=1.2,arrowinset=0.4]{<-}(0.3,0)(1.2,0)}
\multiput(1.5,0)(3,0){3}{\pscircle[linewidth=0.025,fillstyle=solid,fillcolor=mygray](0,0){0.12}\psline[linewidth=0.8pt,arrowscale=1.2,arrowinset=0.4]{->}(0.3,0)(1.2,0)}
\rput(0,0.5){$\Delta_{1,4-s}$}
\rput(1.5,0.5){$\Delta_{1,6-s}$}
\rput(3,0.5){$\Delta_{1,8-s}$}
\rput(4.5,0.5){$\Delta_{1,10-s}$}
\rput(6.0,0.5){$\Delta_{1,12-s}$}
\rput(7.5,0.5){$\Delta_{1,14-s}$}
\rput(9.25,0){$\cdots$}
\end{pspicture}
\qquad &(s \le 1,\, {\rm odd}),\\[0.6cm]
\begin{pspicture}[shift=-0.3](0,-0.2)(9.5,0.7)
\multiput(0,0)(3,0){3}{\pscircle[linewidth=0.025,fillstyle=solid,fillcolor=mygray](0,0){0.12}\psline[linewidth=0.8pt,arrowscale=1.2,arrowinset=0.4]{->}(0.3,0)(1.2,0)}
\multiput(1.5,0)(3,0){3}{\pscircle[linewidth=0.025,fillstyle=solid,fillcolor=black](0,0){0.12}\psline[linewidth=0.8pt,arrowscale=1.2,arrowinset=0.4]{<-}(0.3,0)(1.2,0)}
\rput(0,0.5){$\Delta_{1,s}$}
\rput(1.5,0.5){$\Delta_{1,s+2}$}
\rput(3,0.5){$\Delta_{1,s+4}$}
\rput(4.5,0.5){$\Delta_{1,s+6}$}
\rput(6.0,0.5){$\Delta_{1,s+8}$}
\rput(7.5,0.5){$\Delta_{1,s+10}$}
\rput(9.25,0){$\cdots$}
\end{pspicture}
\qquad &(s \ge 3,\, {\rm odd}).
\end{array}\right.
\label{eq:Fodd}
\ee
We note that the periodicity property $\mathcal F_{r,s} \simeq \mathcal F_{r+1,s+2}$ yields 
$\mathcal F_{1,s} \simeq \mathcal F_{(4-s)/2,2}$ for $s$ even and $\mathcal F_{1,s} \simeq \mathcal F_{(3-s)/2,1}$ for $s$ odd.

We now relate these modules to the ones found in \cref{sec:higher} from the dimer model. The realisation \eqref{eq:finalLk} of the Virasoro modes depends on the parity of $n$. We thus define the (parity-dependent) map
\be
\mathfrak g\; : \; \mathsf H \ \rightarrow \ {\rm End\big(\ctwotimes \infty\big)}
\ee
whose action on the generators is   
\be
\mathfrak g(\mathbf 1) = \mathbb I, 
\qquad \ \ 
\mathfrak g(a_k) = \sideset{}{'}\sum_{q \ge 0} \theta^-_{-q}\theta^+_{q+k} - \sideset{}{'}\sum_{q>0} \theta^+_{-q+k}\theta^-_q + \tfrac12 \delta_{k,0}\, 
\delta_{n, 1 \bmod 2}.
\ee
Here, we recall that primed sums indicate that $q$ runs over integers or half-integers for $n$ even or odd, respectively. Comparing with \eqref{eq:Szlim}, we see that
\be
\mathfrak g(a_0) = \tfrac 12 - S^z.
\ee
One also verifies the commutation relation
\be
[\mathfrak g(a_k),\mathfrak g(a_\ell)] = k\, \delta_{k+\ell, 0}, 
\ee
thus proving that $\mathfrak g$ is an $\mathsf H$-homomorphism. It is then straightforward to show that 
\be
\mathfrak g(a_k) |\Omega_v \rangle = 0, \qquad (k>0), \qquad \mathfrak g(a_0) |\Omega_v \rangle = (\tfrac12-v)|\Omega_v \rangle,
\ee
for all $v$. As a consequence, the map 
\be
\mathfrak h\; :\; \mathcal F_\lambda \rightarrow  E^v, \qquad \lambda = \tfrac 12 -v,
\ee
defined as
\be
\mathfrak h\big(a_{k_1}^{n_1}a_{k_2}^{n_2} \dots a_{k_\ell}^{n_\ell} |\lambda\rangle\big) = \mathfrak g(a_{k_1})^{n_1}\mathfrak g(a_{k_2})^{n_2} \dots \mathfrak g(a_{k_\ell})^{n_\ell} |\Omega_v\rangle,
\label{eq:hdef}
\ee
is a homomorphism of $\mathsf H$-modules. 

Remarkably, the Virasoro and Heisenberg modes satisfy the commutation relations
\be
[\LFF_k,a_q] = -q\, a_{k+q} - \tfrac{q(q-1)}2 \delta_{k+q,0}\,\mathbf{1}, \qquad  [\Ldim_k, \mathfrak{g}(a_q)] = -q\, \mathfrak g(a_{k+q}) - \tfrac{q(q-1)}2 \delta_{k+q,0}.
\label{eq:Lkaqcomms}
\ee 
The similarity between these commutators is crucial for the proof of the next two propositions.
\begin{Proposition}
The map $\mathfrak h$ is bijective.
\label{sec:hbij}
\end{Proposition}
\proof 
Both $\mathcal F_\lambda$ and $E^v$ are infinite-dimensional vector spaces, graded by the eigenvalues of $\LFF_0$ and $\Ldim_0$, respectively. For $\mathcal F_\lambda$, a state of the basis \eqref{eq:Hbasis} has grade $\sum_{i=1}^r k_i n_i$. At grade $\ell$, the number of states is $p(\ell)$, the number of integer partitions of $\ell$. For $E^v$, the number of states at grade $\ell$ is also $p(\ell)$, as is obvious from its character \eqref{eq:Vermachar}. 

The grade of a basis state $|\mathsf w\rangle$ of $\mathcal F_\lambda$ and that of $\mathfrak h(|\mathsf w\rangle)\in E^v$ can be calculated using only the commutators \eqref{eq:Lkaqcomms} at $k = 0$. Because the right-hand sides have the same form, the grades of $|\mathsf w\rangle$ and $\mathfrak h(|\mathsf w\rangle)$ are equal, so $\mathfrak h$ is grade-preserving and thus maps each subspace of $\mathcal F_\lambda$ of grade $\ell$ to the corresponding subspace of $E^v$, also of grade $\ell$. Both subspaces are finite-dimensional, and they have equal dimensions. We now show that $\mathfrak h$ is bijective at each grade.

At a fixed grade $\ell$, the domain and target spaces have equal dimensions, so it suffices to show that $\mathfrak h$ is injective. Suppose there exists a state $|\mathsf v\rangle = \sum_\mathsf w \alpha_{\mathsf w} |\mathsf w \rangle $ in $\mathcal F_\lambda$ such that $\mathfrak h(| \mathsf v \rangle)=0$, with the sum on $\mathsf w$ running over the elements of the basis \eqref{eq:Hbasis} with $\sum_{i=1}^r k_i n_i= \ell$. Then, from \eqref{eq:mw},
\be
\kappa_{\mathsf w}^{-1} \mathfrak g(m_{\mathsf w}) \mathfrak h(|\mathsf v \rangle) = 
\kappa_{\mathsf w}^{-1} \mathfrak h(m_{\mathsf w} |\mathsf v \rangle) = 
\alpha_{\mathsf w} |\Omega_v\rangle = 0 
\quad \rightarrow \quad \alpha_{\mathsf w} = 0,
\ee
so the only possibility is that $|\mathsf v \rangle = 0$. 
\eproof

It follows that $\mathfrak h$ is, in fact, an isomorphism of $\mathsf H$-modules. 
The next proposition shows that this isomorphism property also holds at the level of the Virasoro modules.

\begin{Proposition}
The map $\mathfrak h$ is an isomorphism of Virasoro modules.
\label{sec:hiso}
\end{Proposition}
\proof
By extending $\mathfrak g$ to the universal enveloping algebra of $\mathsf H$ in the obvious way, we see that the operators $\mathfrak g(\LFF_k)$ provide a realisation of the Virasoro algebra with $c=-2$  in the Fock space $\ctwotimes \infty$. We also note, from \eqref{eq:Lkaqcomms}, that $\mathfrak g(\LFF_k)$ all commute with $\mathfrak g(a_0)$ and therefore leave the subspaces $E^v$ invariant. 
In fact,
\be
\big(\mathfrak g(\LFF_k) - \Ldim_k \big) E^v = 0
\label{eq:gLmLe0}
\ee
holds for all $v$ and all $k$. 
The proof of \eqref{eq:gLmLe0}
uses, as a basis for $E^v$, the elements $\mathfrak h(|\mathsf w\rangle)$ with $|\mathsf w\rangle$ of the form \eqref{eq:Hbasis}, and is inductive on the grade, with starting point $|\Omega_v\rangle$ at grade zero. For $k>0$, $\mathfrak g(\LFF_k)|\Omega_v\rangle = 0$ and $\Ldim_k|\Omega_v\rangle = 0$. For $k=0$,
\be
\mathfrak g(\LFF_0)|\Omega_v\rangle = \Ldim_0  |\Omega_v\rangle = \Delta_v |\Omega_v \rangle
\ee
with $\Delta_v$ defined in \eqref{eq:confdim}. For $k<0$, we have
\be
\mathfrak g(\LFF_{-k}) |\Omega_v \rangle = \sum_{\mathsf w} \alpha_{\mathsf w} \,\mathfrak h(|\mathsf w \rangle), \qquad \Ldim_{-k} |\Omega_v \rangle = \sum_{\mathsf w} \beta_{\mathsf w}\, \mathfrak h(|\mathsf w \rangle), 
\ee
for some $\alpha_{\mathsf w}, \beta_{\mathsf w} \in \mathbb C$. These coefficients can be calculated using
\be
\alpha_{\mathsf w} |\Omega_v \rangle = \kappa_w^{-1}\Big[\mathfrak g(m_{\mathsf w}), \mathfrak g(\LFF_{-k})\Big] |\Omega_v \rangle, \qquad \beta_{\mathsf w} |\Omega_v \rangle = \kappa_w^{-1}\Big[\mathfrak g(m_{\mathsf w}), \Ldim_{-k}\Big] |\Omega_v \rangle.
\ee
Because $[\mathfrak g(\LFF_{-k}), \mathfrak g(a_q)] = [\Ldim_{-k}, \mathfrak g(a_q)]$, we also have that $[\mathfrak g(\LFF_{-k}), \mathfrak g(m_{\mathsf w})] = [\Ldim_{-k}, \mathfrak g(m_{\mathsf w})]$ for any word $m_{\mathsf w}$, and therefore, $\alpha_{\mathsf w} = \beta_{\mathsf w}$, completing the proof at grade $\ell = 0$.

For the higher grades, any basis state $|\mathsf s \rangle$ of $E^v$ in the basis \eqref{eq:hdef} takes the form of $\mathfrak g(a_{q})|\mathsf s' \rangle$ for some $q<0$, implying that the respective grades $\ell$ and $\ell'$ of $|\mathsf s\rangle$ and $|\mathsf s'\rangle$ satisfy $\ell > \ell'$. Then,
\be
\big(\mathfrak g(\LFF_{k})-\Ldim_k\big) |\mathsf s \rangle = \big(\mathfrak g(\LFF_{k})-\Ldim_k\big) \mathfrak g(a_{q}) |\mathsf  s' \rangle = 
\big[\big(\mathfrak g(\LFF_{k})-\Ldim_k\big),\mathfrak g(a_{q})\big] |\mathsf s' \rangle +  \mathfrak g(a_{q})\big(\mathfrak g(\LFF_{k})-\Ldim_k\big) |\mathsf  s' \rangle =0 
\ee
where the inductive assumption was used at the last equality to set the second term to zero.
\eproof

It follows from this proposition that $E^v \simeq \mathcal F_\lambda$ for $\lambda = \frac12 -v$. Comparing with \eqref{eq:c=-2stuff}, we find that the integer and half-integer values of $v$ correspond to $\lambda = \lambda_{1,s}$ with $s = 2(1-v)$ odd and even, respectively. The module structure of $E^v$ can therefore be directly read off from \eqref{eq:Feven} and \eqref{eq:Fodd}, thus completing the proof of \cref{thm:structurethm}.

The structure results for the dimer Fock spaces in the $c=-2$ description have interesting repercussions for critical dense polymers. The structures of the Virasoro modules of the scaling limit of the modules $\stan_n^d$ were conjectured in \cite{R11} for logarithmic minimal models $\mathcal{LM}(1,p')$, and more generally in \cite{MDRR15} for $\mathcal{LM}(p,p')$. Critical dense polymers corresponds to $\mathcal{LM}(1,2)$, and in this case from \cref{thm:structurethm}, the limiting structures of $\stan_n^d$ can be proved explicitly to be given by
\be
\stan_n^d \quad \xrightarrow{n \rightarrow \infty} \quad 
\left\{
\begin{array}{ll}
\begin{pspicture}[shift=-0.25](-0.6,-0.2)(0.2,0.7)
\pscircle[linewidth=0.025,fillstyle=solid,fillcolor=black](0,0){0.12}
\rput(0,0.5){\small $\Delta_{\frac{d-1}2}$}
\end{pspicture}& \qquad   n {\rm \ odd, \quad \ } n {\rm \ even\ with \ } d= 0,\\[0.4cm]
\begin{pspicture}[shift=-0.3](-0.6,-0.2)(1.7,0.7)
\pscircle[linewidth=0.025,fillstyle=solid,fillcolor=mygray](0,0){0.12}\psline[linewidth=0.8pt,arrowscale=1.2,arrowinset=0.4]{->}(0.3,0)(1.2,0)
\multiput(1.5,0)(3,0){1}{\pscircle[linewidth=0.025,fillstyle=solid,fillcolor=black](0,0){0.12}}
\rput(0,0.5){\small $\Delta_{\frac{d-1}2}$}
\rput(1.5,0.5){\small $\Delta_{\frac{d+1}2}$}
\end{pspicture}&
\qquad n {\rm \ even \ with \ } d>0.
\end{array}
\right.
\label{eq:stanstructures}
\ee

To prove this, we first recall that the limiting character of $\stan_n^d$ is written in terms of irreducible ones in \eqref{eq:char}. For $n$ odd, as well as for $n$ even with $d=0$, the limiting structure of $\stan_n^d$ is completely determined by its character and is irreducible.

For $n$ even and $d>0$, the character is the sum of two irreducible ones, so the Virasoro module has two composition factors, and its structure is not completely fixed by the character. Recalling that $\beta$ is the fugacity of loops in $\tl_n(\beta)$, an injective intertwiner $f_{n,d}$ was introduced in \cite{MDRRSA15}, for all $\beta \in \mathbb R$, and shown to intertwine $\rho_d(\mathcal H)$ and a spin-chain Hamiltonian $\mathbb H \in {\rm End}(\ctwotimes{n-1})$. At $\beta = 0$, $\mathbb H$ becomes the dimer Hamiltonian $H$, and one can show that $f_{n,d}$ actually acts as an intertwiner for each of the $\tl_n(0)$ generators:
\be
f_{n,d} \cdot \rho_d(e_j) = \tau_v(e_j)\cdot f_{n,d}, \qquad (v = \tfrac{1-d}2, \ j = 1, \dots, n-1),
\ee 
where $\tau_v$ denotes the restricted action of $\tau$ on the subspace $E_{n-1}^v$.

The injectivity of $f_{n,d}$ was proven in \cite{MDRRSA15} and implies that, as a $\tl_n(0)$-module, $\stan_n^d$ is a submodule of $E_{n-1}^{v}$, $v = \frac{1-d}2$. This is true for all $n$ and all $d = n \bmod 2$. Because the Virasoro mode approximations are constructed from $\tl_n(0)$ tangles, the same inclusion, $\stan_{n,d} \subseteq E_{n-1}^{v}$ for $v = \frac{1-d}2$, also applies to the corresponding Virasoro modules in the scaling limit. Comparing the two characters, we find that the structure of $E^v$ for $n$ even can be read off directly from \eqref{eq:struceven} and is indeed highest weight, as in \eqref{eq:stanstructures}.

%%%%%%%%%%%%%%%%%%%%%%%%%%%%%%
%
\section{Module structures at $\boldsymbol{c = 1}$}
\label{sec:iso1}
%
%%%%%%%%%%%%%%%%%%%%%%%%%%%%%%

This appendix provides a proof of \cref{thm:structurethm1}, building on the known structures of Feigin-Fuchs modules at $c=1$, see \cref{sec:FFs}. 
The corresponding data is
\be
(p,p') = (1,1), \qquad Q = 0,
\qquad \lambda = \lambda_{r,s} = \frac{s-r}{\sqrt{2}}, \qquad \Delta_{r,s} = \frac{(r-s)^2}4.
\label{eq:c=1stuff}
\ee
For $c=1$, all the entries of the Kac table are corner entries, so $\mathcal F_{r,s}$ is completely reducible for all $r,s \in \mathbb Z$. In particular, we have
\begin{alignat}{2}
\mathcal F_{1,1} \;&:\; \qquad
\begin{pspicture}[shift=-0.25](0,-0.2)(6.4,0.7)
\multiput(0,0)(1.5,0){4}{\pscircle[linewidth=0.025,fillstyle=solid,fillcolor=black](0,0){0.12}}
\multiput(0.60,-0.1)(1.5,0){4}{$\oplus$}
\rput(0,0.5){$\Delta_{1,1}$}
\rput(1.5,0.5){$\Delta_{1,3}$}
\rput(3,0.5){$\Delta_{1,5}$}
\rput(4.5,0.5){$\Delta_{1,7}$}
\rput(5.95,0){$\cdots$}
\end{pspicture}\label{eq:F11}
\\[0.3cm]
\mathcal F_{\lambda} \;&:\; \qquad 
\begin{pspicture}[shift=-0.25](0,-0.2)(0.4,0.7)
\pscircle[linewidth=0.025,fillstyle=solid,fillcolor=black](0,0){0.12}
\rput(0,0.5){$\Delta_{\lambda}$}
\end{pspicture}
\qquad \quad (\Delta_\lambda {\rm \ not \ in \ Kac \ table}).\label{eq:Firred}
\end{alignat}

As discussed in \cref{sec:higherc=1}, the dimer Fock space $\ctwotimes \infty$ splits as the direct sum of subspaces $\hat E^x$. The corresponding bases are given by states of the form \eqref{eq:Xbasis} with $j-i = x$ for $n$ odd and $j-i = x+\frac 12$ for $n$ even. Using \eqref{eq:Lk2ftheta} at $k=0$ and the form of the basis, we find that the character over $\hat E^x$ is a Verma module character,
\be   
{\rm ch}[\hat E^x](q) = \frac{q^{\Delta^{\!\rm{2f}}_x-\frac c{24}}}{\prod_{j=1}^\infty (1-q^{j})},
\ee
where $c=1$ and $\Deltaff_x$ is defined in \eqref{eq:deltax}. Crucially, for $x \in \frac12 \mathbb Z \setminus\{0\}$, there are no integers $r$ and $s$ such that $\Deltaff_x = \frac {x^2}2$ equals $\Delta_{r,s} = \frac{(r-s)^2}4$. In these cases, $\Deltaff_x$ {\it is not} in the $c=1$ Kac table, and the character of $\hat E^x$ is irreducible, thus proving the part of \cref{thm:structurethm1} pertaining to $x \neq 0$.

What remains is to prove the case $x = 0$, for which $n$ is odd and the fermionic labels take half-integer values. We proceed by showing that $\hat E^0 \simeq \mathcal F_0$ by explicitly constructing the isomorphism. In fact, the proof presented below works in full generality, showing
that $\hat E^x \simeq \mathcal F_\lambda$ for $\lambda = -x$. As in \cref{sec:iso}, we define two maps
\be
\hat {\mathfrak g}\; : \; \mathsf H \ \rightarrow \ {\rm End\big(\ctwotimes \infty\big)} \qquad {\rm and} \qquad \hat {\mathfrak h}
\; :\; \mathcal F_\lambda \rightarrow  \hat E^x, \qquad (\lambda = -x),
\ee
defined as
\be
\hat {\mathfrak g}(\mathbf 1) = \mathbb I, 
\qquad \ \ 
\hat {\mathfrak g}(a_k) = \sideset{}{'}\sum_{q \ge 0} \hat\theta^-_{-q}\hat\theta^+_{q+k} - \sideset{}{'}\sum_{q > 0} \hat\theta^+_{-q+k}\hat\theta^-_q - \tfrac12 \delta_{k,0}\, 
\delta_{n, 0 \bmod 2}
\ee
and
\be 
\hat {\mathfrak h}\big(a_{k_1}^{n_1}a_{k_2}^{n_2} \dots a_{k_\ell}^{n_\ell} |\lambda\rangle\big) = \hat {\mathfrak g} (a_{k_1})^{n_1} \hat {\mathfrak g}(a_{k_2})^{n_2} \dots \hat {\mathfrak g}(a_{k_\ell})^{n_\ell} |\hat\Omega_x\rangle.
\ee
We find that
\be
[\hat {\mathfrak g}(a_k),\hat {\mathfrak g}(a_\ell)] = k\, \delta_{k+\ell, 0}, \qquad \hat {\mathfrak g}(a_0) = - \mathcal X, \qquad \hat {\mathfrak g}(a_k) |\hat \Omega_x \rangle
 = 0 \qquad (k>0),
\ee
implying that $\hat {\mathfrak g}$ is an $\mathsf H$-homomorphism and $\hat E^x$ a highest-weight $\mathsf H$-module. This is almost the same as
what was found in \cref{sec:iso}. One key distinction is that the constant terms in $\mathfrak g(a_0)$ and $\hat{\mathfrak g}(a_0)$ are different. At the level of $\mathsf H$, 
these constants are irrelevant, as any $\mathsf H$-homomorphism $\mathfrak g$ is defined up to an additive constant in $\mathfrak g(a_0)$. The choice of this constant is, however, 
important in the Virasoro picture. Here, we find that $\Delta^{\textrm{\!\tiny FF}}_\lambda = \Delta^{\textrm{\!\tiny 2f}}_x$ for $\lambda = -x$, and
\be
[\LFF_k,a_q] = -q\, a_{k+q}, \qquad  [\Lff_k, \hat {\mathfrak g}(a_q)] = -q\, \hat {\mathfrak g}(a_{k+q}).
\label{eq:Lkaqc1}
\ee 
We note that modifying the constant in $\hat {\mathfrak g}(a_0)$ would alter the second relation. We now
have all the ingredients to show the next proposition.
\begin{Proposition} 
\label{Prop:C3}
The map $\hat{\mathfrak h}$ is bijective and it is an isomorphism of Virasoro modules.
\end{Proposition}
The proofs of the two statements in \cref{Prop:C3} are identical to those of Propositions \ref{sec:hbij} and \ref{sec:hiso} in \cref{sec:iso}. It follows that $\hat E^x \simeq \mathcal F_{-x}$ and that the module structures of $\hat E^x$ can be read off from \eqref{eq:F11} and \eqref{eq:Firred}, thus completing the proof of \cref{thm:structurethm1}.

%%%%%%%%%%%%%%%%%%%%%
%

\end{document}